\documentclass[useAMS,usenatbib]{mn2e}

\usepackage{threeparttable}

\usepackage{dpfloat}

\usepackage{amssymb}
\usepackage{mathptmx}       
\usepackage{helvet}         
\usepackage{courier}        
\usepackage{type1cm}        
\usepackage{graphicx}        
\usepackage{multicol}        
\usepackage[bottom]{footmisc}

\usepackage{natbib}
\usepackage{aas_macros}  

\newcounter{subfigure}
\usepackage{sidecap}
\usepackage{color}
\usepackage[dvipsnames]{xcolor}

\newcommand{\re}{$\mathrm{R_e}$}
\newcommand{\hb}{H$\beta$}
\newcommand{\hbo}{H$\beta_o$}
\newcommand{\sig}{$\sigma$}
\newcommand{\atlas}{ATLAS$^{\mathrm{3D}}$}
\newcommand{\mgb}{Mg\,$b$}

\title[The hELENa project - I.]{The hELENa project -- I. Stellar populations of early-type galaxies linked with local environment and galaxy mass} 

\author[Agnieszka Sybilska et al.]{A. Sybilska$^{1}$\thanks{E-mail: arys@eso.org}, T. Lisker$^{2}$, H. Kuntschner$^{1}$, A. Vazdekis$^{3,4}$, G. van de Ven$^{5}$, \newauthor R. Peletier$^{6}$,  J. Falc\'{o}n-Barroso$^{3,4}$, R. Vijayaraghavan$^{7,8}$, J. Janz$^{9}$\\ 
$^{1}$European Southern Observatory, Karl-Schwarzschild-Strasse 2, 85748 Garching bei M\"{u}nchen, Germany\\
$^{2}$Astronomisches Rechen-Institut, Zentrum f\"{u}r Astronomie der Universit\"{a}t Heidelberg, M\"{o}nchhofstrasse 12-14, D-69120 Heidelberg, Germany\\
$^{3}$Instituto de Astrof\'isica de Canarias, V\'ia L\'actea s/n, E-38205 La Laguna, Tenerife, Spain\\
$^{4}$Departamento de Astrof\'isica, Universidad de La Laguna, E-38205 La Laguna, Tenerife, Spain\\
$^{5}$Max Planck Institute for Astronomy, K\"{o}nigstuhl 17, 69117 Heidelberg, Germany\\
$^{6}$Kapteyn Astronomical Institute, University of Groningen, Postbus 800, 9700 AV Groningen, the Netherlands\\
$^{7}$NSF Astronomy \& Astrophysics Postdoctoral Fellow\\
$^{8}$Department of Astronomy, University of VIrginia, 530 McCormick Road, Charlottesville, VA 22904, USA\\
$^{9}$Centre for Astrophysics \& Supercomputing, Swinburne University, Hawthorn, VIC 3122, Australia
}

\begin{document}

\date{}


\maketitle

\label{firstpage}

\begin{abstract}
We present the first in a series of papers in T\textbf{h}e role of \textbf{E}nvironment in shaping \textbf{L}ow-mass \textbf{E}arly-type \textbf{N}earby g\textbf{a}laxies (hELENa) project. In this paper we combine our sample of 20 low-mass early types (dEs) with 258 massive early types (ETGs) from the {\atlas} survey -- all observed with the SAURON integral field unit (IFU) -- to investigate early-type galaxies' stellar population scaling relations and the dependence of the population properties on local environment, extended to the low-{\sig} regime of dEs. The ages in our sample show more scatter at lower {\sig} values, indicative of less massive galaxies being affected by the environment to a higher degree. The shape of the age-{\sig} relations for cluster vs. non-cluster galaxies suggests that cluster environment speeds up the placing of galaxies on the red sequence. While the scaling relations are tighter for cluster than for the field/group objects, we find no evidence for a difference in \textit{average} population characteristics of the two samples. We investigate the properties of our sample in the Virgo cluster as a function of number density (rather than simple clustrocentric distance) and find that dE ages negatively correlate with the local density, likely because galaxies in regions of lower density are later arrivals to the cluster or have experienced less pre-processing in groups, and consequently used up their gas reservoir more recently. Overall, dE properties correlate more strongly with density than those of massive ETGs, which was expected as less massive galaxies are more susceptible to external influences. 
\end{abstract}


\section{Introduction}

\begin{figure*}
\centering
\includegraphics[width=0.66\columnwidth]{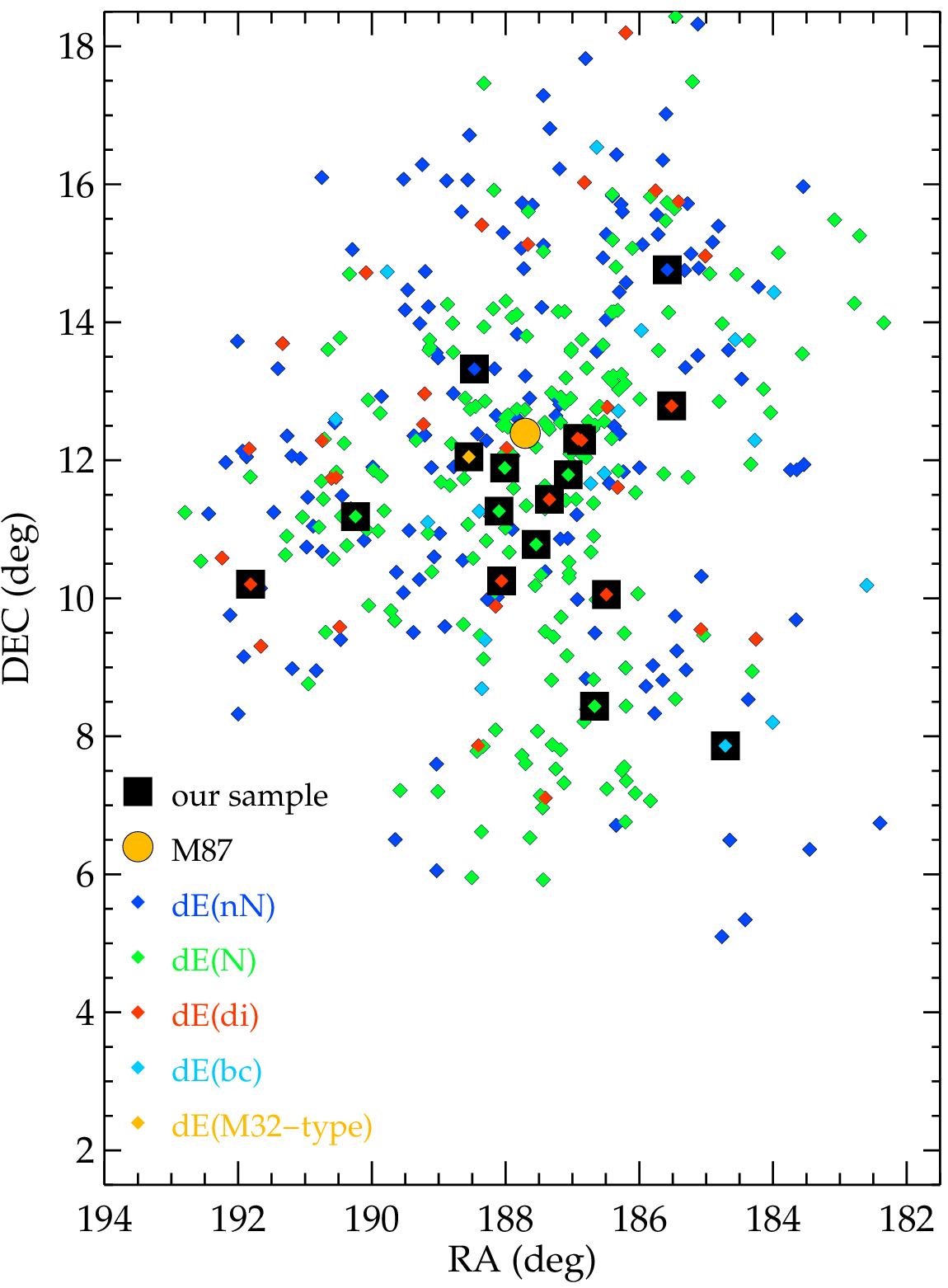}
\includegraphics[width=0.66\columnwidth]{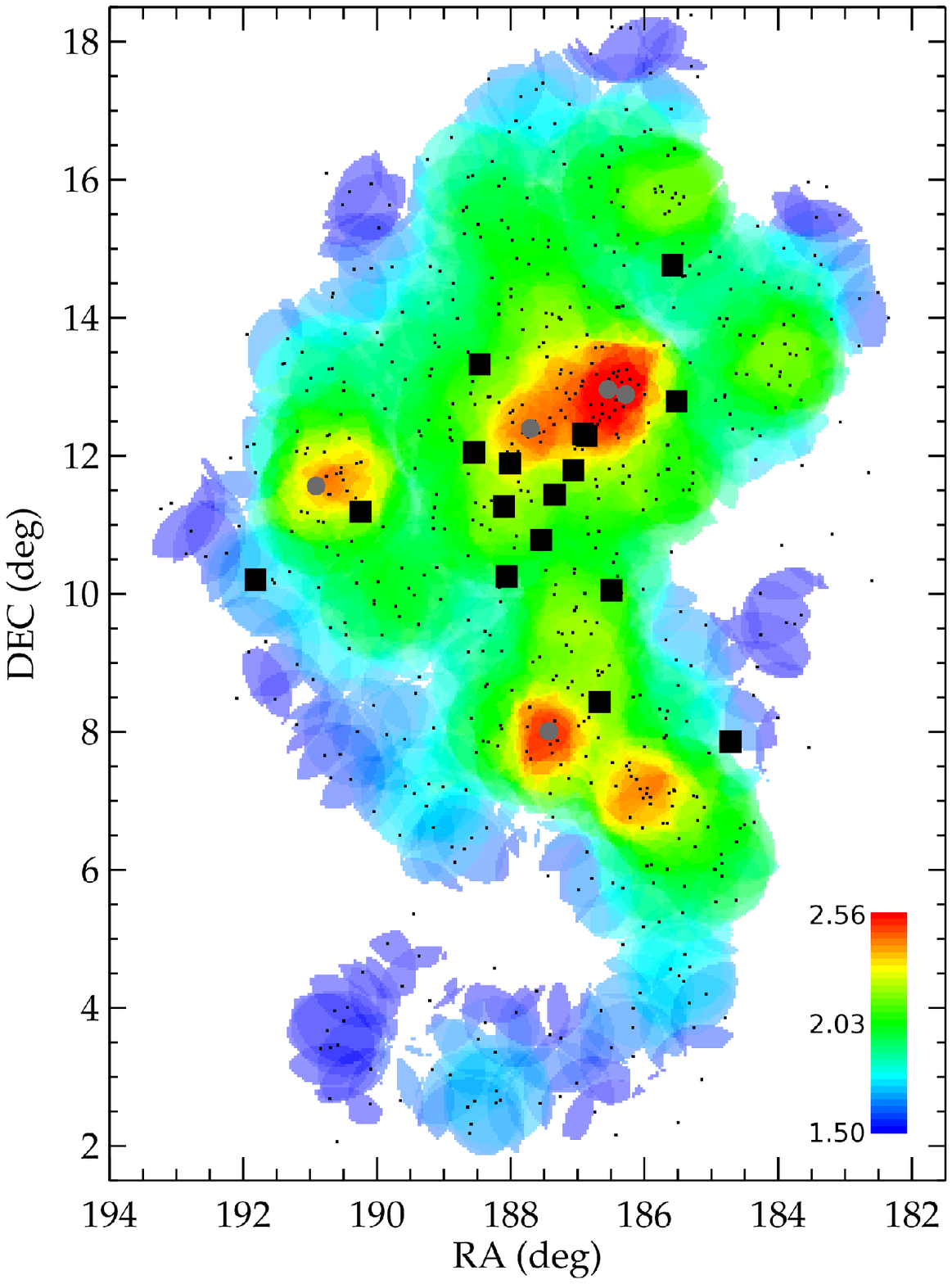}
\includegraphics[width=0.66\columnwidth]{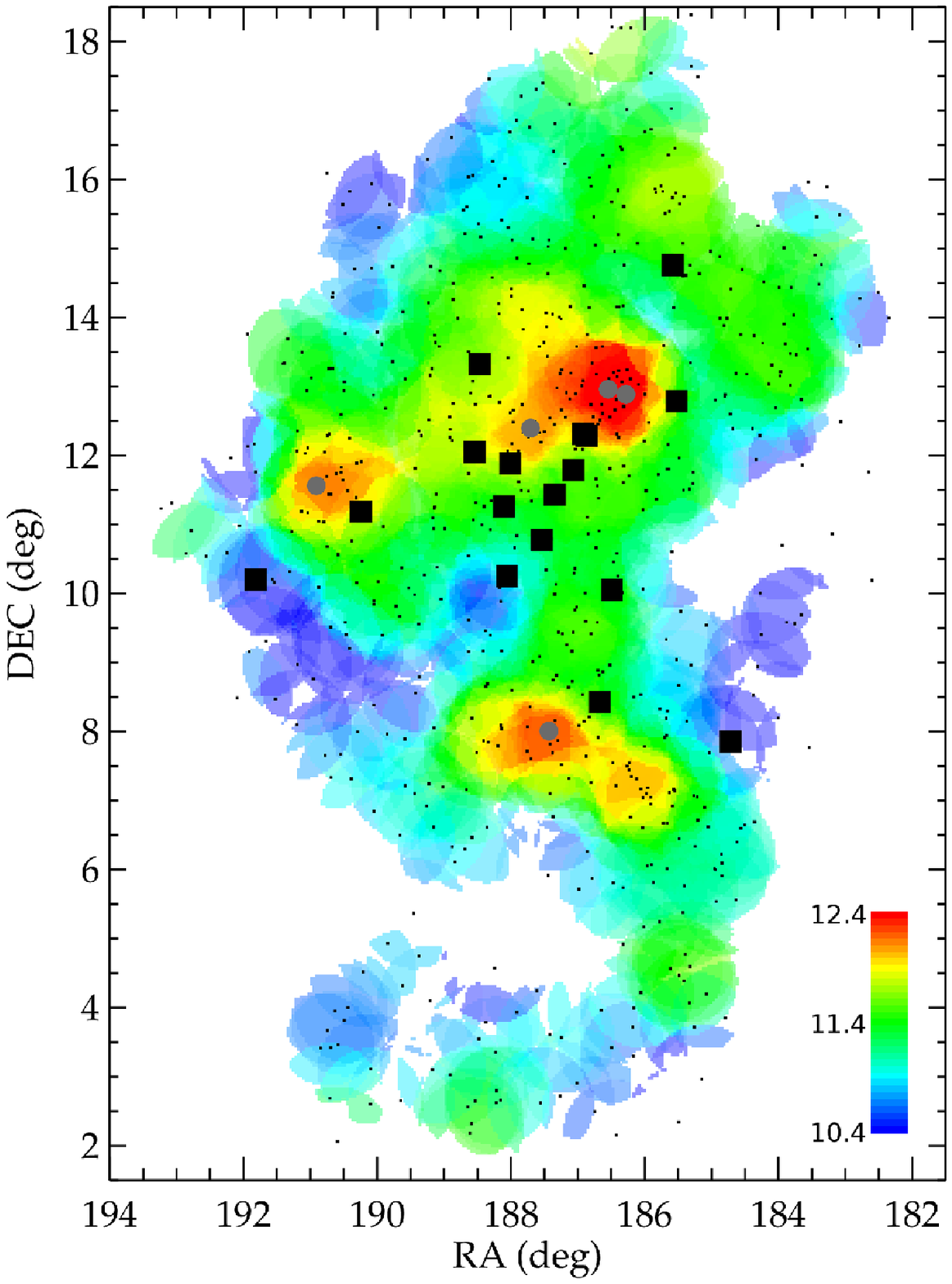}
\caption{\textit{Left:} observed sample (only Virgo objects, i.e. excluding field/group galaxies) plotted on top of the photometric sample of \protect\cite{lisker:2006a}. See the legend for symbols explanation. \textit{Middle and right:} the sample plotted on top of a Virgo number density and luminosity density maps, respectively, produced by including all certain/possible member galaxies down to Mr\,$\leqq$\,-15.2\,mag. Logarithmic density is converted linearly into RGB color (see color bars). Small dots are all the 666 galaxies used in the density calculations (see Sec.~\ref{density-maps}), grey filled circles mark the five brightest early-type galaxies of the cluster (from east to west: M60, M87, M49, M86, M84), and large black squares mark our SAURON sample. Note that the few (nN) galaxies in the lower right corner of the left panel belong to the so called W group (see \protect\citealt{gavazzi:1999}) known to lie far in the background and are as such excluded from the density calculations and maps shown in the remaining two panels.}
\label{sample}  
\end{figure*}

Early-type galaxies (ellipticals and lenticulars) show strong correlation of stellar population parameters and velocity dispersion (e.g. \citealt{peletier:2007}). Compared to late-type galaxies they also contain little gas and dust, whose presence tends to complicate stellar population analysis. The 2000s and early 2010s showed a significant influx of high-quality, large-field IFU spectroscopic data for massive late- and early-types, first through the SAURON survey (\citealt{dezeeuw:2002}) and later the {\atlas} project (\citealt{cappellari:2011}) and the CALIFA survey (\citealt{sanchez:2012}). High signal-to-noise (S/N) IFU data was essential to providing accurate kinematic and stellar population parameters, free from limitations typical of long-slit data such as slit losses, as well as radial information not biased in any particular direction. It also made it possible to compute reliable dynamical properties (dynamical masses and mass-to-light ratios) through mass-modeling of the data (e.g. \citealt{cappellari:2008}). The more recently launched MaNGA (\citealt{bundy:2015}) and SAMI (\citealt{cortese:2014}) IFU surveys have also started significantly adding to our understanding of galaxy formation and evolution. 

The success of the SAURON survey and the fact that the lower-mass regime remained (and still does) largely unexplored, made us initiate a project aimed at obtaining large-field IFU data for dwarf early-types (dEs). The initial sample (presented in \citealt{rys:2013}) consisted of 12 galaxies. During the course of this as well as the related  \textit{SMAKCED} project (e.g. \citealt{janz:2014}, \citealt{toloba:2014}) yet more evidence was found to complement earlier findings (e.g. \citealt{lisker:2006a}, \citealt{toloba:2009}) that the environment plays a significant role in shaping dwarf early-type galaxy properties. Still, the dependence of those properties on the location in the Virgo cluster is a complex one and goes beyond simple projected (or even 3D) clustrocentric distances. Projected clustrocentric distance has been commonly used in the literature as a proxy for the environment density under the assumption that the closer a galaxy appears to be to a cluster's centre, the longer it has resided in it (see, e.g. \citealt{lisker:2013}) and have been subjected to stronger environmental effects. The necessary underlying assumption is that the cluster under study is a relaxed one with local density scaling with the aforementioned distance. This is, however, not true in the case of the Virgo cluster for which the luminosity and number density maps (Fig.~\ref{sample}) clearly show an uneven density distribution within the cluster, with some of the highest density regions clearly offset from the central M87.\looseness-1

Since environmentally-induced galaxy transformation is a combination of time spent under a given influence as well as its strength, we expect certain galaxy properties to scale with the local density, bearing in mind, though, that they can also be influenced by pre-processing in prior environments, e.g. in infalling groups (see, e.g. \citealt{conselice:2001}) or ``post-processing'', where infalling group galaxies experience additional strong ram pressure during orbital pericentre passage because of high infall velocities as well as compressed dense intracluster and/or intragroup medium hot gas (\citealt{vijayaraghavan:2013}). Also, observational studies show different subclasses of cluster galaxies to have differing line-of-sight velocity distributions in addition to varying spatial distribution (e.g. \citealt{derijcke:2010}).

\begin{table*}
\caption{(1) Galaxy names as in the Virgo Cluster Catalog (VCC) of \protect\cite{binggeli:1985}, New General Catalog (NGC), or as a GALEX identification number (ID), (2,3) J2000 coordinates (from \textit{NASA Extragalactic Database (NED)}, http://ned.ipac.caltech.edu/), (4) for Virgo galaxies: projected distances from the cluster's centre \protect\citep{lisker:2007}, (5) morphological types (\protect\citealt{lisker:2007} for the Virgo objects, \protect\citealt{michielsen:2008} / \textit{NED} database for the field/group galaxies), (6) photometric substructure from \protect\citealt{janz:2014}, (7) ellipticities at 1 effective radius from \protect\cite{lisker:2007}, (8) r-band effective radii determined by fitting circular aperture photometry profiles on Sloan Digital Sky Survey (SDSS) images with R$^{1/n}$ growth curves, as detailed in \protect\cite{falcon:2011b},  (9) r-band apparent AB magnitudes, (10) and total on-source exposure times, (11) radial extent of the data in units of {\re}, measured in annuli such that for a given (elliptical) annulus our data covers at least 50\% of the points corresponding to it.} 
 \begin{threeparttable}
\centering
\begin{tabular}{|r|r|r|r|r|r|r|r|r|r|r|}
\hline
object    & RA      & DEC    & M87dist & type    	&substructure?&$\epsilon$	&{\re}	& m$_r$ 	& $t_{exp}$& r$_{max}$ \\ 
          & (deg)   & (deg)  & (deg)   &         	&			  & 			&$('')$	&$(mag)$&(hrs)     & (r/{\re})    \\ 
(1)       &(2)      &(3)     &(4)      &(5)       &(6)          &(7)          &(8)        &(9)   			&(10)      & (11)       \\          
\hline
ID\,0650  &211.0660 & 4.1122  & -     &dE/S0    	&N/A		&0.10	&20.1	&13.73	&6.0& 0.65	\\
ID\,0918  &224.7030 & 2.0235  & -     &dE/E     	&N/A		&0.27	& 6.4	&13.79	&5.0& 2.50	\\
NGC\,3073 &150.2170 & 55.6188 & -     &dE/dS0   	&N/A		&0.15	&16.1	&12.98	&5.5& 1.18	\\  
VCC\,0308 &184.7125 & 7.8642  & 5.42  &dE(di;bc)	&2C, NC, Sp	&0.07	&18.7	&13.14	&5.0& 0.96 	\\  
VCC\,0523 &185.5171 & 12.7844 & 2.21  &dE(di)   	&2C			&0.29	&27.9	&12.52	&5.0& 0.79 	\\  
VCC\,0543 &185.5813 & 14.7556 & 3.17  &dE(nN)   	&2C			&0.46	&19.0	&13.35	&2.5& 0.89 	\\
VCC\,0856 &186.4913 & 10.0661 & 2.63  &dE(di)	  	&1C, NC, Sp	&0.13	&17.0	&13.38	&2.5& 0.88 	\\
VCC\,0929 &186.6688 & 8.4447  & 4.09  &dE(N)    	&1C, NC		&0.11	&22.1	&12.51	&5.0& 0.86 	\\
VCC\,1010 &186.8642 & 12.2908 & 0.84  &dE(di)	  	&Bar		&0.40	&17.9	&12.72	&2.5& 1.17	\\
VCC\,1036 &186.9221 & 12.3114 & 0.78  &dE(di)   	&-			&0.56	&17.2	&12.94	&5.0& 1.40	\\
VCC\,1087 &187.5429 & 10.7694 & 0.88  &dE(N)    	&Lens?, NC	&0.31	&28.6	&12.59	&4.0& 0.66	\\
VCC\,1183 &187.0621 & 11.7875 & 1.02  &dE(di)	  	&Lens, NC	&0.40	&18.6	&13.27	&2.5& 0.81	\\
VCC\,1261 &187.3438 & 11.4339 & 1.62  &dE(N)    	&2C, NC		&0.42	&19.7	&12.62	&5.0& 1.22	\\
VCC\,1407 &188.0113 & 11.8841 & 0.59  &dE(N)	  	&1C, NC		&0.17	&11.7	&14.14	&2.5& 0.94	\\
VCC\,1422 &188.0592 & 10.2539 & 6.75  &dE(di)	  	&2C			&0.13	&23.5	&12.78	&2.5& 0.68	\\
VCC\,1431 &188.0979 & 11.2564 & 1.19  &dE(N)    	&1C, NC		&0.03	& 9.6	&13.37	&5.0& 1.56	\\
VCC\,1528 &188.4650 & 13.3311 & 1.20  &dE(nN)	  	&2C			&0.09	&11.2	&13.67	&2.0& 1.07	\\
VCC\,1545 &188.5479 & 12.0367 & 0.89  &M32-type 	&2C, NC		&0.30	&11.9	&13.97	&2.5& 0.80	\\
VCC\,1861 &190.2442 & 11.1997 & 2.79  &dE(N)    	&Lens,NC	&0.01	&20.1	&13.22	&4.5& 1.33	\\
VCC\,2048 &191.8138 & 10.2042 & 4.63  &dE(di)   	&-			&0.48	&16.5	&12.99	&5.0& 1.01	\\
\hline
\end{tabular}
\label{observations} 
\textit{Notes to column (7)}: NC: nucleus component used in decomposition; 1C: 1-component; 2C: 2-component; Sp: hints of spiral arms in residual.
\end{threeparttable}
\end{table*}

The environment in which a galaxy evolves affects the rate of star formation (SF), through quenching (the removal of gas that can form stars), star formation induced by compression due to ram pressure forces (\citealt{tonnesen:2009}), or -- possibly -- inducing a SF burst through compressed gas inflow to the galaxy centre in the case of tidally harassed galaxies. We are therefore interested in studying the relation of population gradients not only comparing galaxies of various masses but also investigating the environment in which they find themselves. That massive early-types predominantly exhibit negative metallicity gradients has been established in the literature (e.g. \citealt{kuntschner:2010}, \citealt{loubser:2012}, \citealt{oliva-Altamirano:2015}, \citealt{wilkinson:2015}). However, no such firm consensus appears to exist for low-mass galaxies. \cite{denbrok:2011} showed that color gradients (primarily reflecting metallicity gradients if an outside-in formation scenario is assumed) are continuous between early-type galaxies in the Coma cluster across the entire probed mass range, with dwarfs having nearly flat to mildly negative metallicity gradients. \cite{koleva:2009b}, on the other hand, from their long-slit spectroscopic dataset including (mostly) Fornax cluster dEs, find strongly negative gradients, with the exception of disky galaxies, which have nearly flat profiles. In \cite{rys:2015} we show that the majority of studied dEs have close to null gradients.  

In this work we aim to study the efficiency of star formation across cosmic time in our low-mass early-type galaxies and also see how it changes with increasing mass. We use the available line-strength indices as well as the kinematic results for the sample to investigate stellar populations gradients and 1\,{\re} integrated values of dEs in search of similarities and/or differences within the sample. Subsequently, we place our dEs on the scaling relations of giant early types, as well as investigate the dependence of their stellar population properties on the local environment. We aim to look for (dis)similarities between the properties of early-type objects across the largest to date mass range.

The paper is structured as follows. A summary of the sample selection, observations, and data reduction is presented in section 2. In section 3 we describe the methods used in the analysis. Section 4 presents our results which are then discussed in section 5. Section~6 provides a summary of our findings.

\section[Data]{Data}
\label{label:data}

\begin{figure*}
\begin{leftfullpage}
\begin{center}
\hspace{0.3cm}Intensity \hspace{1.5cm} V \hspace{1.9cm} {\sig} \hspace{1.9cm} {\hb} \hspace{1.9cm} Fe5015 \hspace{1.5cm} {\mgb}
\includegraphics[width=0.89\textwidth]{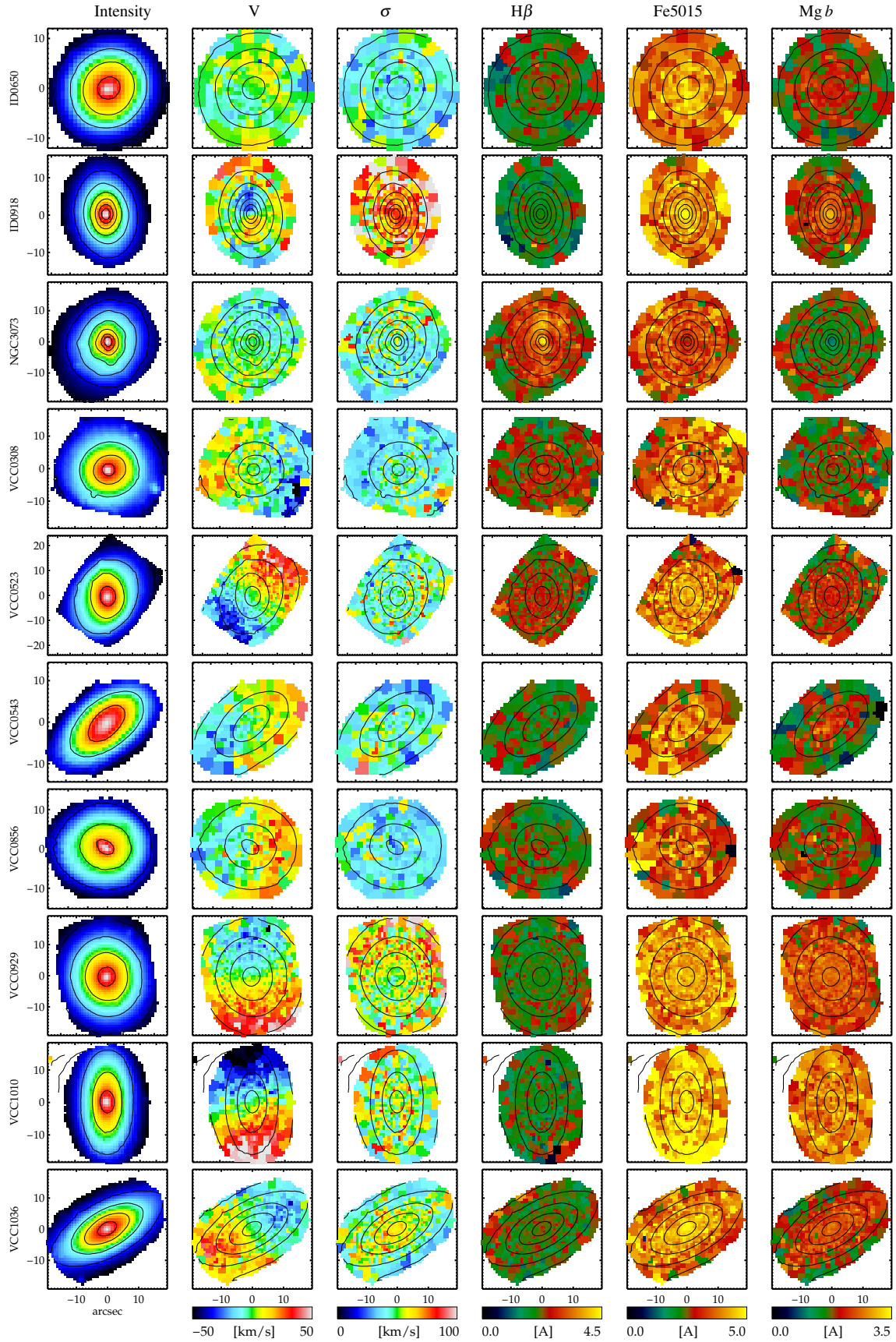}
\end{center}
\caption{Intensity, stellar velocity, velocity dispersion, \hb, Fe5015, and {\mgb} lines maps. The maps shown here comprise the re-analyzed sample of \protect\cite{rys:2013} together with the new data presented here for the first time.}
\label{kinlsmaps}
\end{leftfullpage}
\end{figure*}

\addtocounter{figure}{-1}
\addtocounter{subfigure}{1}

\begin{figure*}
\begin{fullpage}
\begin{center}
\hspace{0.3cm}Intensity \hspace{1.5cm} V \hspace{1.9cm} {\sig} \hspace{1.9cm} {\hb} \hspace{1.9cm} Fe5015 \hspace{1.5cm} {\mgb}
  \includegraphics[width=0.89\textwidth]{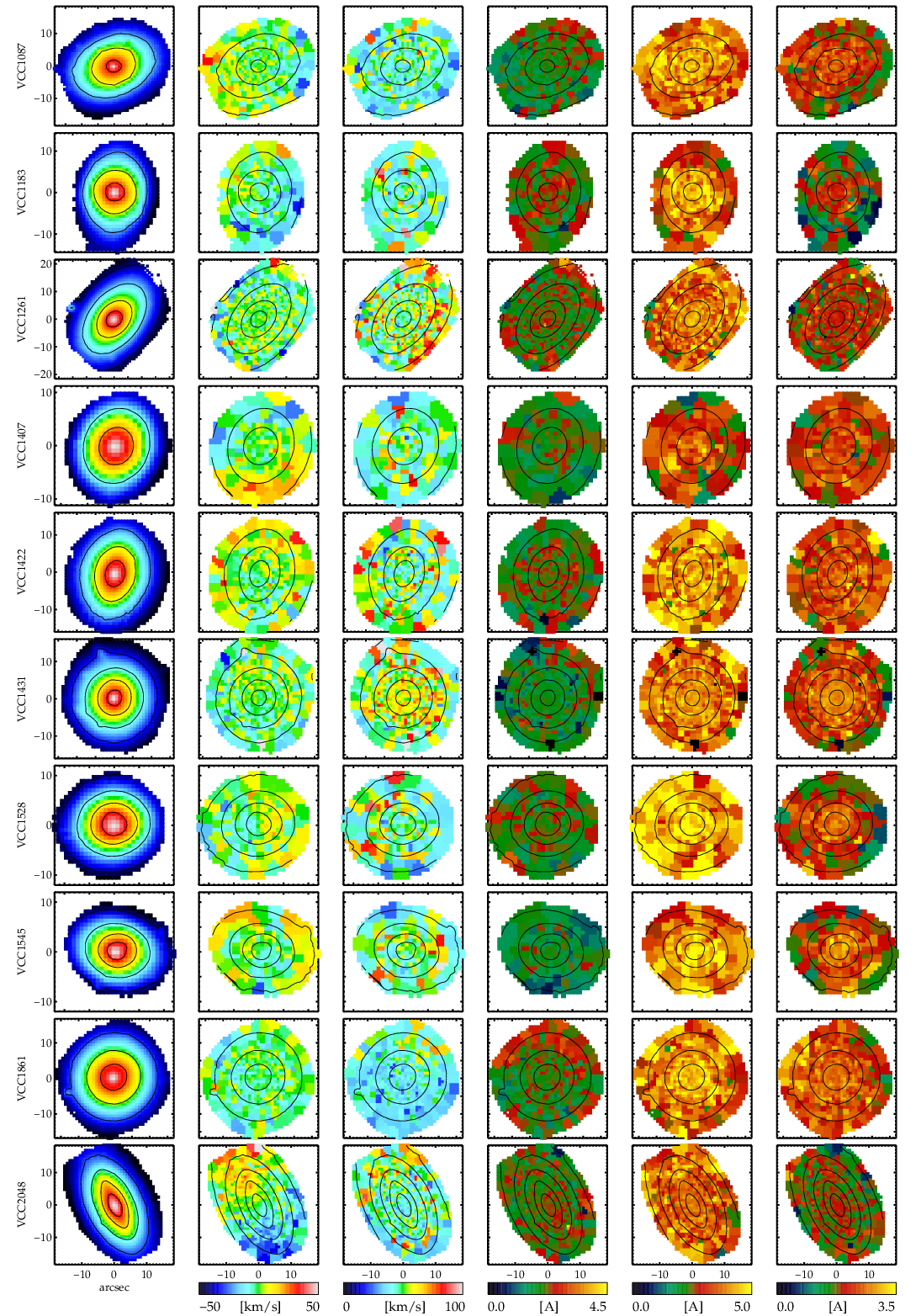}
\end{center}
\end{fullpage}
\caption{Continued.}
\end{figure*}

Our sample combines the re-analyzed dataset of \cite{rys:2013} with a new sample of eight Virgo dEs, obtained -- also with the SAURON IFU -- in 2014. The stellar kinematic and line-strength maps of the combined sample of 20 dEs are presented here (Fig.~\ref{kinlsmaps}. Table~1 gives information on the location, photometric parameters and observing details for all galaxies. In Fig.~\ref{sample} we show the location of our galaxies in the Virgo cluster. The plot on the left shows the sample distribution color-coded by subtype, with the sample of \cite{lisker:2006a} overplotted. The middle and right plots show the locations of our dEs on a number density and a luminosity density map, respectively (see section~\ref{density-maps} for details). 

Details on data selection, observations and data reduction of the \cite{rys:2013} sample are presented therein. The eight galaxies presented here for the first time have been chosen to provide a better 2D spatial coverage of the Virgo cluster, with preference given to objects for which line-of-sight distances were available. The sample was observed in April 2014 during four nights, with each galaxy typically exposed for 2.5\,hrs. This meant a 50\% decrease in exposure time as compared to the sample of \cite{rys:2013}. We did so to maximize the number of observed galaxies, having earlier performed checks on already available data and comparing kinematics obtained using 5\,hrs and 2.5\,hrs exposure times (combining ten and five 0.5\,hr frames, respectively). Keeping the same strict criteria for minimum S/N per spaxel, we do lose some spatial coverage when compared with the 5\,hr exposure time -- thanks to that, however, we make sure we recover highest-quality kinematic and line-strength measurements.

For the extraction and calibration of the data we followed the procedures described in \cite{bacon:2001} using the specifically designed {\sc xsauron} software developed at the Centre de Recherche Astrophysique de Lyon (CRAL). As described in \cite{rys:2013}, we filtered out all individual spaxels with S/N\,$<$\,7 to make sure that our final binned spectra were not contaminated by low-quality measurements. The value was chosen as a result of having run a series of tests, aimed to exclude regions where the spread of measured bin values exceeded the actual measurement errors, and roughly corresponds to surface brightness of $\mu_V\approx23.5$\,mag at the edge of the fields. The final maps extend on average out to 0.96~effective radii (min: 0.65, max: 2.50\,{\re}, see column 11 of Table~\ref{observations}).

\section{Methods}

\subsection{Density maps}
\label{density-maps}
To produce the density maps shown in Fig.~\ref{sample} we have used all 666 Virgo cluster galaxies that are certain and possible cluster members (based on the Virgo Cluster Catalog of \citealt{binggeli:1985} but with membership updates as outlined in \citealt{weinmann:2011}), are brighter than an r-band magnitude of 16.0, and do not meet the position \emph{and} velocity criterion of \cite{gavazzi:1999} for belonging to the M or W cloud behind Virgo. Projected number density is derived for each galaxy by adopting the distance to the 10$\mathrm{^{th}}$ nearest neighbor galaxy as  the radius of a circle, and dividing 11 by the area of that circle in Mpc$^2$ (\citealt{dressler:1980}). For the projected luminosity density, we first convert the absolute r-band magnitudes of all galaxies to solar luminosities (with $M_{r,\odot}=4.68$ mag, \citealt{sparke:2006}), using a distance modulus of $m-M=31.09$ ($d=16.5$ Mpc, \citealt{blakeslee:2009}) or, where available, an individual distance published from surface brightness fluctuation analyses (\citealt{blakeslee:2009}, \citealt{jerjen:2004}, \citealt{tonry:2001}). We then divide, for each galaxy, the total luminosity within a circle of 0.2 Mpc projected radius by the area of that circle. At each position of the diagram, we compute the average of the logarithmic density from at least 5 and at most 12 galaxies within a circle of at most 0.8 degrees. The circle is smaller if the number 12 is reached earlier, allowing a better resolution. If the minimum number of 5 is not reached, the point is left white.
 
\subsection{Stellar kinematics}
\label{methods-stekin}

Stellar absorption-line kinematics were derived for each galaxy by directly fitting the spectra in pixel space using the penalized pixel-fitting method (pPXF) of \cite{cappellari:2004}. The method fits a stellar template spectrum convolved with a line-of-sight velocity dispersion (LOSVD) to the observed galaxy spectrum in pixel space, logarithmically binned in wavelength. The algorithm finds the best fit to the galaxy spectrum and returns the mean velocity $V$, velocity dispersion \sig, as well as higher order Gauss-Hermite moments $h_3$ and $h_4$, if requested. The procedure of creating an optimal stellar template was repeated for each bin using MILES single stellar population (SSP) scaled-solar models of \cite{vazdekis:2015} which cover an age range of 0.03-14\,Gyr and a metallicity range of -2.27 to +0.40. We binned our data to reach a signal-to-noise ratio of 30/{\AA} to ensure reliable velocity dispersion recovery (see section 3.2 of \citealt{rys:2013}). Error estimates were obtained through performing Monte-Carlo (MC) simulations and measuring kinematics of the different realizations of the input spectra with added noise.

\subsection{Line-strength measurements}
\label{ls-mes}

While recovering reliable stellar kinematic information requires high signal-to-noise ratios (see e.g. \citealt{toloba:2009} or \citealt{rys:2013}), imperfections in the stellar continuum shape (resulting from poorer data quality or imperfect data reduction) can be dealt with by applying additive polynomials. Line strength (LS) analysis, however, requires most careful tools to be applied to high quality data: only when the continuum has been accurately determined is the derived LS information reliable. As a result, few dE studies have attempted to go that route beyond obtaining central/integrated values for low-mass galaxies, and those that did try to provide spatially resolved LS results often show high uncertainties in the derived parameters (\citealt{michielsen:2008}). The quality of the sample presented here has made the task possible and allows us to show not only LS integrated values but also radial profiles and maps. 

To perform our line-strength analysis we use the Line Index System (LIS) of \cite{vazdekis:2010} and the MILES models of \citealt{vazdekis:2015} which are based on the MILES stellar library presented in \cite{sanchez-blazquez:2006} and \cite{falcon:2011a}, containing 985 stars covering a wide parameter space (T$_{eff}$, log (g), [Fe/H]), particularly at lower metallicities.\looseness-2

When observing Virgo objects with SAURON we have 3 traditional Lick/IDS indices available: \hb, {\mgb} and Fe5015, of which {\hb} is the  age-sensitive index and the other two are metallicity indicators. We broadened our data to match the LIS 5\,\AA\,resolution and then calculated the index values. The errors on the indices were calculated through Monte-Carlo simulations according to the method described in \cite{rys:2013}. For each realization, we perturbed the spectra to incorporate the following error sources: $V$ measurement, which affects the position of the lines; {\sig} measurement, which affects the lines' depth; and systematic errors (using as the error spectrum the difference between the data and the best-fit model spectra).\looseness-1

To correct for the remaining imperfections in the continuum shape we followed the approach of \cite{kuntschner:2006}. We applied a continuum correction (see their section~3.1.2 for details) and to account for the resulting added uncertainties in the measurements added a constant value of 0.1\,{\AA} to our error estimates for the {\hb} and {\hbo} indices (for comparison: the errors after the correction for the integrated values are in the 0.16-0.37 range, see Table \ref{resultstable}). Before the LS calculations were performed, we also ran the {\sc gandalf} software of \cite{sarzi:2006} to clean the data from the effect of the emission lines. The only significant {\hb} emission is present in one of our field galaxies (NGC\,3073, see Fig.~\ref{example-spectra}), with a few more objects showing moderate amounts of {\hb} emission and/or presence of [O{\textsc{iii}}] lines. The majority of galaxies are essentially emission-free.

\subsection{Aperture measurements}

\begin{figure}
\begin{center}
\includegraphics[width=0.91\columnwidth]{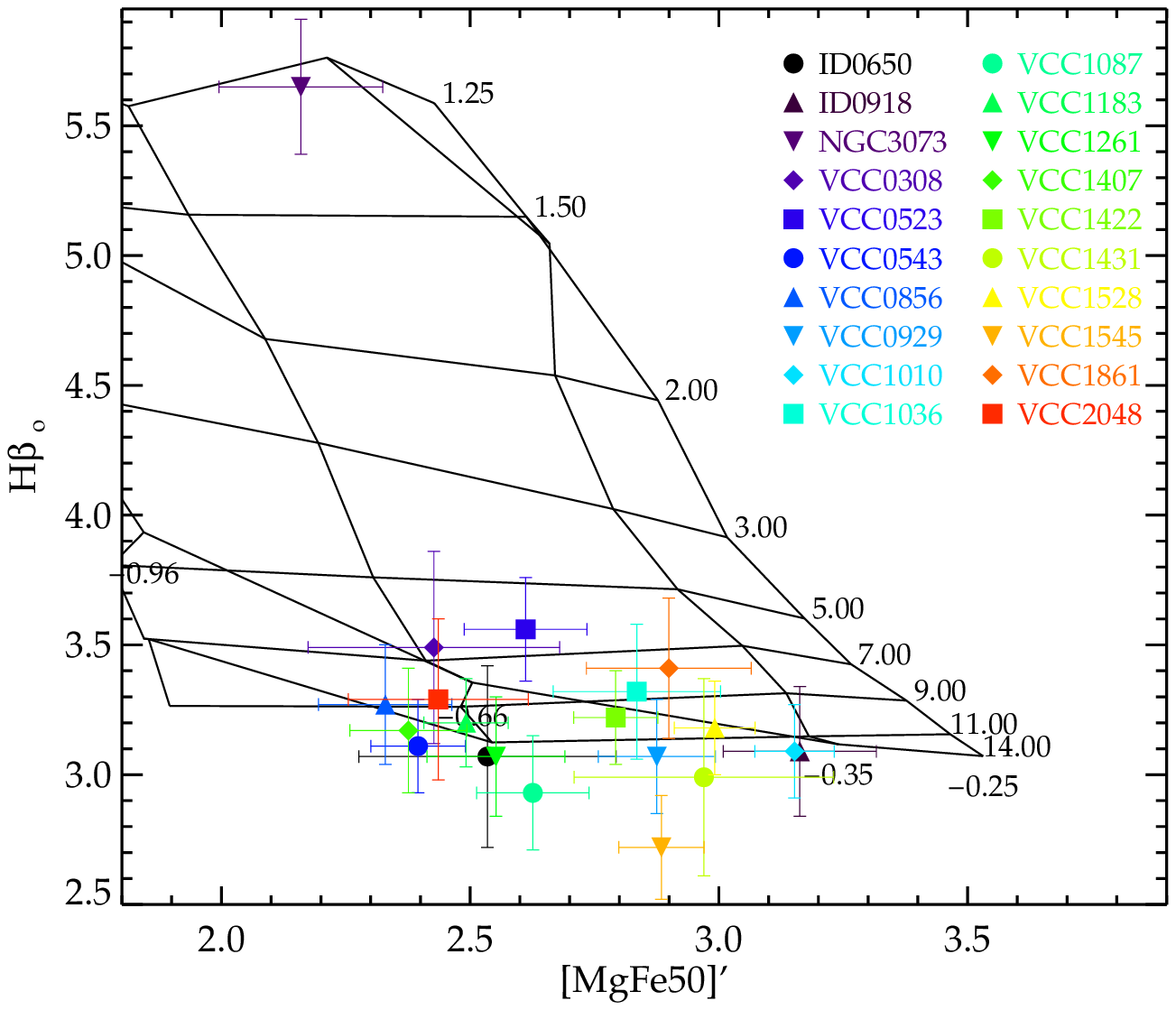}
\includegraphics[width=0.91\columnwidth]{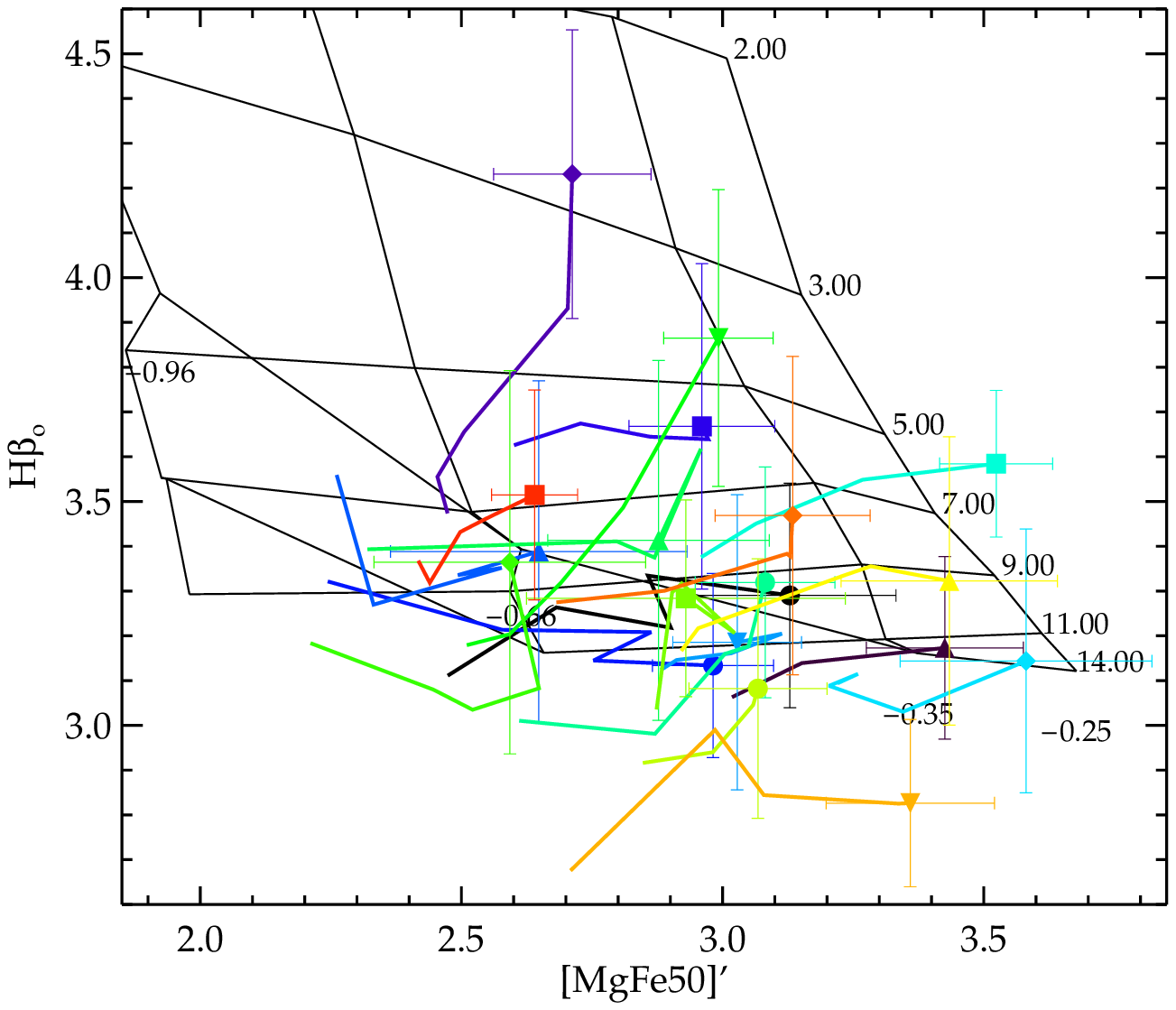}
\includegraphics[width=0.91\columnwidth]{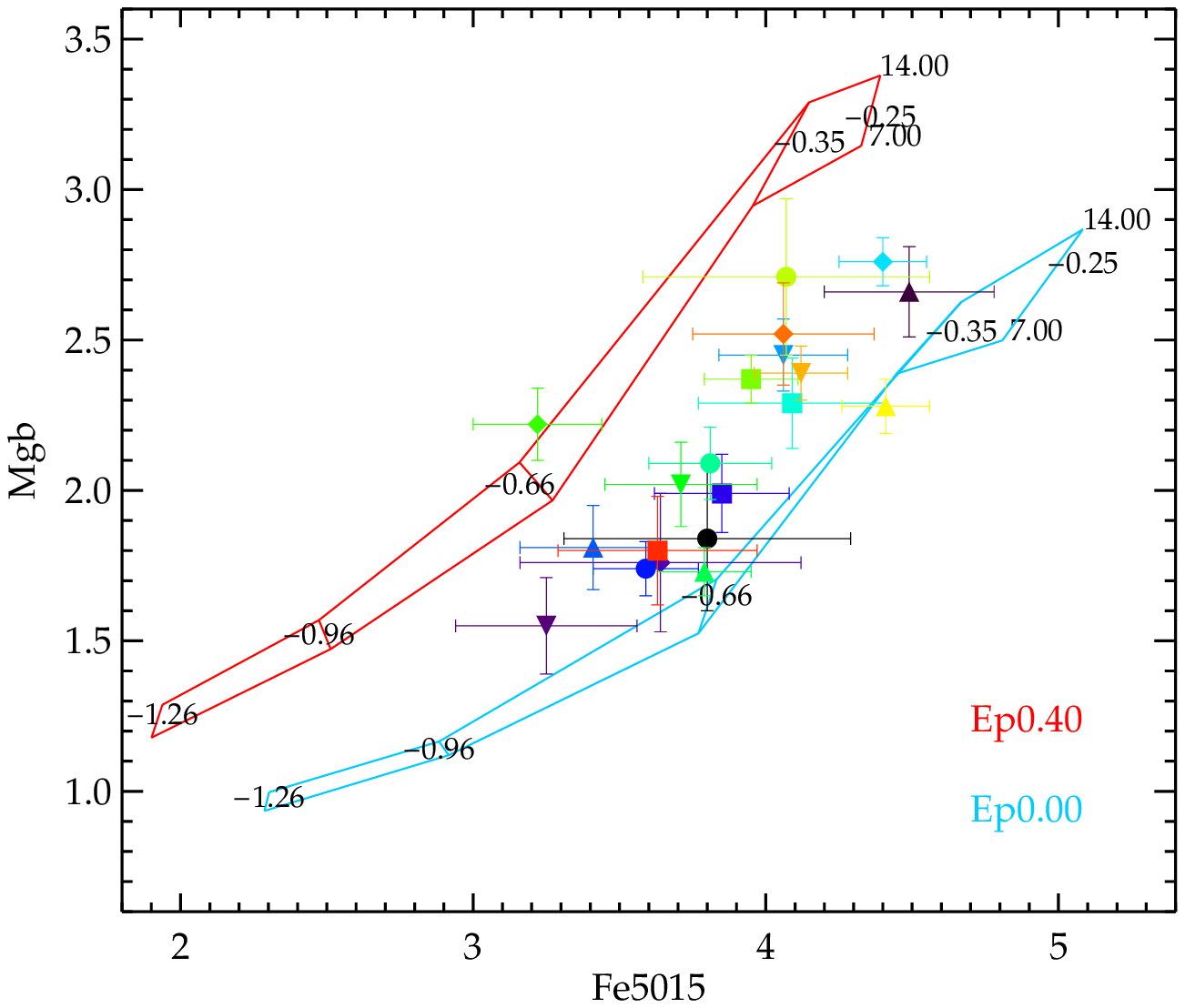}
\end{center}
\caption{\textit{Top and middle:} {\hbo} vs. [MgFe50]' line-strength 1\,{\re} values and profiles, respectively, overplotted on the grid of MILES SSP models predictions. For the profiles the filled symbols indicate central values and the lines connect index values for different radial bins in a given galaxy. Note that in the middle panel the y-range has been restricted for clarity, hence the NGC\,3073 profile is not shown. \textit{Bottom:} {\mgb} vs Fe5015 index relation: SSP scaled-solar (blue, lower grid) and $\alpha$-enhanced (red, upper grid) models are overplotted for example 7 and 14\,Gyr ages and -1.26 to -0.25 metallicites. Note that the plotted ages are only indicative of typical values of the sample and the actual [Mg/Fe] ratios for each object or radius (in case of profiles) were calculated using the best-fit interpolated age and metallicity values. See Sec.~\ref{mgfe-calc} for details.}
\label{grids}
\end{figure}

We have derived our integrated aperture measurements by averaging (luminosity-weighted) spectra within 1\,{\re} apertures, and remeasuring line-strength indices. The  aperture values are used to place our galaxies on the scaling relations of giant early type galaxies from the ATLAS$^{3D}$ survey (\citealt{mcdermid:2015}, Kuntschner et al., in prep.).

For those of our objects for which the data do not extend out to 1\,{\re} we applied an aperture correction. We calculated integrated values at a few distances below 1 {\re} (for multiples of 1/8\,\re) and then fitted a linear function of the form $y = A + B \times x$ to the log(r)-log(value) relation. The values at 1\,{\re} were obtained directly by taking the offset $A$ of the fitted relation and were determined separately for each index and object. 

The median of the relative correction (i.e. $(val_{uncorr}-val_{corr})/val_{uncorr}$, where $val_{uncorr}$ is the measured value at maximum available radial extent and $val_{corr}$ is the calculated value at 1\,\re) was less than 2\% for {\sig} and less than 1\% for the three indices -- well below the measurement errors for those quantities. This is both due to the fact that the values in question are measured for radii not much below 1\,\re, as well as because the relations themselves are relatively flat (cf. Fig.~\ref{population_gradients1}).

\subsection{Radial profiles and gradients}

To obtain radial profiles of {\sig} and line strengths we averaged the bin values in elliptical annuli (i.e. along the lines of constant surface brightness) of increasing width, equal in log space. The same method was used by \citeauthor{kuntschner:2006} (\citeyear{kuntschner:2006}, \citeyear{kuntschner:2010}) for the SAURON survey and subsequently by \cite{scott:2013} and Kuntschner et al. (in prep.) for the ATLAS$^{3D}$ sample, with which we compare our results here. The errors on the indices were obtained by taking a standard deviation of the values corresponding to the individual bins included in a given annulus.

Gradients were then obtained assuming a linear relation between the log(age)\,/\,[M/H]\,/\,[Mg/Fe] and the radius $log(r/R_{e})$, i.e.: 
\begin{equation}
 value[r/R_e] = value_{R_e} + \nabla_{value} \cdot log(r/R_e).
\end{equation}
We used the observed points between 1\,arcsec and the effective radius or the farthest available data point, whichever was applicable.

\subsection{Derived stellar population parameters}

When calculating LS-derived stellar population parameters we use the abundance-ratio insensitive index combination [MgFe50]' of \cite{kuntschner:2010} as well as the optimized {\hbo} index defined by \cite{cervantes:2009} which is less dependent on metallicity than the traditional {\hb} index. We then derive age and metallicity ([M/H]) with the help of MILES stellar population models of \cite{vazdekis:2015}, linearly interpolating between the available model predictions. The index values for integrated 1\,{\re} apertures as well as the derived profiles overplotted on the model predictions are shown in Fig.~\ref{grids}. To reduce the effects of grid discretization, we oversample the original models using a  linear interpolation in the age-[M/H]-[index] space, creating a finer grid with 0.05\,Gyr step in age and 0.01\,dex step in metallicity. Then, we obtained best-fit population parameters by effectively computing the ``distance'' from our measured indices to all the predicted values of those indices on the finer grid, and finding the age-[M/H] combination with the minimum total distance.

We note here that the MILES models are degenerate for ages below ca. 7-11\,Gyr (depending on metallicity) at the typical [M/H] of our objects (see top and middle panel of Fig.~\ref{grids}). This necessarily translates into larger error bars in age and age gradients for points falling within the affected region. It does not affect the derived [M/H] or [Mg/Fe] values. Those points that are outside of the grid on the low-{\hb} end are consistent with the MILES models predictions within a $\sim 3 \sigma$ certainty level. 

\subsection{Determining [Mg/Fe] abundance ratios}
\label{mgfe-calc}

In this work we obtain [Mg/Fe] abundance ratios directly by using MILES scaled-solar and $\alpha$-enhanced models of \cite{vazdekis:2015}  and interpolating between the predictions of both models on the {\mgb}\,-\,Fe5015 plane. This allows us to not only more accurately assess total metallicities but also get insight into the lengths of the star formation episodes as the production of different elements requires different formation timescales. 

For each line strength measurement we extracted {\mgb} and Fe5015 index pairs from both sets of models, corresponding to the best-fitting SSP age and metallicity derived in an earlier step. We then calculated the distance between the two points as well as the distance between the points and our measured value to obtain the estimate of the [Mg/Fe] enhancement. A similar approach was applied in e.g. \cite{thomas:2005}. See the bottom panel of Fig.~\ref{grids} for a depiction of the method. 

\subsection{Virial and stellar masses}

To compute estimates of the dynamical masses for the dE sample, we used the formula of \cite{cappellari:2006} where a virial mass can be estimated using the following formula:
\begin{equation}
M_{vir} = 5.0 \cdot \sigma_e^2 \cdot R_e/G
\end{equation}
with 5.0 being a scaling factor shown to also be applicable to low-mass early type galaxy by \cite{rys:2014}. Dynamical masses for the {\atlas} sample were taken from \cite{cappellari:2013a}.

To obtain stellar mass estimates for both samples, we used the photometric tables based on the MILES models predictions available at the project's website\footnote{http://miles.iac.es}, where for each calculated age and metallicity pair a stellar mass estimate can be read off as a fraction of the total galaxy mass. 

\subsection{ATLAS$^{3D}$ data} 

In order to ensure that the values presented in the paper -- our dEs as well as the {\atlas} sample -- are derived in as homogeneous a way as possible, we adpoted the following strategy. 

We first transformed the {\atlas} line-strength measurements to the LIS system used in our own analysis. To this end we applied the following correction to the {\atlas} measurements:
\begin{equation}
I_{LIS} = a_0 + a_1 \cdot I_{Lick/IDS} + a_2 \cdot I_{Lick/IDS}^2 + a_3 \cdot I_{Lick/IDS}^3
\end{equation}
where in the case of a transformation to the LIS-5.0 {\AA} resolution (used in our analysis of the dE sample) for $H\beta$ the four coefficients are (0.116, 1.001, -0.003, 0.001), for Fe5015 -- (-0.154, 1.162, -0.025, -), and for {\mgb} -- (0.116, 0.950, 0.009, -0.001) -- see the project's website for more details. We note  that the main contribution to the difference between the systems comes from the offset of the relation, thus the effect on the gradient analysis is minimal and errors introduced by comparing the two samples are well within the measurement errors for the line-strength values themselves. 

Subsequently, we used the same MILES models of \cite{vazdekis:2015} and the derivation methods as in the case of our dE data to derive SSP ages, metallicities and abundance ratios for the {\atlas} sample. For reference, in App.~\ref{appendix-comparison} and Fig.~\ref{miles-schiavon} we show a comparison of the resulting values presented in the current work with the published ones from \cite{mcdermid:2015}, where the offsets and errors as well as trend sources  in the comparison are described in detail. We note here that the offsets between the two results seen in the figure are predominantly due to the differences in the stellar populations models used (MILES used in this work vs. Schiavon~\citeyear{schiavon:2007} used by \citealt{mcdermid:2015}) and not the Lick-to-LIS transformation.

\section{Results}

\begin{figure*}
\centering
\includegraphics[width=2.09\columnwidth]{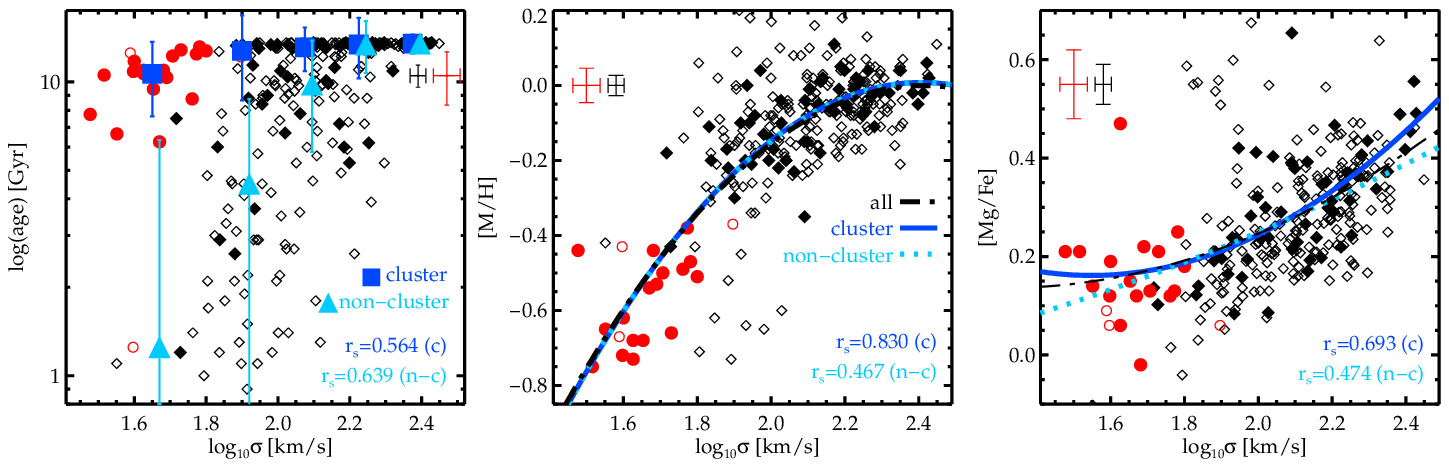}
\includegraphics[width=2.09\columnwidth]{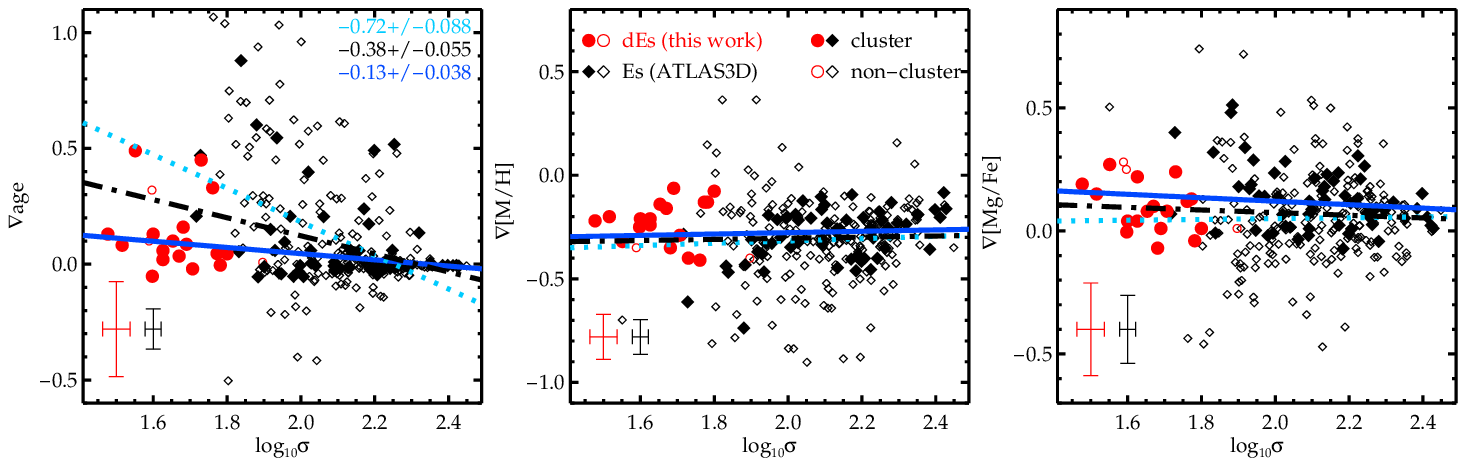}
\hspace{0.5cm}
\includegraphics[width=1.0\columnwidth]{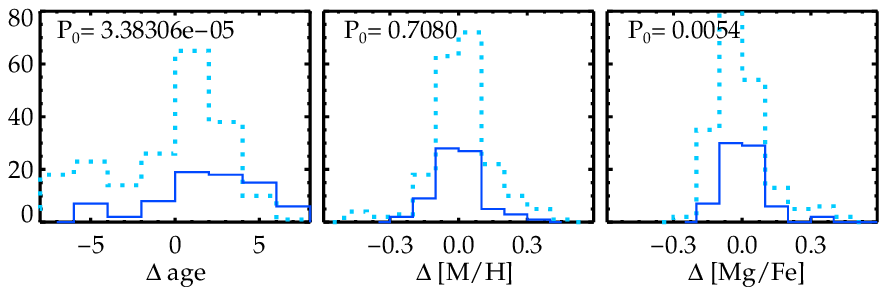}
\includegraphics[width=1.0\columnwidth]{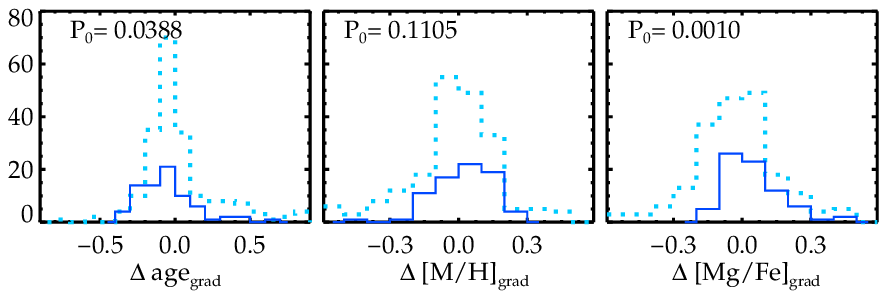}
\centering
\vspace{0.25cm}
\begin{tabular}{|r|r|r|r|r|r|r|}
\hline
 & age & [M/H] & [Mg/Fe] & $\nabla$age & $\nabla$[M/H] & $\nabla$[Mg/Fe] \\
cluster sample rms    & 2.909 & 0.103 & 0.093 & 0.179 & 0.125 & 0.121 \\
non-cluster sample rms & 3.410 & 0.181 & 0.128 & 0.239 & 0.199 & 0.213 \\
\hline
\end{tabular}

\caption{SSP-derived stellar population parameters (age, [M/H] and [Mg/Fe]) vs velocity dispersion \sig: integrated values in the top row and their gradients in the middle row. Filled symbols indicate Virgo cluster objects and open ones - non-cluster, i.e. field/group galaxies. Red and black crosses indicate average error bars for the dE and ETG samples, respectively. In the age-{\sig} panel we show median age values and their associated standard deviations for five log({\sig}) bins (1.50-1.80, 1.80-2.00, 2.00-2.15, 2.15-2.30, 2.30-2.45) for cluster (dark-blue squares) and non-cluster (light-blue triangles) objects, with the latter slightly shifted in x to avoid overlap. Best straight-line fits to the cluster (solid dark-blue), non-cluster (dotted light-blue), and all (dotted-dashed black) objects are shown for the gradient relations in the bottom row and second-degree polynomial fits are shown for the [M/H]-{\sig} and [Mg/Fe]-{\sig} relations. Additionally, slopes of the fitted lines are quoted with their errors for $\nabla$age-{\sig} in the upper right corner of the panel. See App.~\ref{appendix-comparison} for a discussion on the saturation at maximum ages allowed by the MILES models. Spearman correlation coefficients for cluster (c) and non-cluster (n-c) samples are quoted for the integrated relations in the upper panels. The bottom row histograms show residuals of best-fitting lines to all galaxies, separated into cluster and non-cluster objects; line- and color-coding as in the panels above. $P_0$ values give the probabilities that the two histograms are drawn from the same population, as derived from a K-S test. Values of $P_0>0.05$ are typically assumed to support the hypothesis that the two samples come from the same distribution. The table at the bottom of the figure lists rms deviations of the residuals around the respective best-fit relations as a measure of the amount of scatter present in the relations.} 
\label{SSP_relation}  
\end{figure*}

Our dE maps extend out to a maximum of  2.5\,{\re}, with a median at 0.96\,{\re} (see Table~\ref{sample} for individual values for each galaxy). This puts us on par with the ATLAS$^{3D}$ sample of 258~ massive early type galaxies in terms of radial coverage, so we combine the two samples to study early-type galaxies' scaling relations.

We would like to briefly comment on the possible selection effects (or lack thereof) in the combined samples. The ATLAS$^{3D}$ parent sample is complete out to D\,=\,42\,Mpc for ETG galaxies brighter than $M_K$=\,-21.5\,mag (restricted in declination to $|\delta-29^o| < 35^o$ due to observing site location, see \citealt{cappellari:2011}). Our dE sample is, on the other hand, at best representative of the population of the brightest dEs in the Virgo cluster, chosen to cover varying clustrocentric distances, ellipticities and internal features. In the case of both samples, belonging to the early-type class was decided based on morphology: \citeauthor{lisker:2006a} (\citeyear{lisker:2006a}, themselves basing the nomenclature on \citealt{binggeli:1985}) in the case of dEs and a visual inspection of SDSS DR7 (or when lacking -- DSS) images for the ATLAS$^{3D}$ sample. As for the latter, other characteristics such as changing bulge/disk ratio were ignored when separating the ETGs from spirals, in line with e.g. \cite{sandage:1975}.

\begin{figure}
\centering
\includegraphics[width=0.99\columnwidth]{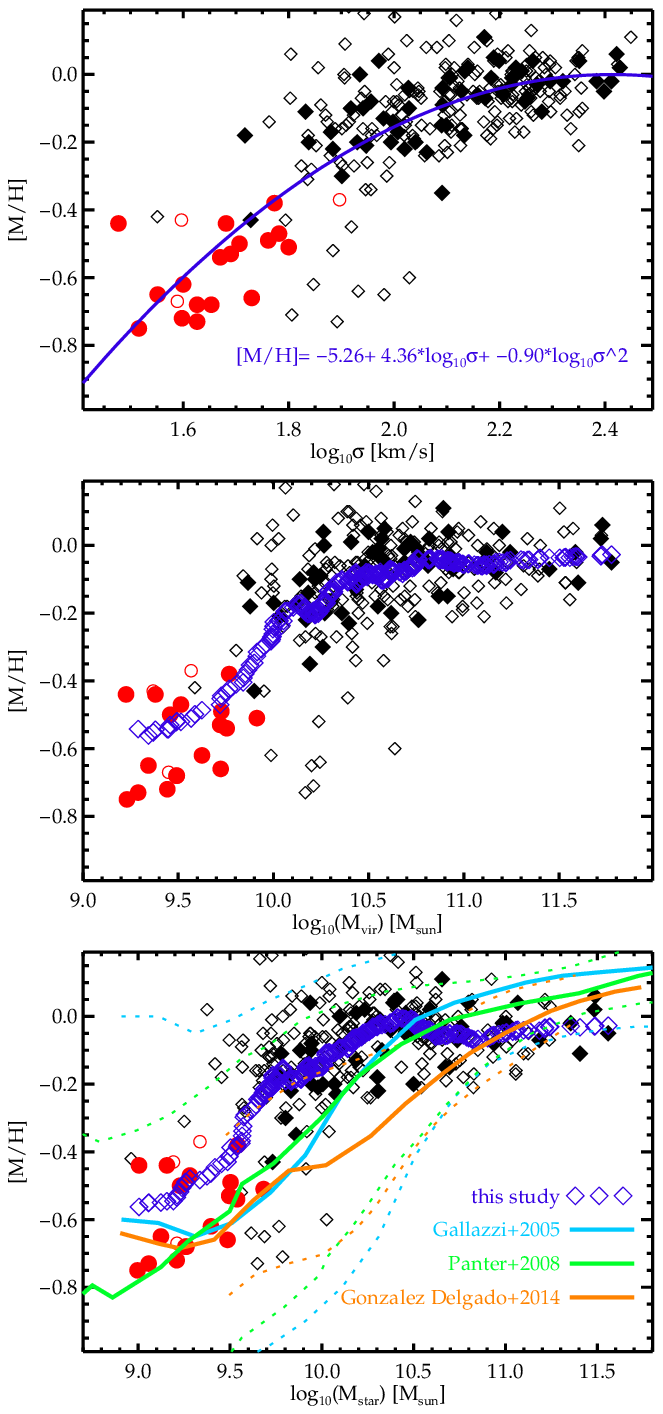}
\caption{Mass-metallicity relation shown for three mass proxies: $\sigma$ (\textit{top}), $M_{vir}$ (\textit{middle}), and $M_{*}$ (\textit{bottom}). Individual symbols as in previous figures. In the top panel a best-fit second-degree polynomial is shown for the combined dE-ETG sample (solid purple line) for which also the best-fit coefficients are given, which are later used to correct for the mass in the $\Delta$[M/H] vs. number density plot in Fig.~\ref{density_relations}. Moving averages are shown with purple open diamonds in the bottom two panels. \textbf{The bottom plot additionally includes average trends found for field+cluster early- and late-type CALIFA galaxies of \protect\cite{gonzalezdelgado:2014b} and the SDSS samples of \protect\cite{gallazzi:2005} and \protect\cite{panter:2008} (solid orange, light-blue and green lines, respectively; the corresponding 16th and 84th percentiles are shown using dotted lines).}}
\label{mass-met-relation}  
\end{figure}

\subsection{Scaling relations}

\subsubsection{Integrated SSP-$\sigma$ scaling relations}

In Fig.~\ref{SSP_relation} we show the SSP age, metallicity and [Mg/Fe] abundance ratio values as well as their gradients as a function of velocity dispersion (see Table~\ref{resultstable-ssp} for the tabulated values and Appendix~\ref{app-lines} for the same LS--{\sig} relations). Our results qualitatively agree with earlier ones on the early-types in the literature in that there exists a positive correlation between all three population parameters and {\sig} (see, e.g. \citealt{graves:2009}, \citealt{kuntschner:2010}, \citealt{johansson:2012}, \citealt{conroy:2014}).


We quantify the amount of scatter in the age-{\sig} relation for non-cluster objects by taking standard deviations of age values in five log({\sig}) bins (see the top left-hand panel of Fig.~\ref{SSP_relation}). The deviation values are 5.08, 4.23, 4.03, 2.70, and 0.55 for 1.50-1.80, 1.80-2.00, 2.00-2.15, 2.15-2.30, 2.30-2.45 bins, respectively. We see that the scatter in the relation increases with decreasing {\sig} in a way that cannot be attributed to measurement errors alone. This can already be seen in \citeauthor{mcdermid:2015} (\citeyear{mcdermid:2015}, Fig. 8) for the same dataset, though the authors only note specifically the increased scatter at lower {\sig} for their line-strength relations. This ``cone'' (rather than a linear trend) created by the combined cluster and non-cluster objects becomes clear with the addition of our dE galaxies at the lowest {\sig}, as it was earlier seen as a curved relation in {\hb} - {\sig} relation in \cite{peletier:2007}. The median lower age of group/field vs. cluster galaxies is due to the fact that the latter sample contains fewer SSP-young objects, and not due to the oldest ages being higher there.

We add our 20 dEs to the low-mass end of the [M/H]-{\sig} and [Mg/Fe]-{\sig} relations and show that these are continuous across the entire probed {\sig} range, from dwarfs to the most massive early-types. Earlier \cite{trager:2000} found only a weak correlation between [M/H] and {\sig} for elliptical galaxies, here, however, the data quality as well as the probed mass range and sample size make the presented relation much more robust.

The scaling relations presented in Fig.~\ref{SSP_relation} are tighter for cluster than for the non-cluster samples, as evidenced by the $rms$ values and the Spearman correlation coefficient values quoted in the figure. The younger ages of field/group objects make them deviate from the combined relations and so one can be inclined to conclude that, given sufficient time, all galaxies will be brought onto the relation. Stellar populations, old at that point, will not be the main contributing factor to the scatter any more. We can relate here to the work of \cite{falcon:2011b} where the tightness of photometric scaling relations is investigated for the SAURON project sample of E, S0 and Sa galaxies. The authors find that correcting for the influence of stellar populations (by replacing luminosity with stellar mass) they are able to significantly reduce the intrinsic scatter in the Fundamental Plane. 

In this context it is useful to point out that the age-{\sig} relation (top left panel of Fig.~\ref{SSP_relation}) shows an interesting feature, hinted at earlier in \cite{sanchezblazquez:2006a}, as well as shown for giant early-types in \cite{mcdermid:2015}. We see that the field/group objects form a stronger age relation with {\sig} such that ages are lower in non-cluster objects for comparable \sig. In other words, we see a steady increase of age as a function of mass for the non-cluster sample, while for the cluster objects the relation shows little to no dependence on {\sig} with a large number of objects approaching highest ages. This indicates that the cluster environment, by affecting star formation, speeds up the placing of galaxies on the red sequence.

\subsubsection{SSP gradients - $\sigma$ scaling relations}

The bottom row of Fig.~\ref{SSP_relation} shows SSP gradients vs. {\sig} relations. We can see age gradients becoming on average steeper (more positive) with decreasing {\sig}, most likely indicative of SF shrinking with time. This, however, does not apply to dEs, which lie somewhat below the ETGs relation and have close to null gradients. More variety is seen in the massive ETG sample: dE gradients have values between -0.05 and +0.49 and ETGs  -0.05 and +1.19, the latter thus covering a three times larger range of values, and the standard deviations for dEs and ETGs are 0.15 and 0.26, respectively. This is mostly due to the presence of a larger number of field galaxies among ETGs which show both negative and strongly positive age gradients. We find no dependence of [M/H] and [Mg/Fe] gradients on {\sig} (seen also by e.g. \citealt{kuntschner:2010} and \citealt{koleva:2011}). 

That [Mg/Fe] gradients scatter around null is an indication that there is, on average, no evidence for a strong variation of the length of SF episodes \textit{within} galaxies. With the exception of a few field objects from the massive ETGs sample the vast majority of both low- and high-mass ETGs have negative [M/H] gradients, i.e. higher metallicities in galaxies' central parts. This is in agreement with earlier results on ETGs (e.g.  \citealt{mehlert:2003} for the Coma cluster, \citealt{rawle:2008} for Abell 3389, \citealt{kuntschner:2010} for Virgo and field, \citealt{bedregal:2011} for Fornax, \citealt{loubser:2012} for BCGs) as well as dEs (e.g. \citealt{koleva:2011}). The rather large scatter around the relation suggests gradients to be very sensitive to the individual evolutionary history of each galaxy. 

\subsubsection{Mass\,-\,metallicity relation}

Fig.~\ref{mass-met-relation} shows the relation between [M/H] and our directly measured mass proxy {\sig}, {\sig}-based total dynamical mass estimate $M_{vir}$, as well as $M_*$ estimates based on the MILES models' predictions. We see in the figure that the relation flattens out at higher masses for all the three quantities. This agrees with earlier findings based on \textit{stellar} metallicity-{\sig} relation as far as the massive early-types regime is concerned (e.g. \citealt{graves:2009} for a sample of $\sim$\,16000 SDSS galaxies). It also reflects the \textit{general} trend found in the literature for the \textit{nebular} metallicities-stellar mass relation (e.g. \citealt{ellison:2009}), though we note that these two quantities  are not directly comparable, given the fundamental differences in what they represent (i.e. total metallicities reflecting the entirety of SF of the galaxy vs. metallicities of the present-day gas component). 

To compare with published stellar mass-metallicity relations we overplot in the bottom panel of Fig.~\ref{mass-met-relation} average trends for the \textbf{SDSS samples of \cite{gallazzi:2005} and \cite{panter:2008} as well as the CALIFA sample of \cite{gonzalezdelgado:2014b}. The first two samples include ca. 175000  and 300000 galaxies} from the SDSS survey which provides aperture spectra of 3'' in diameter, and the third one -- integral field data reaching out up to $\sim$2\,{\re}, with similar coverage in mass and sample size to our combined dE-ETG sample presented in this work. \textbf{While the qualitative agreement between the shown samples is good}, we note that different methods have been employed by the quoted authors in the derivation of the relevant quantities shown in the figure: statistical estimates of ages and metallicities from 3-arcsec SDSS apertures using a set of line strength indices  \textbf{(\citealt{gallazzi:2005}) or parametrized star formations histories (\citealt{panter:2008})} and stellar population spectral synthesis models providing mass-weighted metallicities and stellar mass surface densitites with the assumed Salpeter IMF in the case of CALIFA galaxies. \textbf{Thus, a quantitative comparison of the presented trends is not possible or desired here.} \looseness-1

\subsection{Stellar population parameters vs. local environment density}

\begin{figure*}
\centering
\includegraphics[width=2.05\columnwidth]{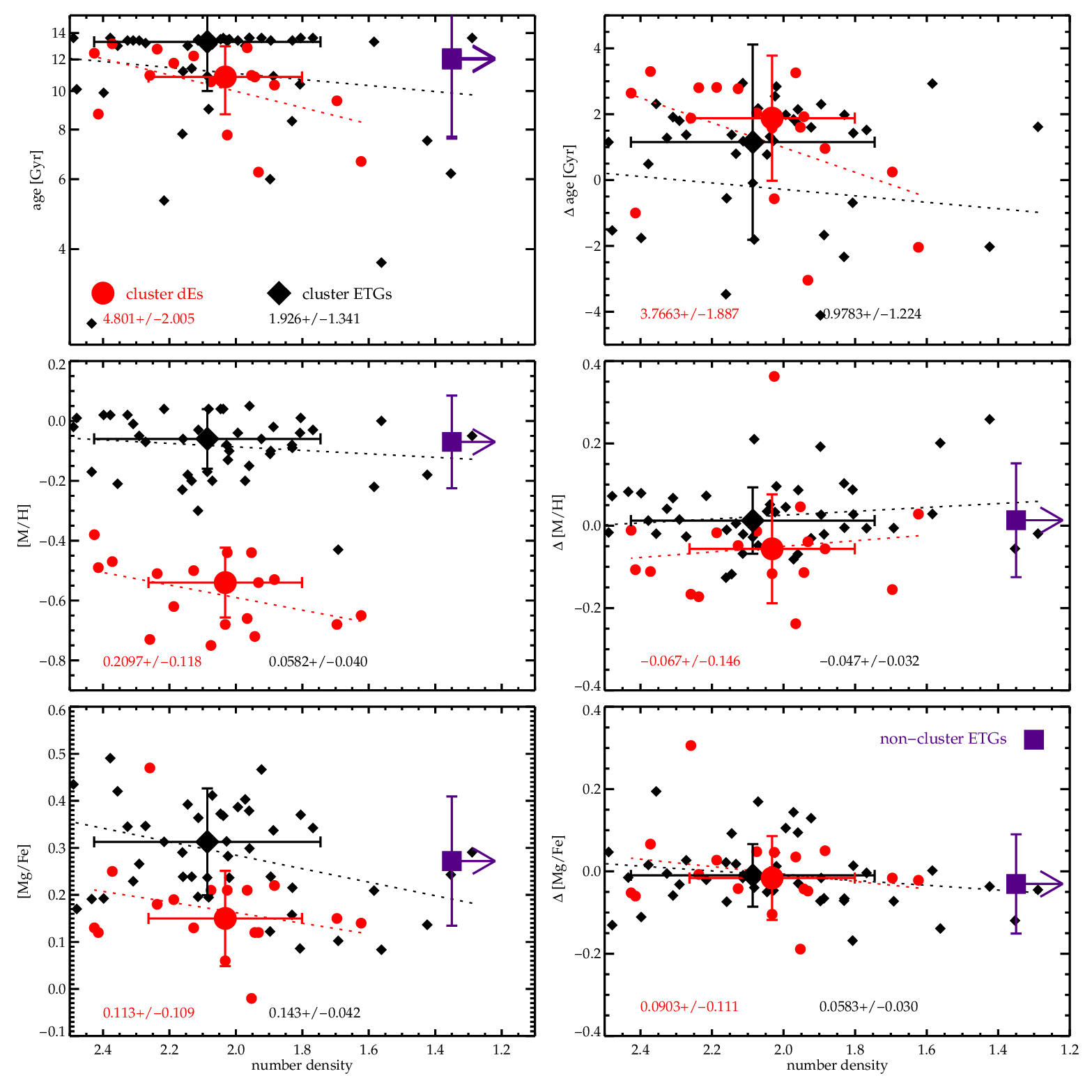}
\caption{SSP parameters as a function of number density: age, [M/H], [Mg/Fe] (left column) and residuals around the age--{\sig} relation for cluster objects and around [M/H]-{\sig} and [Mg/Fe]-{\sig} relations for the combined dE and ETG sample (right column). Median values of density and parameters are shown for the dE cluster sample (large red circles), ETG cluster sample (large black diamonds) for comparison with the non-cluster sample (large violet squares). In each panel we show best straight line fits to the cluster data for dEs and ETGs and provide best-fit slope values with uncertainties. The median $log_{10}${\sig} values for cluster and non-cluster ETGs are 2.13$\pm$0.17 and 2.12$\pm$0.16, respectively. We do not include median values for non-cluster dEs due to their very limited presence in our sample. See Table~\ref{proxies-compared} for Pearson correlation coefficients for the cluster subsamples.}
\label{density_relations}  
\end{figure*}

\begin{table}
\caption{Comparison of the Pearson correlation coefficient values obtained for the three environmental proxies: projected clustrocentric distance from M87, luminosity density and number density. For an easier visual inspection the coefficients have been colored so that weak to moderate correlations (here $|0.25|-|0.50|$) are in yellow and (moderately) strong correlations ($>|0.50|$) are in orange. Note that luminosity and number densities are anticorrelated with D$_{M87}$, hence the opposite signs in the corresponding trends. See Section~\protect\ref{label:data} for details on the calculations and Section~\protect\ref{discussion-env} for the discussion.}
\centering

\begin{tabular}{|r|r|r|r|r|r|r|}
\hline
           &\multicolumn{2}{c}{D$_{M87}$}&\multicolumn{2}{c}{lumdens}&\multicolumn{2}{c}{numdens}\\
           &\multicolumn{1}{c}{E}         &\multicolumn{1}{c}{dE}	&\multicolumn{1}{c}{E} 		 &\multicolumn{1}{c}{dE}	&\multicolumn{1}{c}{E}			&\multicolumn{1}{c}{dE}\\
\hline           
H$\beta$ & 0.11 & \colorbox{yellow}{0.47} & \colorbox{yellow}{-0.25} & \colorbox{yellow}{-0.42} & \colorbox{yellow}{-0.28} & \colorbox{Dandelion}{-0.58} \\
Fe5015 & -0.16 & \colorbox{yellow}{-0.27} & \colorbox{yellow}{0.28} & 0.04 & 0.14 & \colorbox{yellow}{0.39} \\
\medskip
Mg$b$ & -0.15 & \colorbox{yellow}{-0.41} & \colorbox{yellow}{0.35} & \colorbox{yellow}{0.32} & \colorbox{yellow}{0.31} & \colorbox{Dandelion}{0.67} \\
$\nabla$\,H$\beta$ & 0.00 & -0.08 & 0.21 & 0.18 & 0.17 & \colorbox{yellow}{0.28} \\
$\nabla$\,Fe5015 & -0.04 & \colorbox{yellow}{0.43} & 0.09 & \colorbox{Dandelion}{-0.66} & -0.00 & \colorbox{yellow}{-0.26} \\
\medskip
$\nabla$\,Mg$b$ & -0.06 & \colorbox{yellow}{0.40} & 0.01 & \colorbox{yellow}{-0.27} & -0.13 & -0.20 \\
age & -0.09 & \colorbox{yellow}{-0.42} & 0.07 & \colorbox{yellow}{0.25} & 0.20 & \colorbox{Dandelion}{0.53} \\
$[M/H]$ & -0.07 & \colorbox{yellow}{-0.27} & \colorbox{yellow}{0.30} & 0.12 & 0.20 & \colorbox{yellow}{0.41} \\
\medskip
$[Mg/Fe]$ & -0.17 & -0.10 & \colorbox{yellow}{0.41} & \colorbox{yellow}{0.25} & \colorbox{yellow}{0.43} & \colorbox{yellow}{0.26} \\
$\nabla$\,age & 0.02 & \colorbox{yellow}{0.25} & -0.00 & \colorbox{yellow}{-0.31} & -0.19 & \colorbox{yellow}{-0.32} \\
$\nabla$\,$[M/H]$ & -0.20 & 0.20 & \colorbox{yellow}{0.25} & -0.22 & 0.21 & 0.07 \\
$\nabla$\,$[Mg/Fe]$ & 0.16 & 0.19 & -0.24 & 0.12 & -0.13 & -0.16 \\

\hline
\end{tabular}

\label{proxies-compared}
\end{table}

In Fig.~\ref{density_relations} we plot age, [M/H], [Mg/Fe] and residuals around their corresponding best-fitting relations with {\sig} shown in Fig.~\ref{SSP_relation} for the combined dE and ETG sample. We do so because, apart from local density, it is also galaxy mass/compactness which influences to what extent a given object is affected by the local environment. Also, as shown earlier, more massive galaxies tend to be older and have higher [M/H] and [Mg/Fe] ratios. The figure shows the above population parameters as a function of number density (together with best straight-line fits), while in Table~\ref{proxies-compared} we provide correlation coefficients between LS and SSP parameters and all three local density proxies: number and luminosity density as well as projected clustrocentric distance.

In the case of dEs we see that there is a moderate to strong correlation between integrated $H\beta$ and {\mgb} values and all three proxies, with the strongest relation present for number density, and the other two proxies giving similar correlations. In the case of Fe5015 a null (for luminosity density) to weak (for projected distance and number density) correlation is seen. The line-strength and SSP relations understandably agree in that the age indicator {\hb} and the derived ages themselves show moderate to strong correlation with density. Metallicity residuals show no dependence on any of the proxies. The existence of a weak correlation is also visible for [Mg/Fe], from which we can conclude that objects in lower density regions have on average lower abundance ratios in the SSP sense. \looseness-1

When it comes to massive ETGs, on average weaker correlations are seen (though we do note the moderately strong correlation of [Mg/Fe] with luminosity and number densities), i.e. they are affected by the local environment of the cluster to a lesser degree than the low-mass galaxies. For both samples, once corrected for mass (right column of Fig~\ref{density_relations}), the slight trends with number density visible in the [M/H] and [Mg/Fe]-{\sig} relations are gone. We take this as a further proof for the metal content and metal abundance ratios being generated primarily by galaxy mass.

Our results can be compared to those of \cite{roediger:2011b} who investigated correlations of color-based ages and metallicities (and their gradients) with projected clustrocentric distance and galaxy density for a variety of Virgo morphological types (their Fig.~9 and~12). Grouping all early-types together, they did not find any correlation with either of the proxies. We note that their quoted ages never go below $\sim$\,7\,Gyr, unlike the galaxies of our or the ATLAS$^{3D}$ sample. We conclude that the larger age spread and the lower correlation of dE ages with the environment in their study is likely due to the different methods used in the derivation of the population parameters. 

To show how the average population properties compare across the probed environments,  Fig.~\ref{density_relations} includes averages for cluster and non-cluster objects with densities for non-cluster shown with an arrow to indicate that the positions are approximate. We find no evidence for a difference in median age, [M/H] or [Mg/Fe] of the two samples. The fact that we do not find any significant difference between the average metallicities or element abundance ratios of cluster and non-cluster objects (as well as within the cluster environment as a function of number density) is in agreement with the findings of, e.g. \cite{johansson:2012} from an analysis of stellar populations of SDSS galaxies, as well as \cite{cooper:2008} or \cite{ellison:2009} for SDSS gas-phase metallicities.

We note, however, that completely separating the effect of mass from the effect of the environment from our relations is neither straightforward nor desired as mass constitutes just one of the galaxy parameters that can influence the trends and, conversely, the environment accounts for just about 15\% of the scatter in the mass-metallicity relation (e.g. \citealt{cooper:2008}). An in-depth treatment of the issue goes beyond the scope of this paper.

\section{Discussion}

\subsection{Scaling relations: from dwarfs to giants?}

Do scaling relations provide clues with regard to the origin and evolution of galaxies, and if so -- in what way? Extensive debates have taken place in the literature on this topic and on the structural connections (or lack thereof) among various galaxy classes (see, e.g. \citealt{kormendy:2012} and \citealt{graham:2013} for in-depth reviews and a full list of references). The important question is whether a continuity on a given relation necessarily translates into a proof that the galaxies involved belong to one type of objects, or whether it reflects a global property of galaxies, of e.g. certain mass or luminosity. The presence of a broken relation could, in turn, be interpreted as a sign of structural difference and indicate that there exist thresholds below/above which certain (trans)formation processes can or cannot take place. Our dEs follow some of the scaling relations of the normal ETGs, as shown in Fig.~\ref{SSP_relation} and Fig.~\ref{mass-met-relation} for the stellar populations, and earlier in \cite{rys:2014} for their kinematic properties, suggesting, at least in this context, a lack of structural differences between the two groups of galaxies in the probed mass range.

From the [M/H]-{\sig} relation -- continuous across the entire probed {\sig} range, from dwarfs to most massive early-types -- we conclude that galaxy-wide (i.e. driven by total mass) processes are important in determining the total metallicities, with local environment in the cluster playing a limited role (as noted earlier by, e.g. \citealt{thomas:2010} or \citealt{rogers:2010}).

The rather tight correlation of the [Mg/Fe] abundance ratio with {\sig} (a proxy for galaxy mass) is expected in the context of the tightness of the {\mgb}-{\sig} relation (upper right panel of Fig~\ref{gradients_integrated_relation}). The [Mg/Fe] ratio is informative because the two elements are produced by different types of supernovae (SNe) having progenitors of different masses. Higher [Mg/Fe] ratios then indicate relatively more impact from the type II SNe (main $\alpha$-elements producers). Traditionally, lower $\alpha$ elements abundance in low-mass galaxies was taken to mean extended (or showing multiple bursts) SF as compared with the higher mass end. Recently it has also been shown that there is a mass-dependent variability of the initial mass function (IMF, e.g. \citealt{labarbera:2015}, \citealt{spiniello:2016}) which would imply that even shorter SF timescales at the high-mass end are needed to accommodate the observed [Mg/Fe]-{\sig} relation of ETGs. Going into the discussion on IMF variations is, however, beyond the scope of the paper and we refer the reader to the recent papers of e.g. \cite{smith:2014}, \cite{martinnavarro:2015} and~(\citeyear{martinnavarro:2016}), or \cite{labarbera:2015} for an in-depth treatment of the topic.

\subsection{Population gradients in proposed formation scenarios}

Are population gradients expected? Among the most often quoted scenarios for the transformation of dE progenitors into the galaxies we see today are galaxy harassment (hereafter GH, \citealt{moore:1998}) and ram-pressure stripping (hereafter RPS, \citealt{gunn:1972}). GH could in principle create gradients if the tidal interaction woiuld push remaining gas towards the centre of a galaxy which would then induce secondary star formation. RPS timescales for cold disk gas removal, which prevents the galaxy from further star formation, will depend on the galaxy properties and its orbit, and galactic centres of the bright dEs are likely spared when using the \citeauthor{gunn:1972} criterion in the Virgo cluster (\citealt{lisker:2013}). What is unknown, however, is, given the different predicted timescales for the two mechanisms, just how much gas will actually be left in a galaxy when it reaches the regions of the cluster dense enough for the GH to make a significant impact. The answer to this question depends on factors such a galaxy mass (with more massive galaxies able to retain more gas), orbital parameters, gas content/preprocessing, etc. Needless to say, other factors such as the cutoff of gas accretion, the so called starvation (e.g. \citealt{larson:1980}), are also able to play a role in determining observed galaxy properties. Galaxies must have experienced varying degrees of influence from the different transformation mechanisms, therefore we cannot expect them to follow a simple pattern with regard to their gradients since various factors can compensate for each other and smear the gradients out and/or introduce a significant scatter. Also, we are typically unable to show whether an observed pattern has been \textit{caused} by the (environmental) transformation or whether it has been preserved \textit{despite} the environmental effects. No larger samples for Virgo dEs with reliable gradient measurements are yet available, thus the issue is yet to be studied in detail from the observational point of view.

The fact that we see roughly no dependence of [M/H] and [Mg/Fe] gradients on {\sig} and given that {\sig} can be seen a a proxy for galaxy mass, can be understood as a sign that gradient creation does not seem to be regulated by the depth of the galaxy potential well (or at least mass is not the dominant factor there), which leaves external factors as the likely cause (though we note that there are a few exceptions among the {\atlas} field sample).\looseness-1

We can see in our sample that the outside-in star formation scenario is true for nearly all objects in denser environments, with individual differences likely attributable to different merger/encounter/arrival histories (see also \citealt{schaefer:2017} who study star formation as a function of local environment density as part of the SAMI IFU survey). Those of the non-cluster ETGs with negative age gradients which cannot be reconciled with zero to within the errors are likely a result of gas accretion and the resulting star formation occurring in the galaxies' outskirts or (less likely) a result of a minor merger (series) where a given galaxy cleared its surroundings of satellites and their remaining gas\,/\,younger populations contributed to the observed negative age gradient.

Our average dE SSP age gradients are mildly positive, with the majority, however, consistent with zero to within the errors. Metallicity gradients are, on the other hand, negative for the majority of objects, regardless of their mass. When stellar populations are resolved temporally, both the young and old components of our galaxies (\citealt{rys:2015}) exhibit a range of age gradients, with the majority of them, however, again consistent with zero. Earlier, \cite{koleva:2011} noted the presence of gradients in the old components of their sample, which led them to conclude that gradients are robust against environmental effects. It is tempting to take the presence of gradients in old populations as a sign that they have survived the possible environmental transformation. But, have they survived \textit{intact}? Recent simulations of \cite{schroyen:2013} suggest that galaxies preserve their metallicity gradients once these gradients are formed -- \textit{when the galaxies are not perturbed by external forces}. This last assumption does not hold for our objects. 

Last but not least, population gradients are expected to depend on the richness of a given cluster (e.g. \citealt{labarbera:2005}), and the degree of its relaxation (e.g. Coma vs. Virgo where in the former the density and clustrocentric distance are directly related, unlike in Virgo where distinct cluster substructures are clearly visible). It would therefore be useful to compare different cluster environments to look for (dis)similarities in the behavior of population gradients within and among galaxies.

\subsection{Determining the dependence of galaxy properties on the environment}
\label{discussion-env}

We have analyzed the radial as well as luminosity and number density dependence of the population properties having two goals in mind. The first was to try and assess to what degree population parameters and their gradients correlate with the location in the cluster and the local environment. Spectroscopic population gradients have until now been studied only for a handful of early-type dwarfs (e.g. \citealt{koleva:2011}) and never in the context of deriving global (i.e. cluster-wide) relations. 

Our second goal was to try and determine the most reliable parameter to study such environmental factors as the influence of tidal or ram-pressure forces. As argued in e.g. \cite{roediger:2011b} and \cite{rys:2013}, the often used projected clustrocentric distances are less accurate than the 3D estimates (which can be obtained through, e.g. surface brightness fluctuation measurements) since they ignore the depth of the studied cluster. Still, \textit{both} methods necessarily rely on the assumption that the measured distances from the cluster centre more or less directly scale with the local density and with the time a given galaxy has already spent in a cluster.\looseness-1

While a reasonably good first-order approximation, the above assumption is invalid for a number of reasons:
\begin{itemize}
\item galaxies are likely accreted in groups (rather than individually) so the level of their preprocessing depends additionally on the pre-cluster history (see, e.g. \citealt{toloba:2014}),
\item today's location of a galaxy (even if a true 3D one) is not a clear indication of how much time the galaxy has already spent in the cluster: this depends, among other factors, on the orbital parameters of a given galaxy,
\item similarly, today's location of a galaxy is not directly related to how much impact environmentally-driven forces have had on the object: apart from the time spent in the cluster, the initial disk orientation determines the extent of the influence of a given, e.g. tidal, force on the object (see, e.g. \citealt{villalobos:2014}),
\item Virgo is not a relaxed, homogeneous cluster (as shown in Fig.~\ref{sample}) and a few off-centered high-density regions can be identified there, typically associated with other than M87 giant galaxies; environmental (tidal) forces are expected to be stronger there than in lower density regions, regardless of the region's projected distance from the cluster's centre.
\end{itemize}

The comparison of the trends seen for various environmental proxies shown in Tab.~\ref{proxies-compared} reveals that while there are, as expected, no huge discrepancies among the correlations for a given property, there exists no clear trend, either, which, if present, would suggest, e.g. that a projected distance trend was simply a diluted version of an intrinsic relation, with the relation's strength (slope) lowered due to the simplifications made in its measurement. If that were the case, we would see a systematic difference among the different methods, however, the correlation coefficients do not show any particular pattern (e.g. with projected distance always being the strongest or the weakest trend). It may well be the case that different transformation processes vary with the discussed environmental proxies to different degrees: while GH primarily correlates with local density, making the number density likely its most accurate tracer, this may be different in the case of RPS since it depends on gas density which likely scales with clustrocentric distance (see \citealt{gunn:1972}). This only confirms that the quantities do focus on different properties of the cluster: actual local environment properties but without the global picture (luminosity/number densities) or the simple and intuitive clustrocentric distance, which, however, are plagued by all the shortcomings listed in the previous paragraph. \looseness-1

Based on the above, we have decided to use number densities in our analysis as these reflect both current and to, some extent, past local densities, both in the form of the general cluster tidal field as well as proximity to other galaxies, responsible for tidal harassment. Overall, we find weaker correlations with density for ETGs than dEs. This is to be expected as these galaxies are trend\textit{setters}, rather than trend \textit{followers} like dEs: due to their high mass ETGs create local environment conditions (and earlier were likely central galaxies of infalling groups), while dEs, being low-mass objects and as such more susceptible to external influence, are affected by those conditions. This is complementary to the findings of e.g. \cite{chung:2007} who show that the \textit{global} environment does affect even the most massive galaxies. The fact that dE ages anti-correlate with density (both luminosity and number density) is expected as a higher density translates into stronger tides and ram-pressure stripping (at least today), but it also means that, statistically, the galaxies there have been in the cluster for a longer time. Both together explain the younger population age in less dense regions. We can thus conclude that galaxies which have undergone the strongest/longest environmental influence are also the ones with highest SSP age\,/\,lowest star formation levels.

In cosmologically simulated clusters, today's 3D velocity is lower for the early cluster population, and faster for the later infall population (e.g. \citealt{smith:2015}), hence it would also be useful to investigate a relation with $v_{helio}$ (noting the important caveat of it showing only \textit{projected} densities and a line-of-sight portion of the actual true velocities). One can also combine $v_{helio}$ with current location to analyze the galaxies in the observer's phase space, looking, e.g. for signatures of a group infall (see, e.g. \citealt{vijayaraghavan:2015}). On top of the above - and with large enough samples - one should always try to look at any global trends in more than one dimension, plotting them as \textit{maps}, rather than 1D profiles, thus taking advantage of all the available bits of information on the galaxies.

\section{Summary}
We have presented a stellar population analysis of a sample of 20 dwarf elliptical galaxies coming from our SAURON study of Virgo and field/group dEs and combined it with the {\atlas} sample of 258 massive early types to study stellar populations of early-type objects as a function of mass and local environment. Our main findings include:
\begin{itemize}
\item The derived SSP ages show a trend for more scatter at lower {\sig} values in the age-{\sig} relation, indicative of lower-mass galaxies being affected by their environment to a higher degree. The abundance ratios of dEs cluster around $\sim$\,0.15, forming a well-defined relation with {\sig} for the combined dEs and ETGs samples. Population gradients show little to no correlation with {\sig}, suggesting that gradient creation is \textit{not} primarily governed by mass.
\item The SSP-{\sig} relations are tighter for the cluster-only sample, suggesting that the cluster environment speeds up the processes which eventually place the galaxies on a given common relation. Also, when age-{\sig} relations for cluster and non-cluster galaxies are compared, we see a steeper relation of the ages with {\sig} in the latter case, meaning that the cluster environment accelerates the placing of galaxies on the red sequence.
\item We plot mass-metallicity relations using as mass proxies velocity dispersion {\sig}, $M_{vir}$ and $M_{*}$ and show that, qualitatively, the relations follow the general trend found in the literature from stellar as well as gas phase metallicities.
\item We analyze global galaxy properties as a function of number density, motivated by the complex, non-uniform structure of the still forming Virgo cluster where the influence of the environment goes beyond a simple relationship with projected clustrocentric distance. We find that dE SSP ages show on average a moderate correlation with our environmental proxy, such that suggests that star formation was quenched earlier for objects currently in denser environments; metallicity and metal abundance ratios show little to no correlation with number density, suggesting that feedback is influenced by the external forces to a lesser degree. Similarly, no significant correlations are found for population gradients, which can suggest that these are robust against environmental influence or that their creation and/or retention is a more complex function of the environment. \looseness-1
\item Massive ETGs show, on average, weaker correlations of their properties with our environment density proxies, i.e. they are affected by the local environment of the cluster to a much lesser degree than the low-mass galaxies. This is to be expected: due to their high mass ETGs create local environment conditions (and earlier were likely central galaxies of infalling groups), while dEs, being low-mass objects and as such more susceptible to external influence, are the ones affected by those conditions.
\item Comparing average SSP properties for cluster and non-cluster objects we find no evidence for differences in the stellar population properties of the two groups, in agreement with earlier literature results. \looseness-1
\item We suggest that all future studies should rely on detailed relations of the studied galaxies to their local and global environment, taking into account such factors as overall local galaxy density, distance to the nearest large neighbor, their line-of-sight velocities. Only in this way will we be able to provide a more accurate picture of their (trans)formation histories.
\end{itemize}

\begin{footnotesize}
\section*{Acknowledgments}

AS is a postdoctoral fellow of the Alexander von Humboldt Foundation (Germany). We thank Mina Koleva, Sven de Rijcke and Eric Emsellem for comments on a draft version of this paper, Marja K. Seidel and Carolin Wittmann for their help during the 2014 observing campaign, and Francesco la Barbera for helpful comments and suggestions on determining abundance ratios of the dE sample. We thank the anonymous referee whose comments and suggestions helped improve the presentation of our results. TL acknowledges the hospitality of the IAC while preparing the 2014 observing campaign. AV and JFB acknowledge support from grant AYA2016-77237-C3-1-P from the Spanish Ministry of Economy and Competitiveness (MINECO). RV is supported by a National Science Foundation Astronomy and Astrophysics Postdoctoral Fellowship under award AST-1501374. JJ thanks the ARC for financial support via DP130100388. The paper is based on observations obtained at the William Herschel Telescope, operated by the Isaac Newton Group in the Spanish Observatorio del Roque de los Muchachos.
\end{footnotesize}

\bibliography{biblio}

\appendix

\section{Example spectral fits.}
\label{appendix-spectra}

In Fig.~\ref{example-spectra} we show central (i.e. collapsed within an aperture of 1.5 arcsec radius) spectra for our dE sample, together with the best-fit models from the pPXF routine, residuals between the data and the models, and fitted emission lines where necessary. The only galaxy for which a significant emission correction was necessary was the field galaxy NGC\,3073, both in the {\hb} as well as [O{\textsc{iii}}]. 

NGC\,3073 was also observed as part of the {\atlas} survey, with 2.5\,hrs exposure time (vs. 5.5\,hrs in our program). The comparison of calculated 1{\re} integrated line-strengths shows a discrepancy for the {\hb} and Fe5015 lines, with our measurements being ca. 0.7\,{\AA} higher, which constitutes a significant difference. However, given the galaxy's particularly strong emission, the removal of which is the largest source of uncertainty in the derived line strengths, which we estimate can be up to 0.5\,{\AA}, combined with the fixed continuum correction uncertainty of 0.1\,{\AA} (assumed after \citealt{kuntschner:2010}) and the calculated random error of 0.12\,{\AA}, the two measurements can be reconciled. In any case, the fact that the other 19 objects are much less emission-affected, as well as given the quality of the spectra and fits presented in Fig.~\ref{example-spectra}, we are confident that the overall results presented here are robust.

\begin{figure*}
\includegraphics[width=0.99\columnwidth]{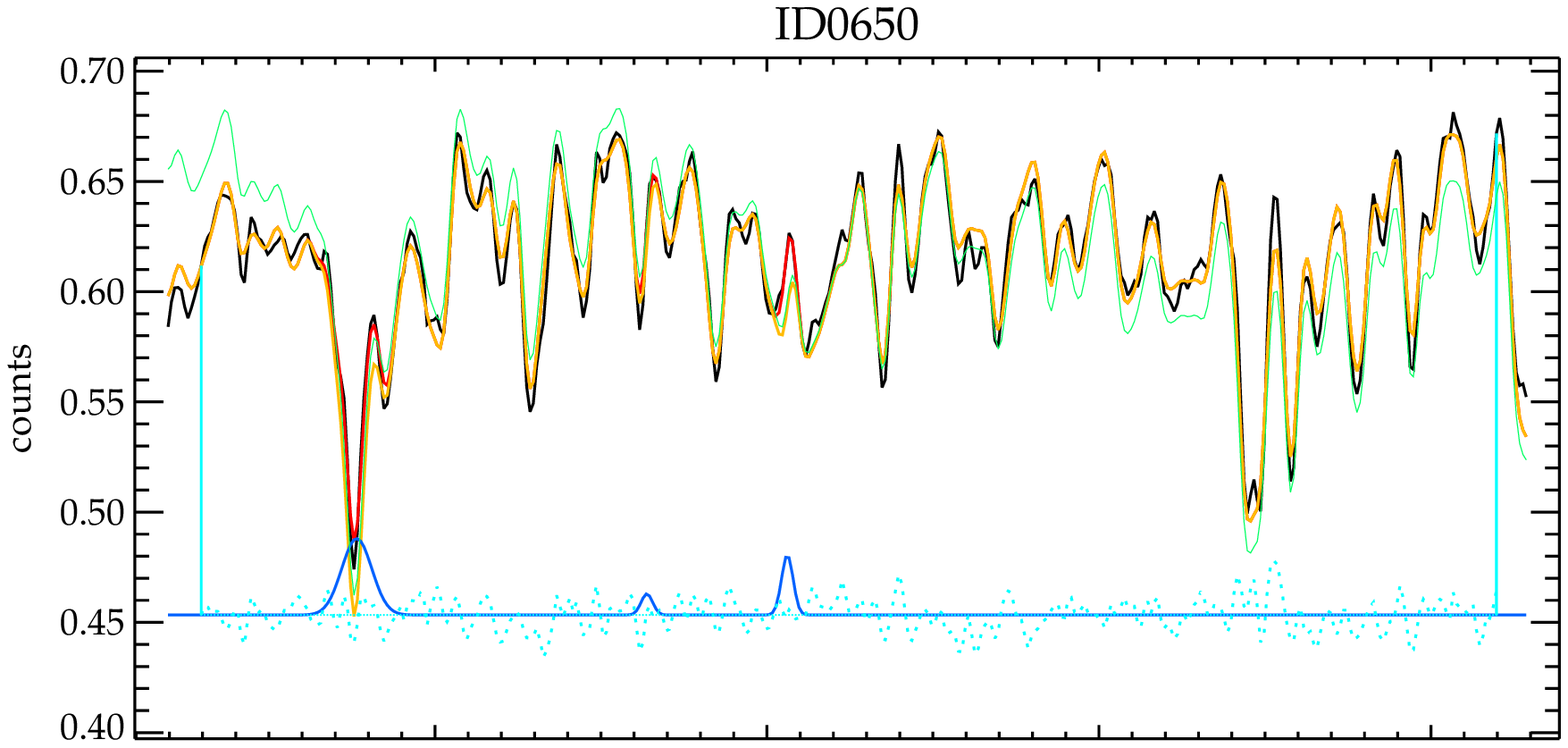}
\includegraphics[width=0.99\columnwidth]{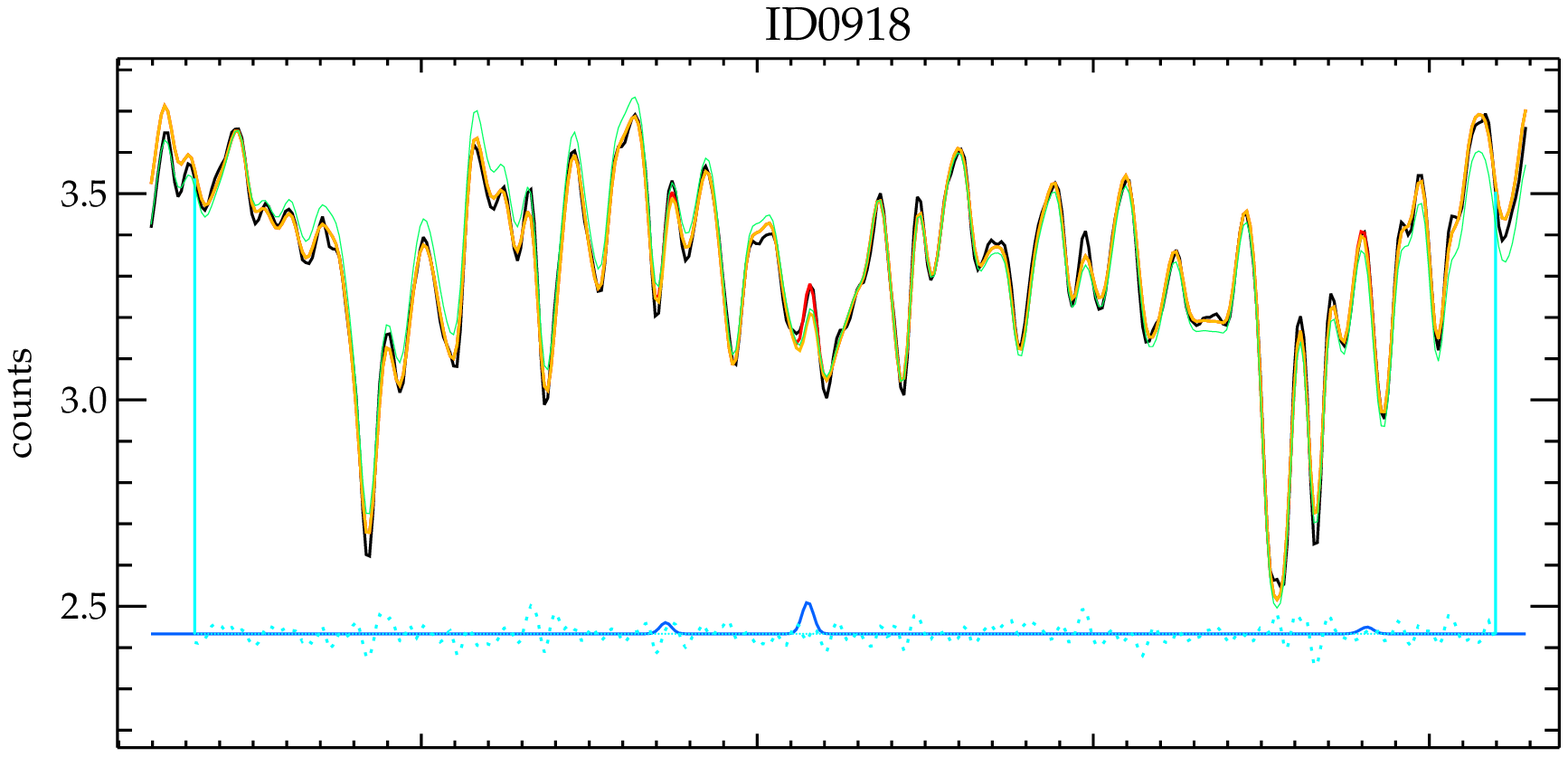}
\includegraphics[width=0.99\columnwidth]{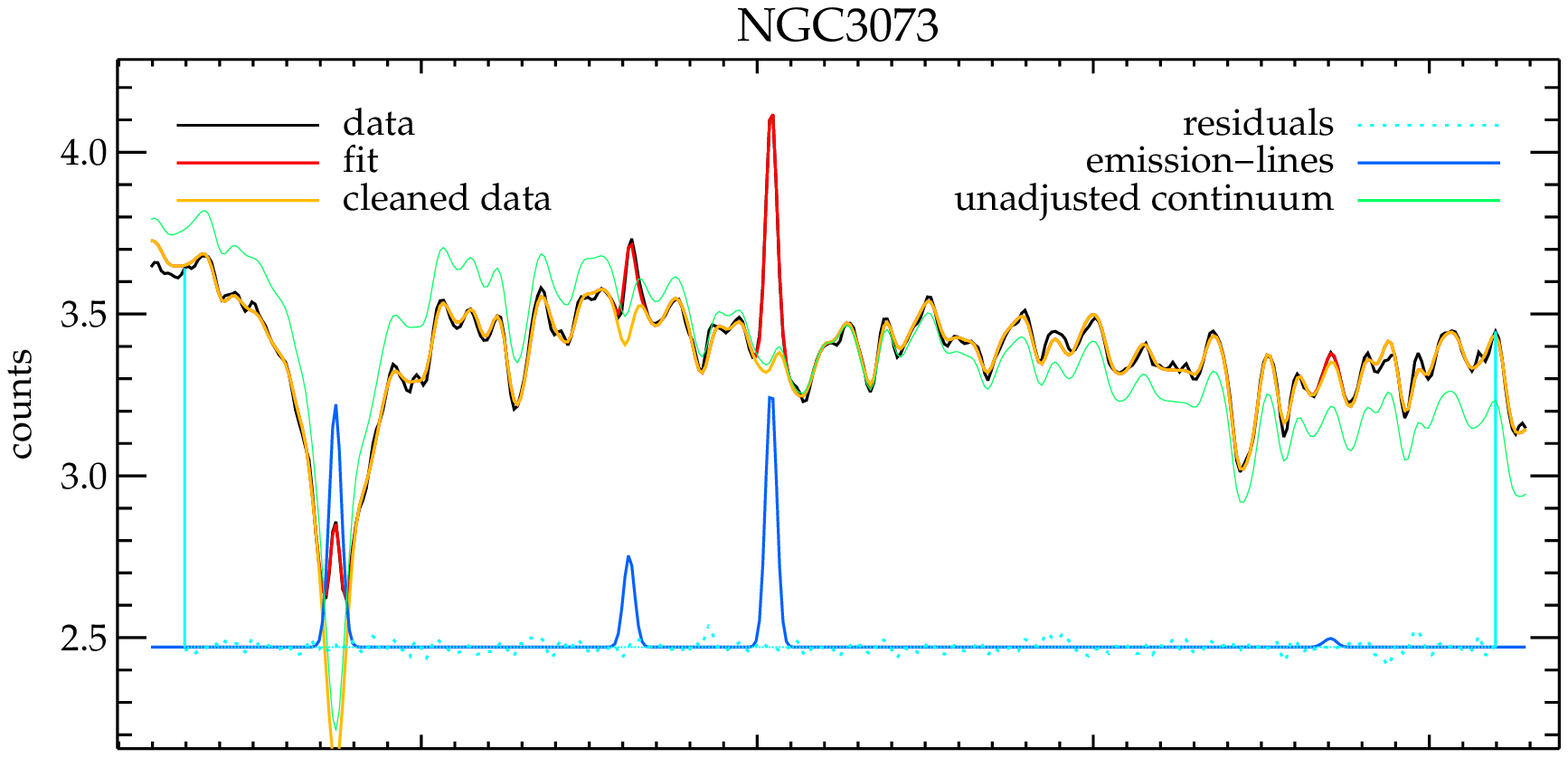}
\includegraphics[width=0.99\columnwidth]{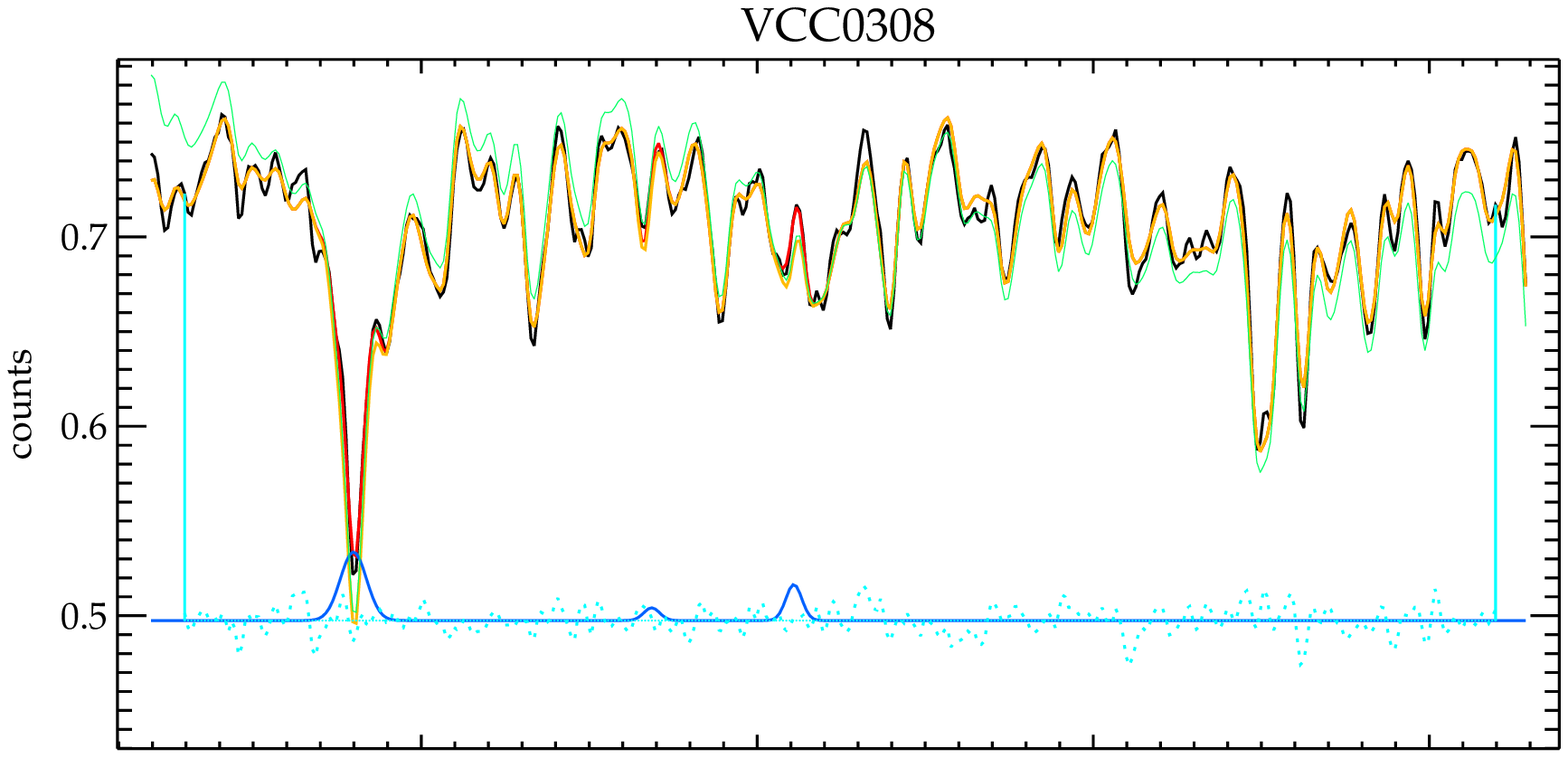}
\includegraphics[width=0.99\columnwidth]{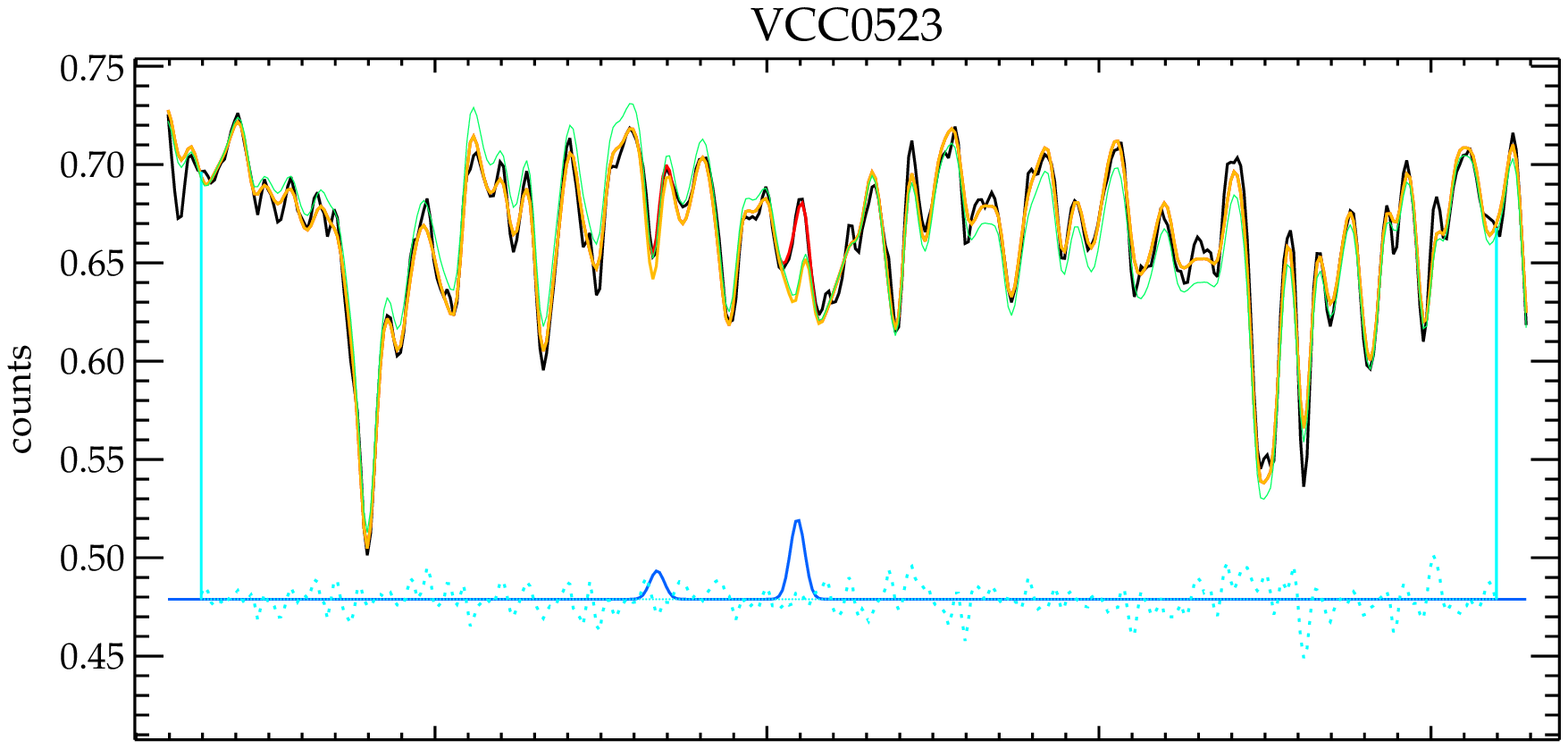}
\includegraphics[width=0.99\columnwidth]{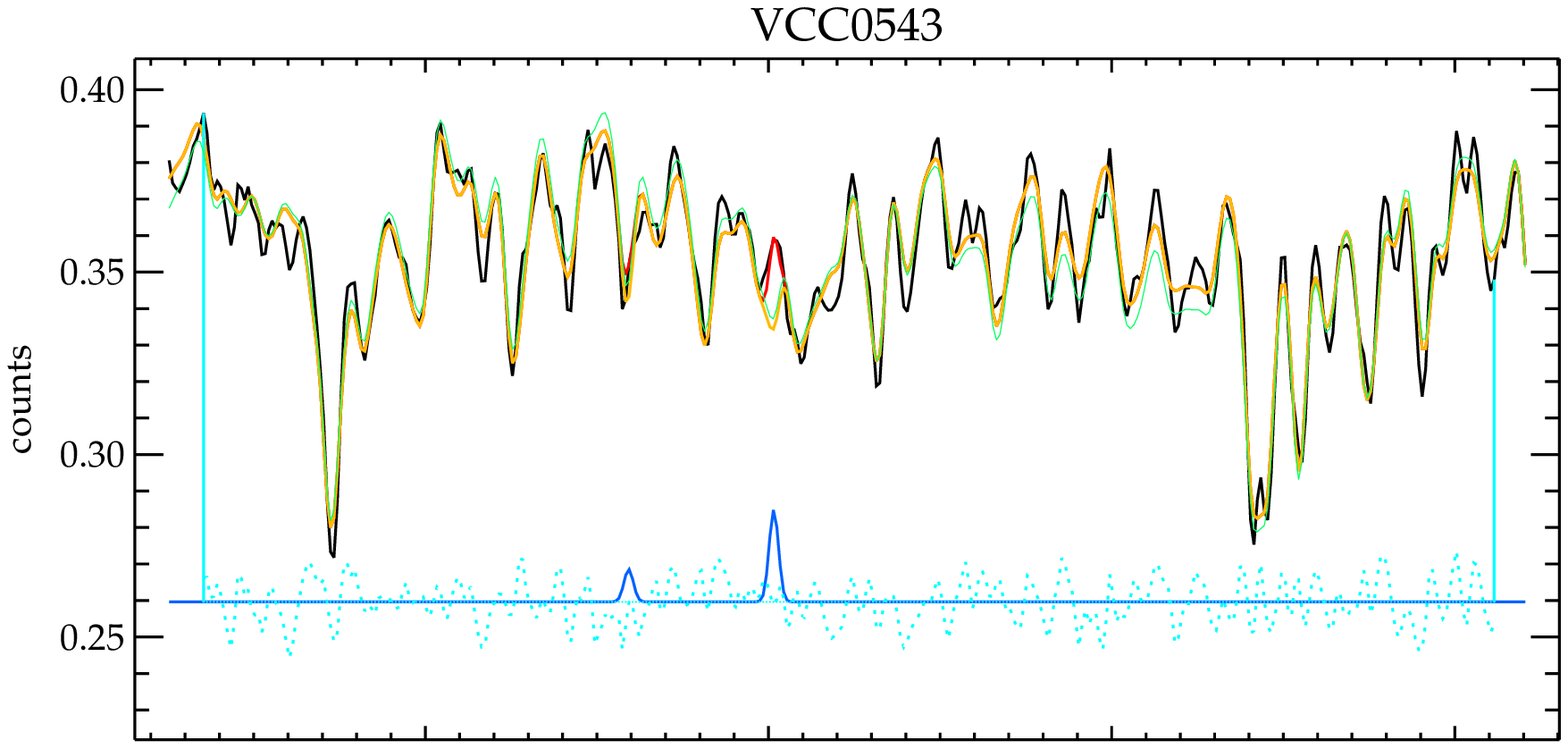}
\includegraphics[width=0.99\columnwidth]{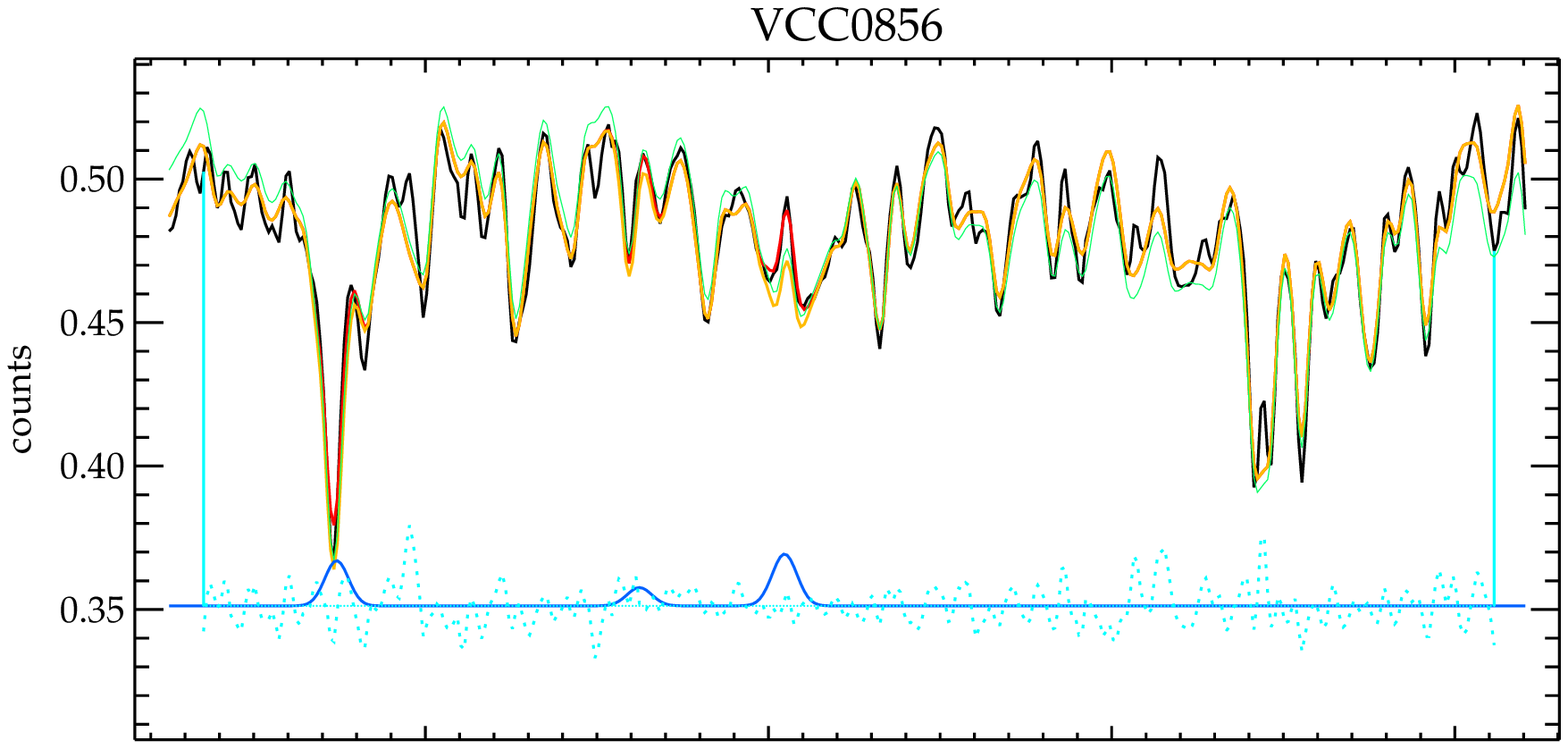}
\includegraphics[width=0.99\columnwidth]{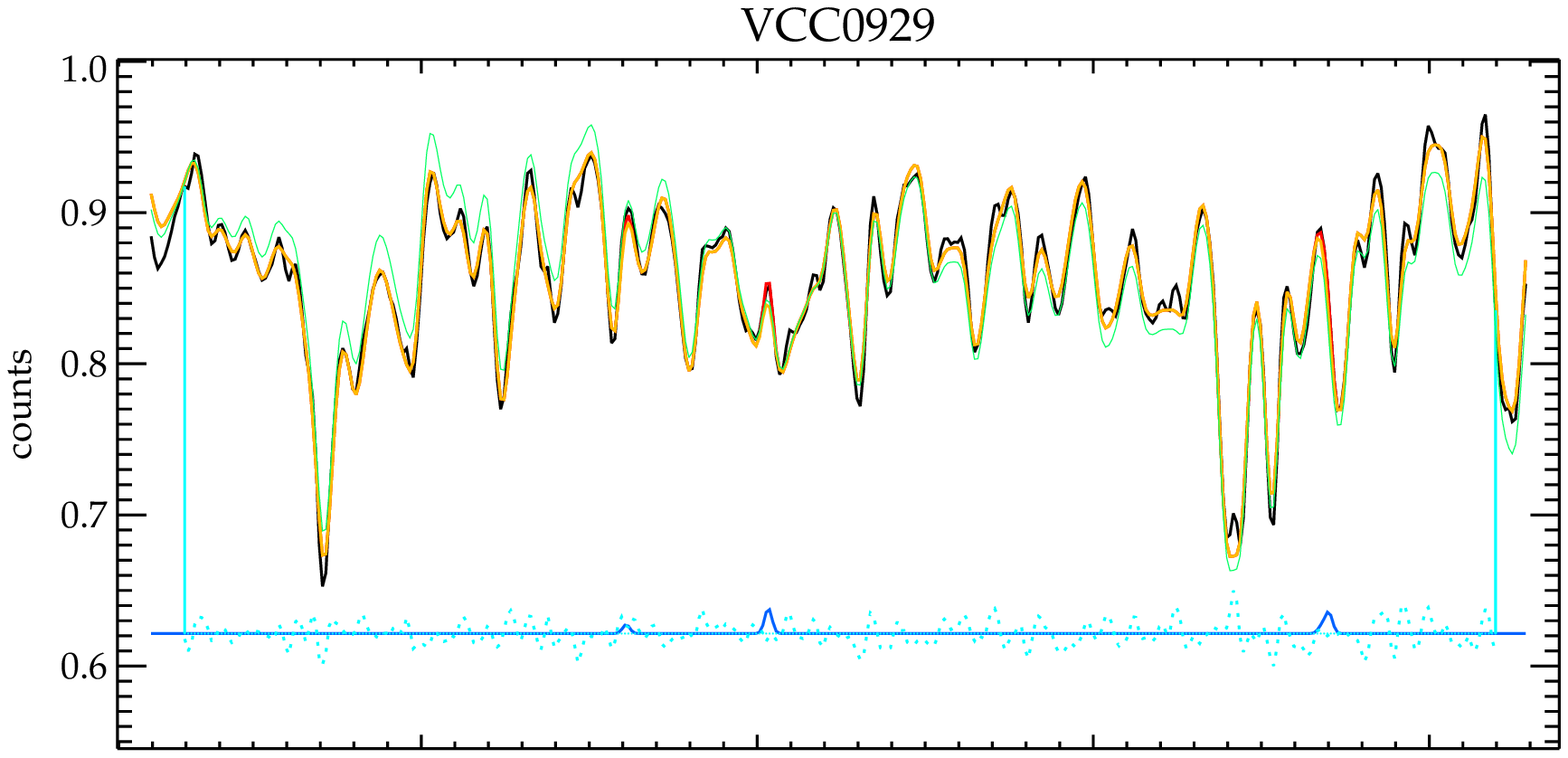}
\includegraphics[width=0.99\columnwidth]{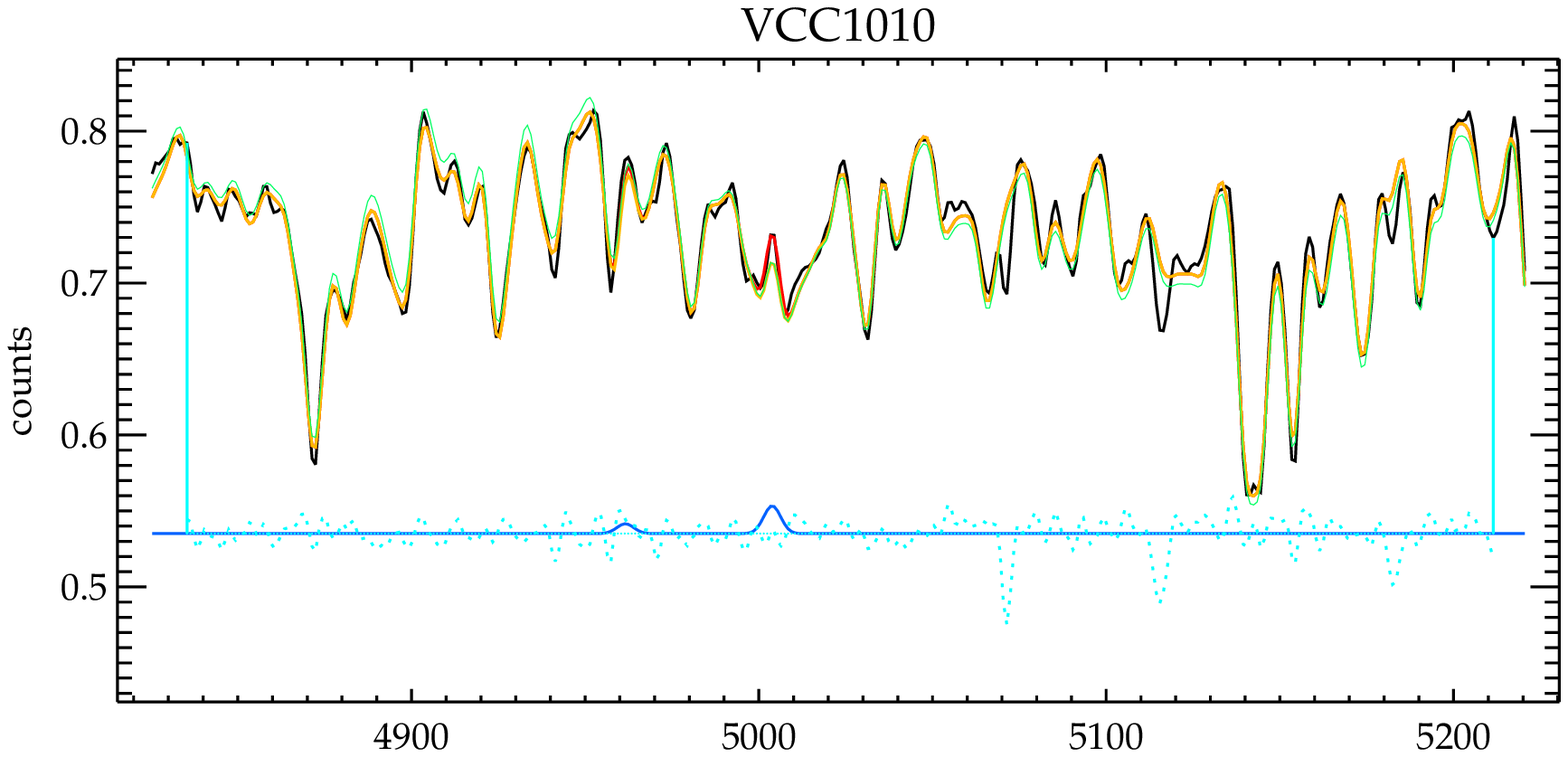}
\includegraphics[width=0.99\columnwidth]{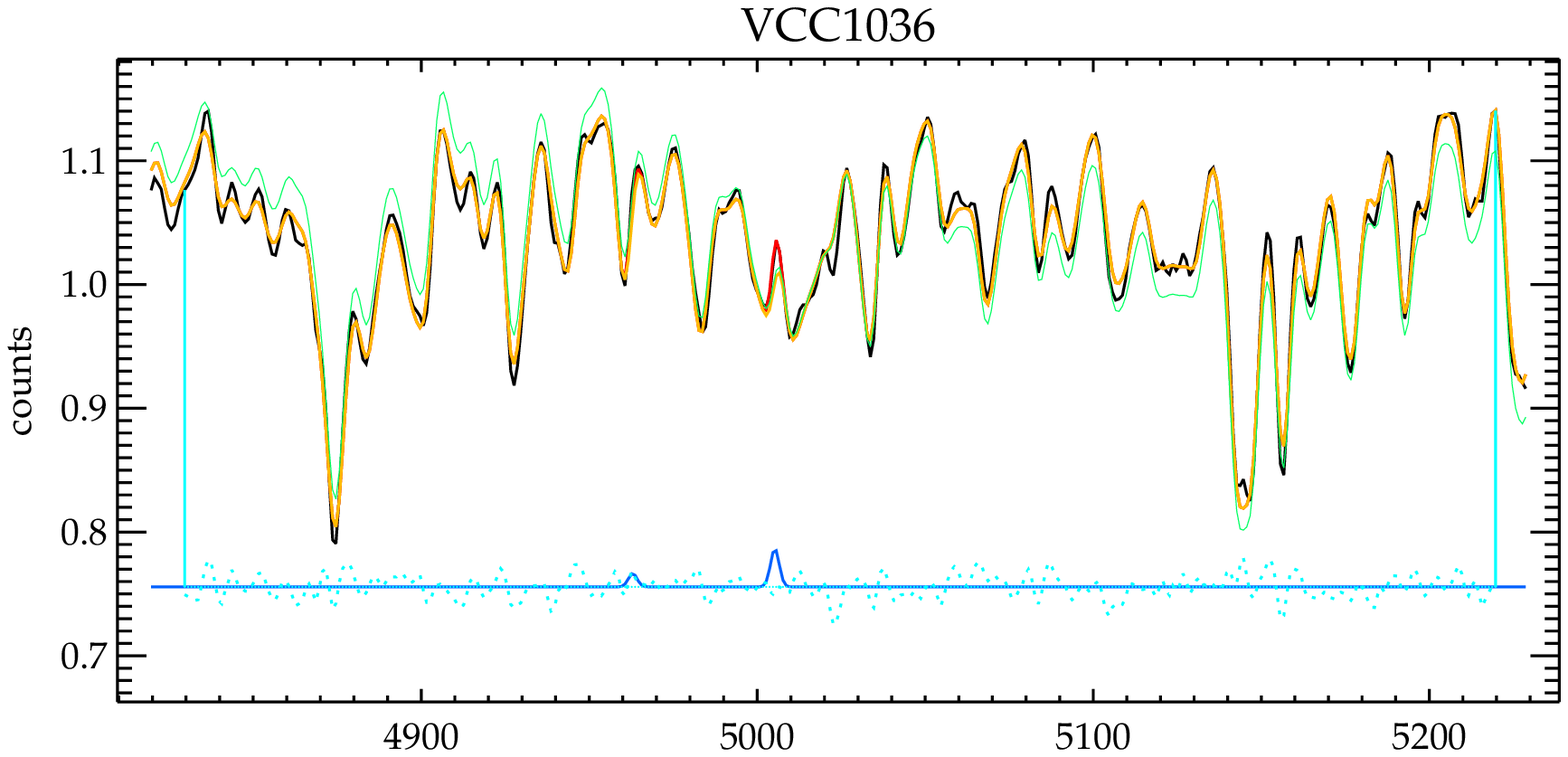}
\begin{scriptsize}
\hspace{12cm} $\lambda$ [\AA] \hspace{8cm} $\lambda$ [\AA] \hspace{-1cm} 
\end{scriptsize}
\caption{Central ($<$1.5'') spectra of each dE galaxy in our sample, shown here with best-fit models, and fitted emission lines where applicable. See the legend in the panel for NGC\,3073 for the lines' description.}
\label{example-spectra}  
\end{figure*}

\addtocounter{figure}{-1}
\addtocounter{subfigure}{1}

\begin{figure*}
\centering
\includegraphics[width=0.99\columnwidth]{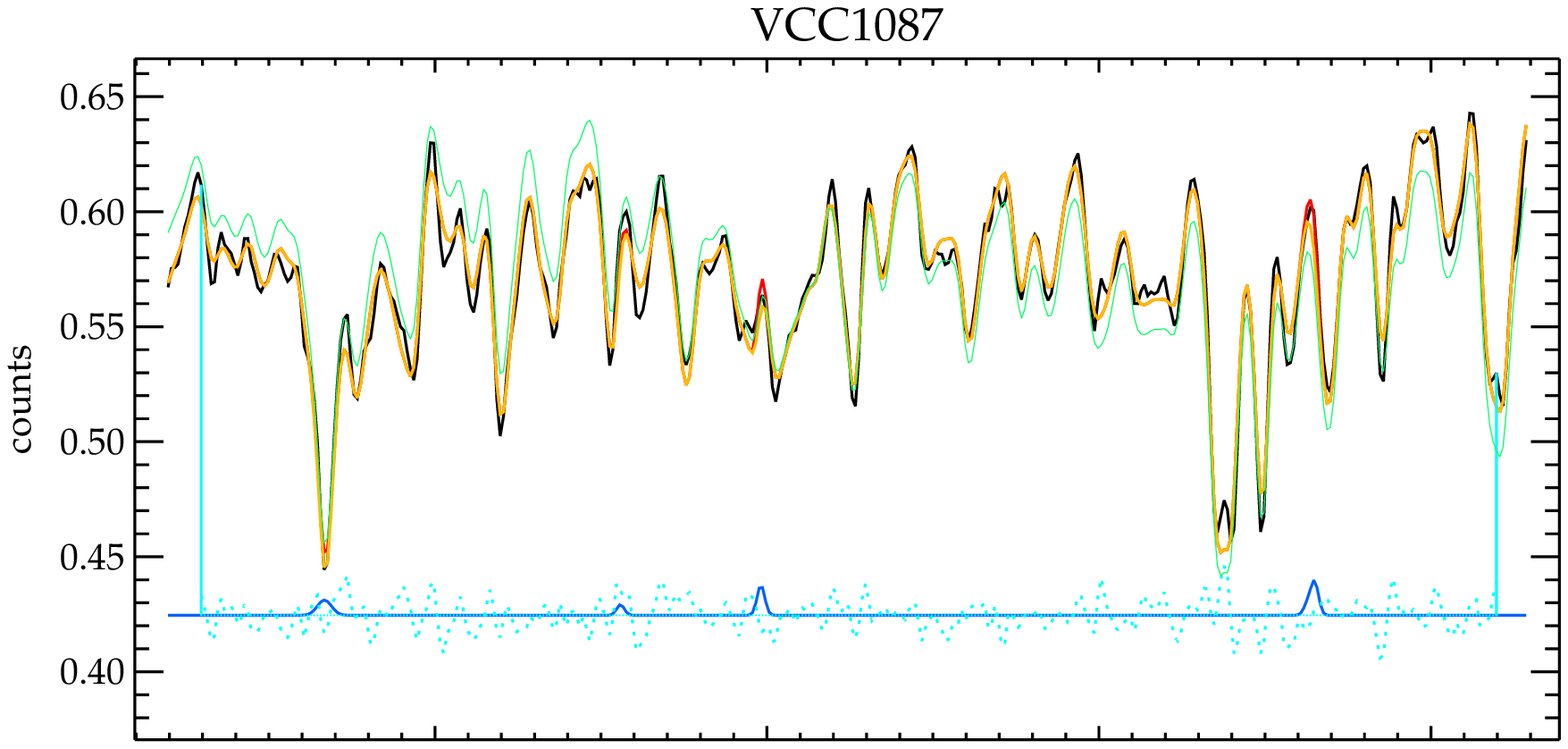}
\includegraphics[width=0.99\columnwidth]{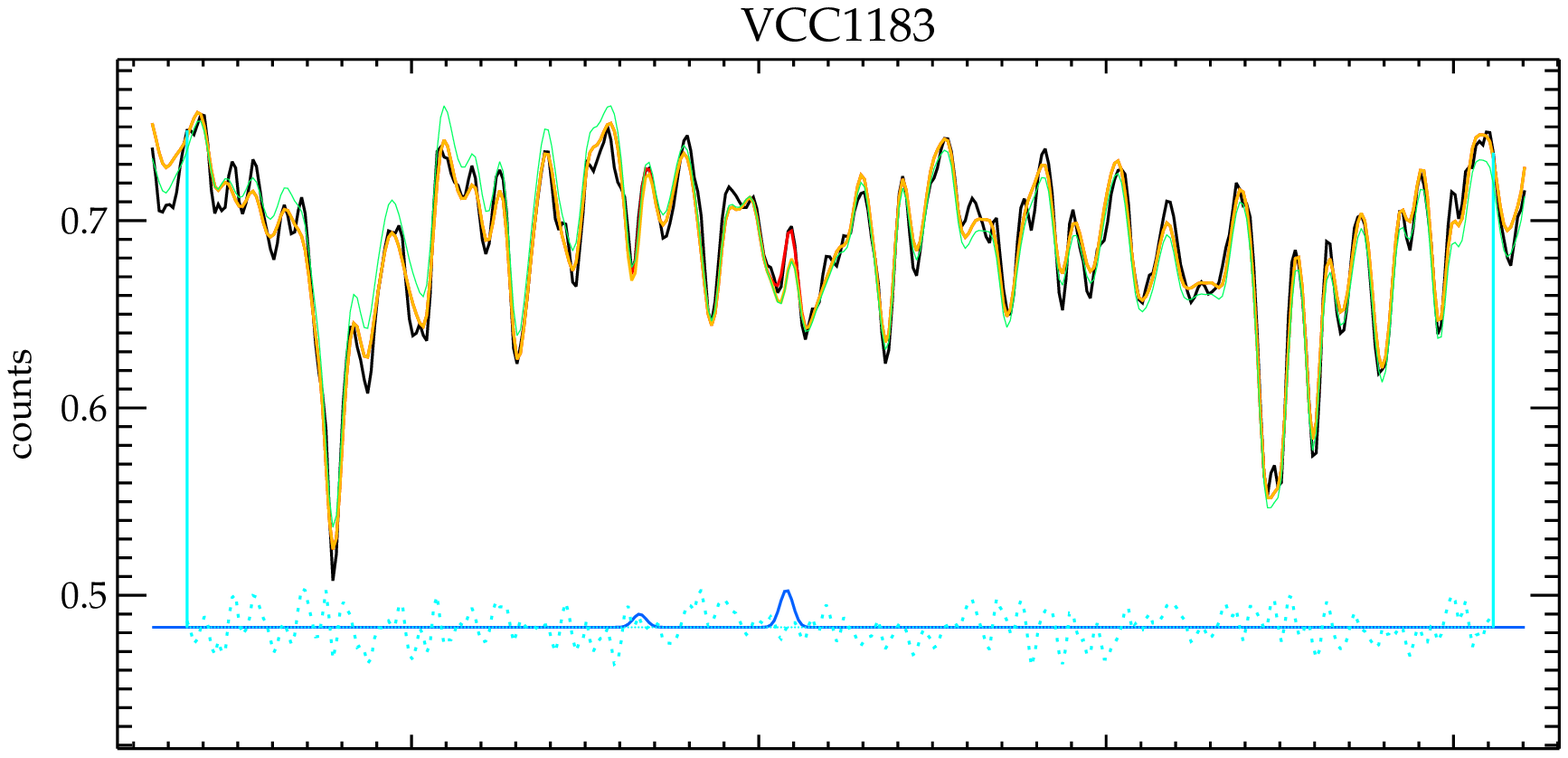}
\includegraphics[width=0.99\columnwidth]{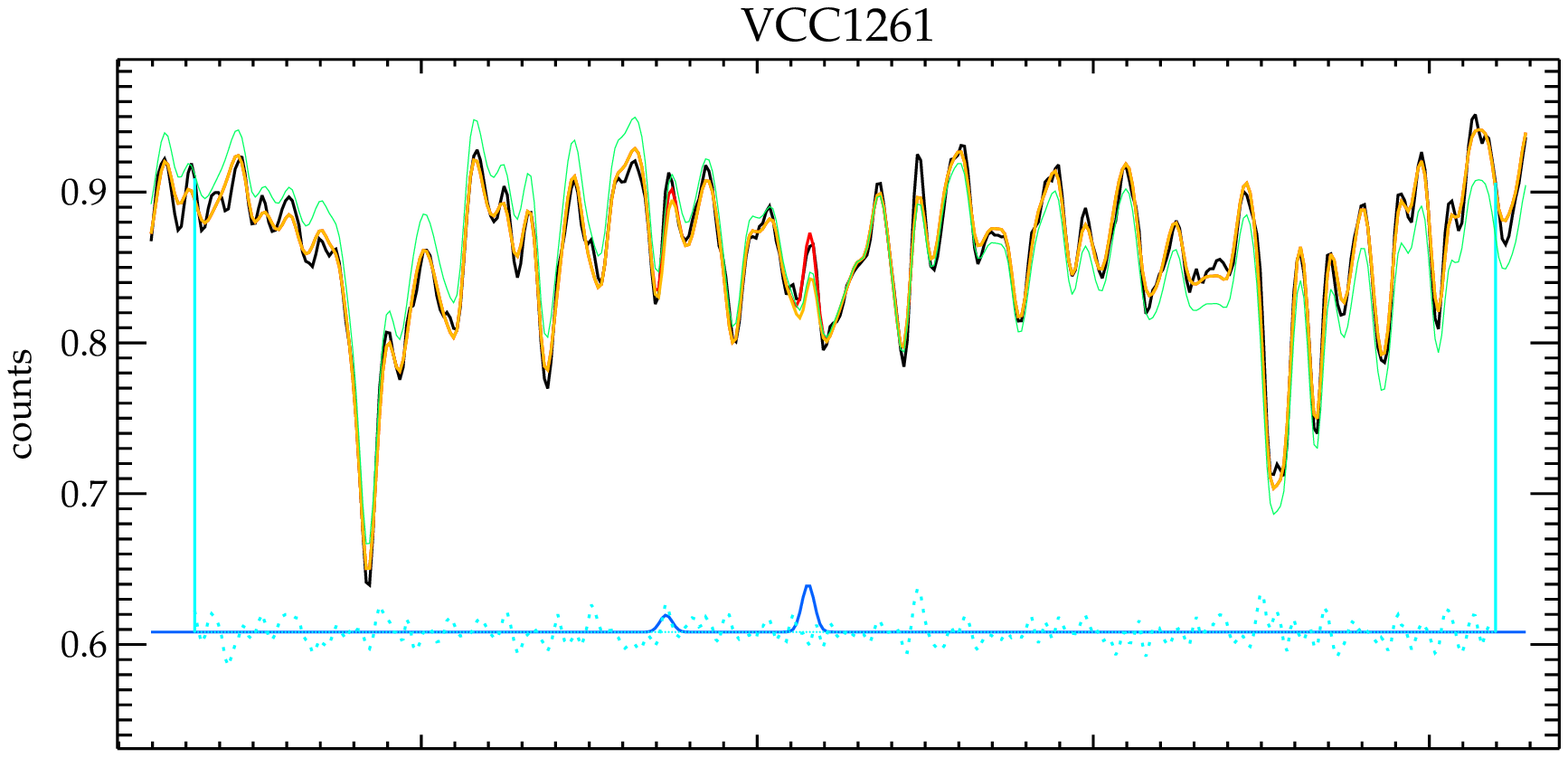}
\includegraphics[width=0.99\columnwidth]{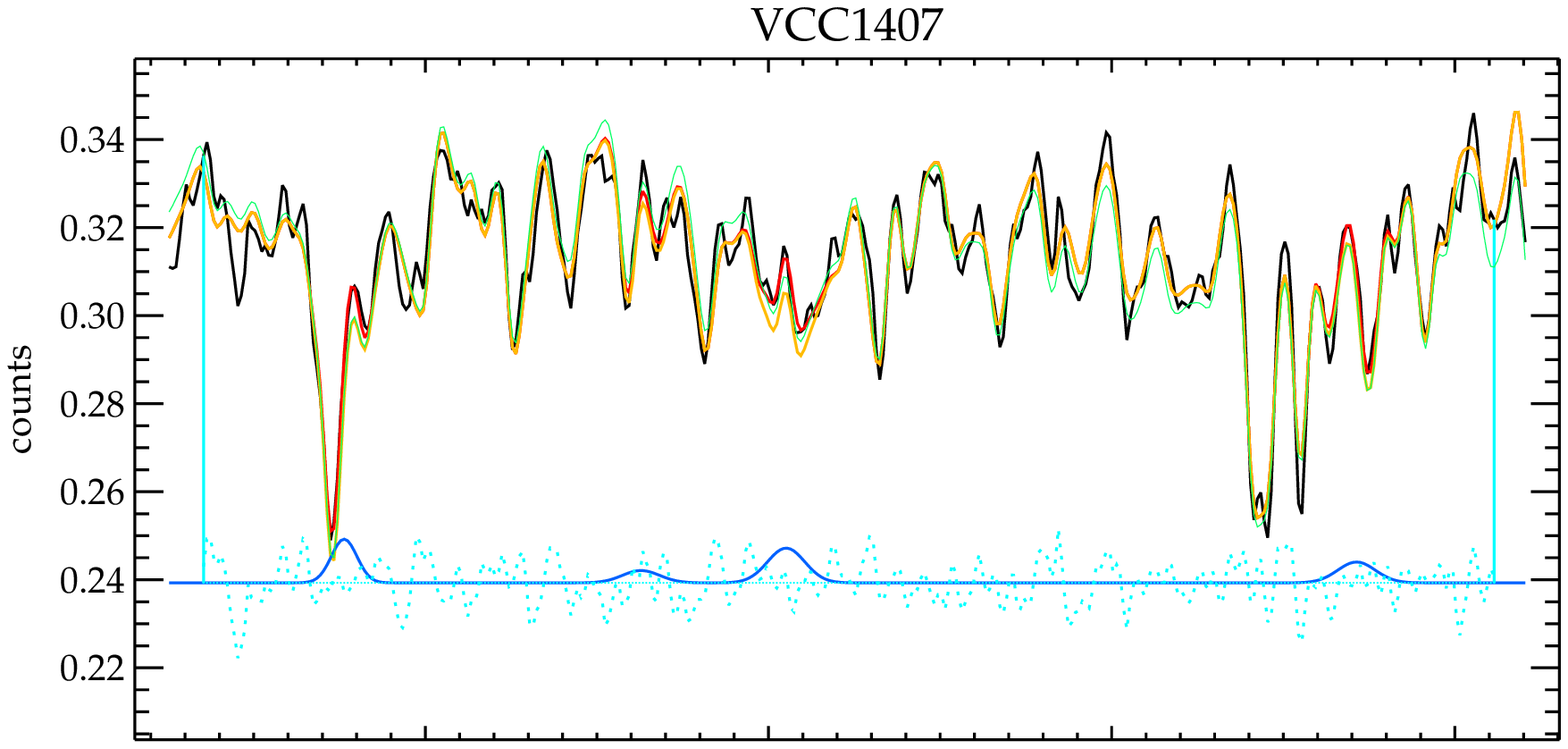}
\includegraphics[width=0.99\columnwidth]{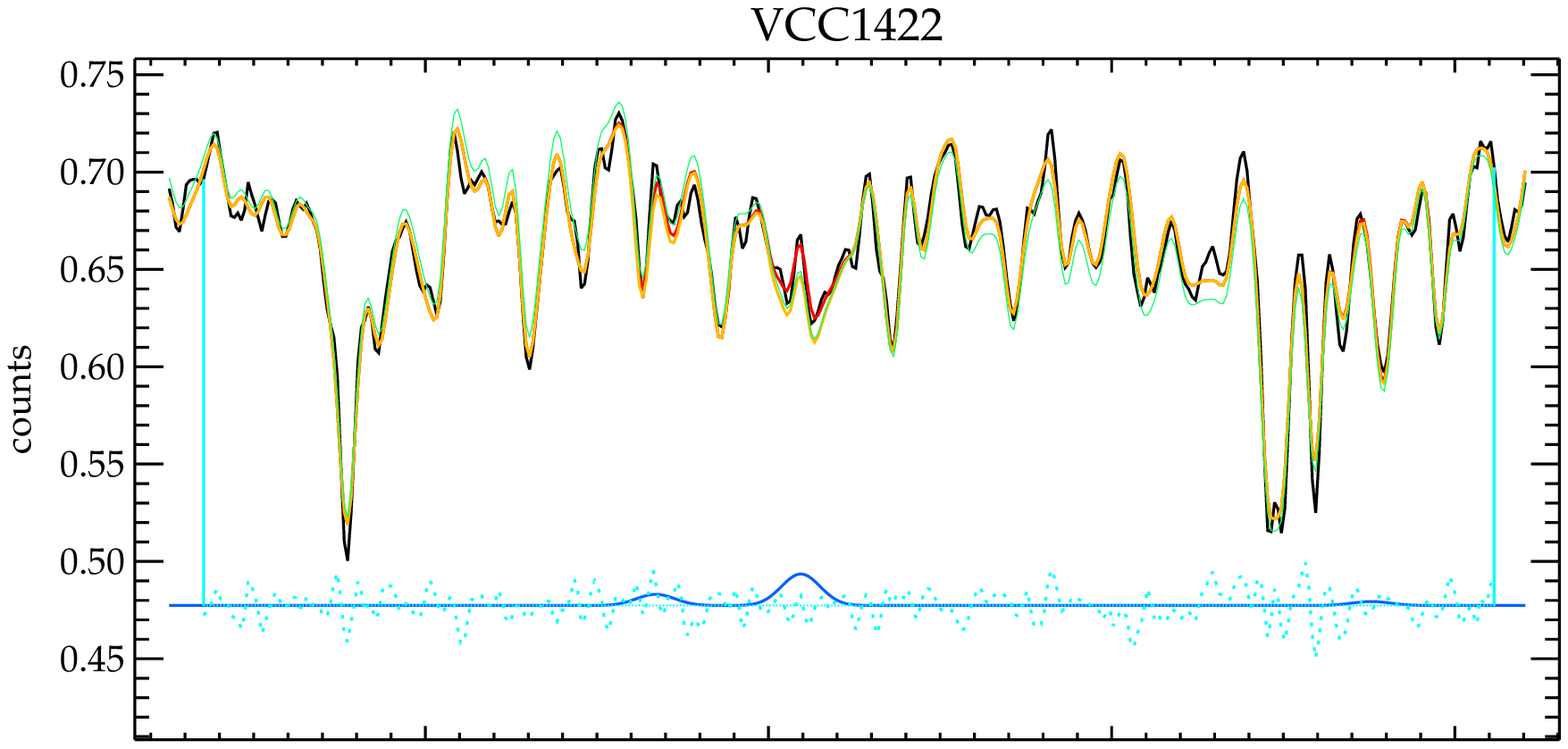}
\includegraphics[width=0.99\columnwidth]{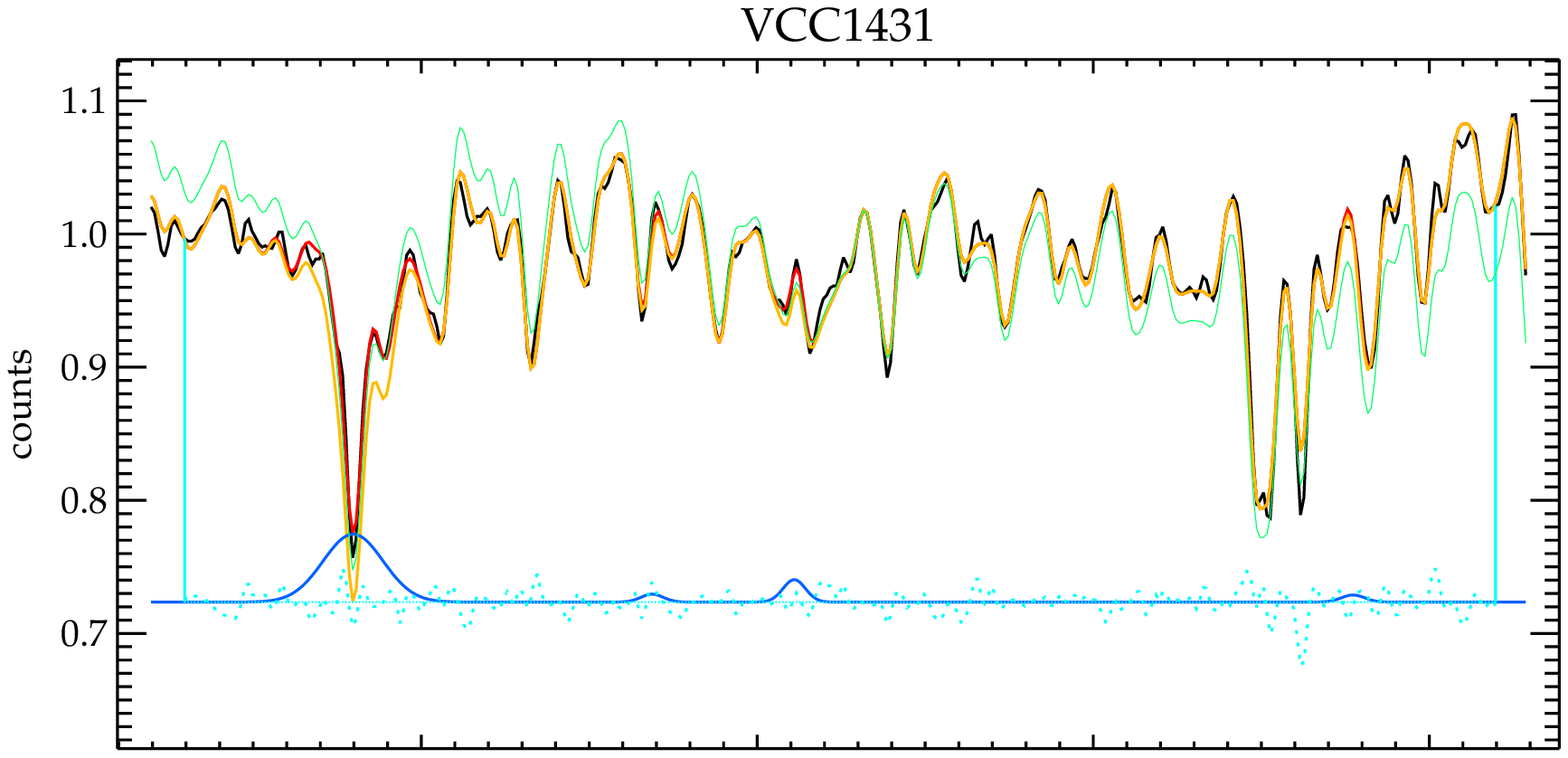}
\includegraphics[width=0.99\columnwidth]{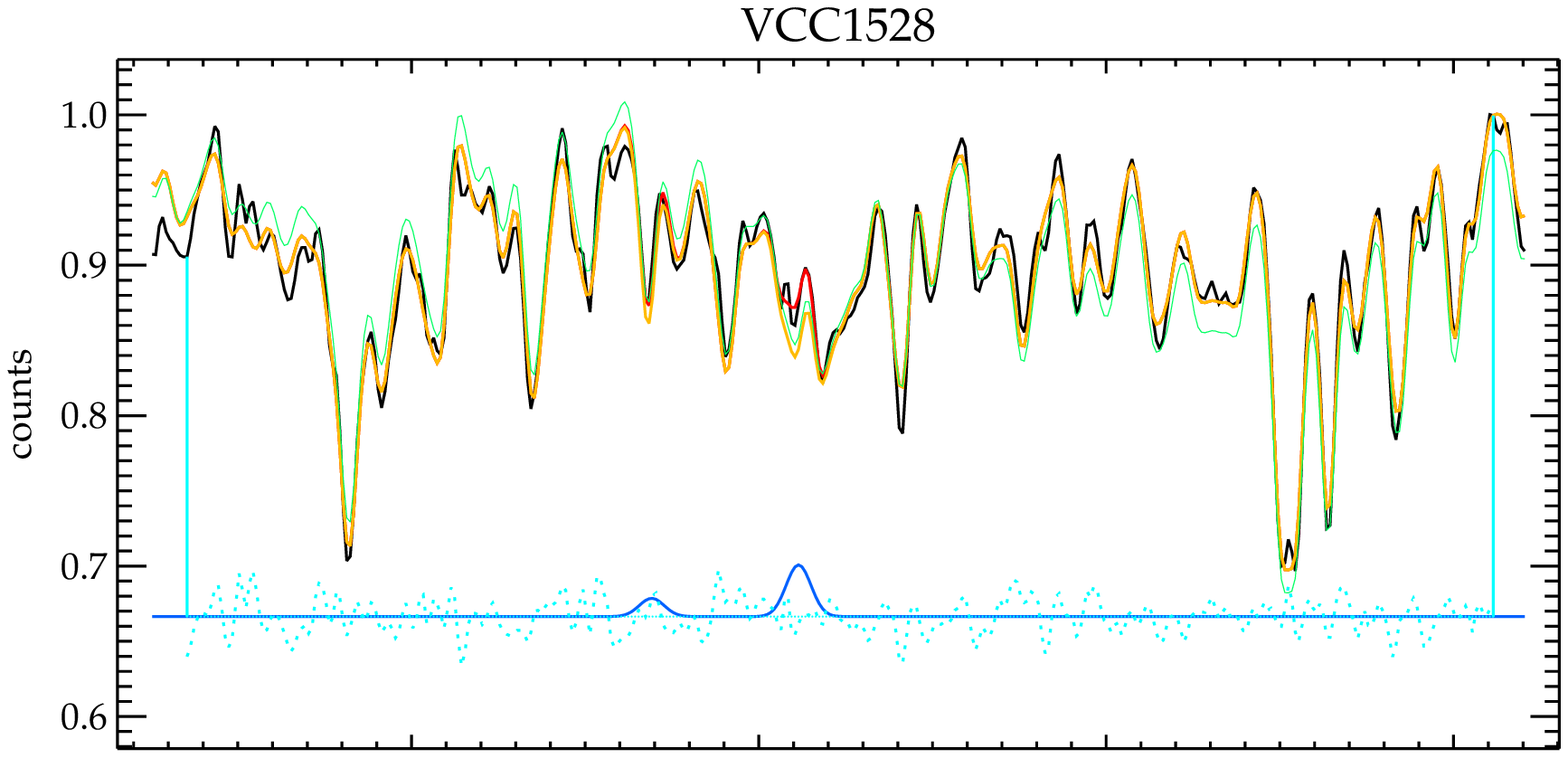}
\includegraphics[width=0.99\columnwidth]{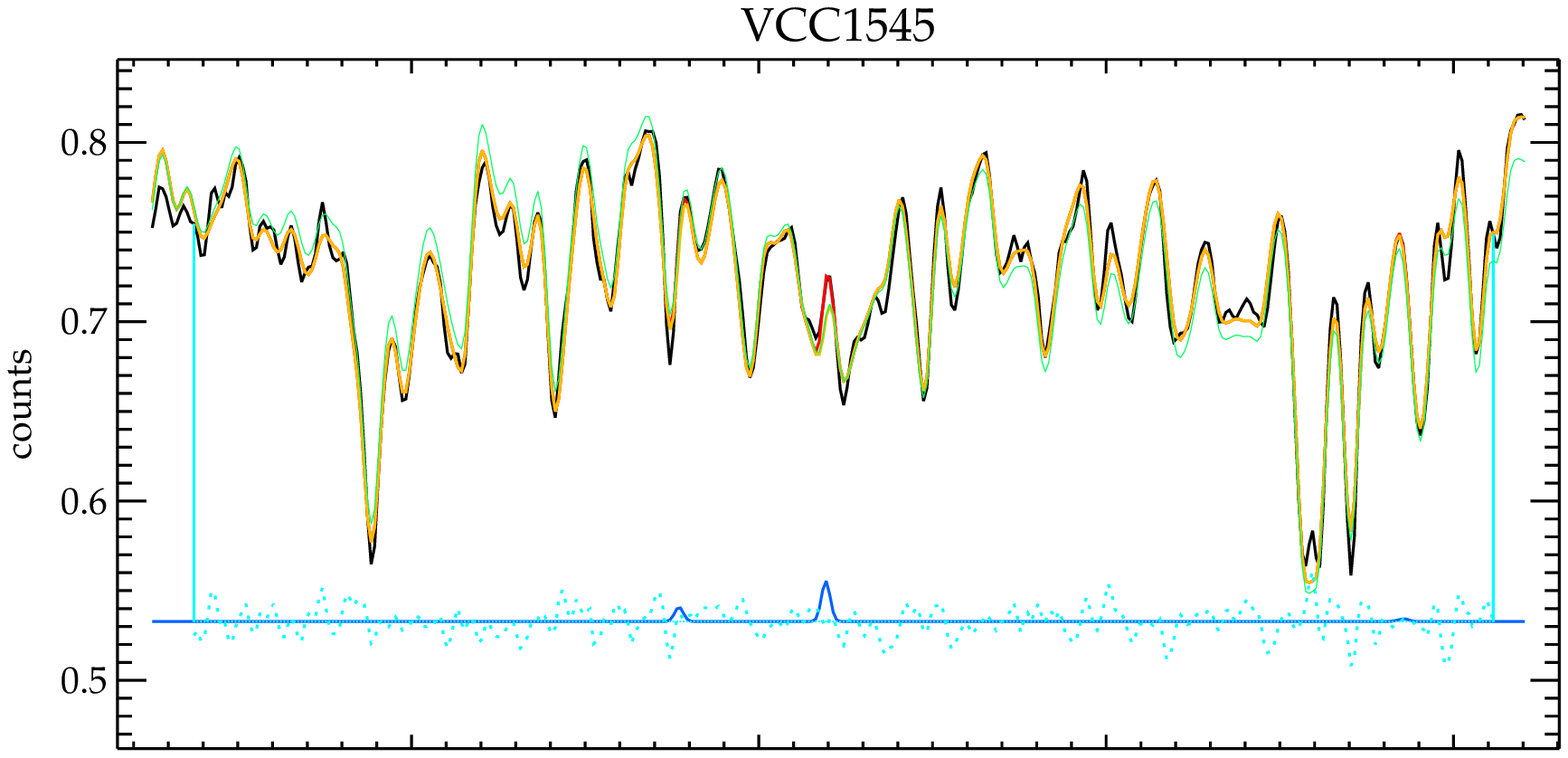}
\includegraphics[width=0.99\columnwidth]{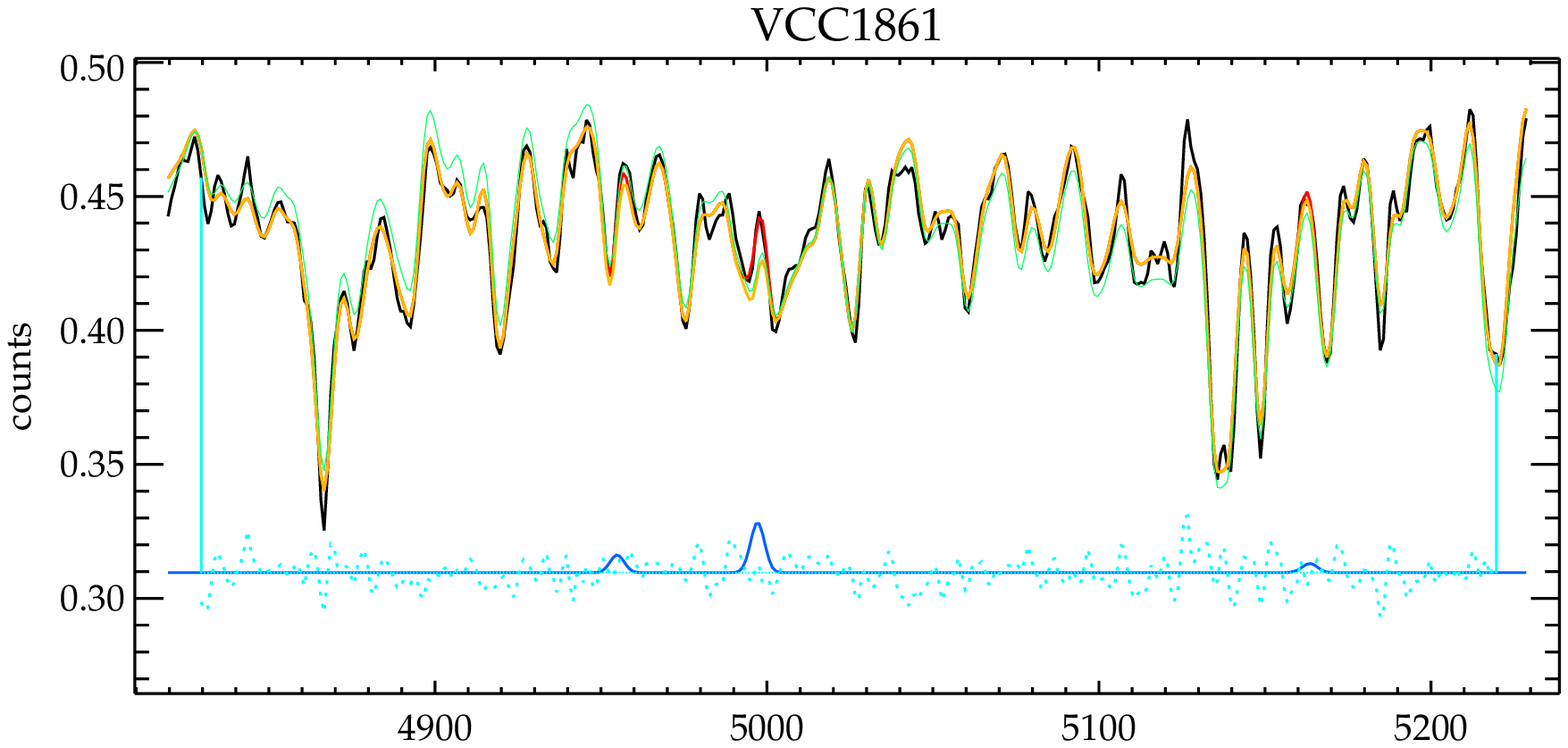}
\includegraphics[width=0.99\columnwidth]{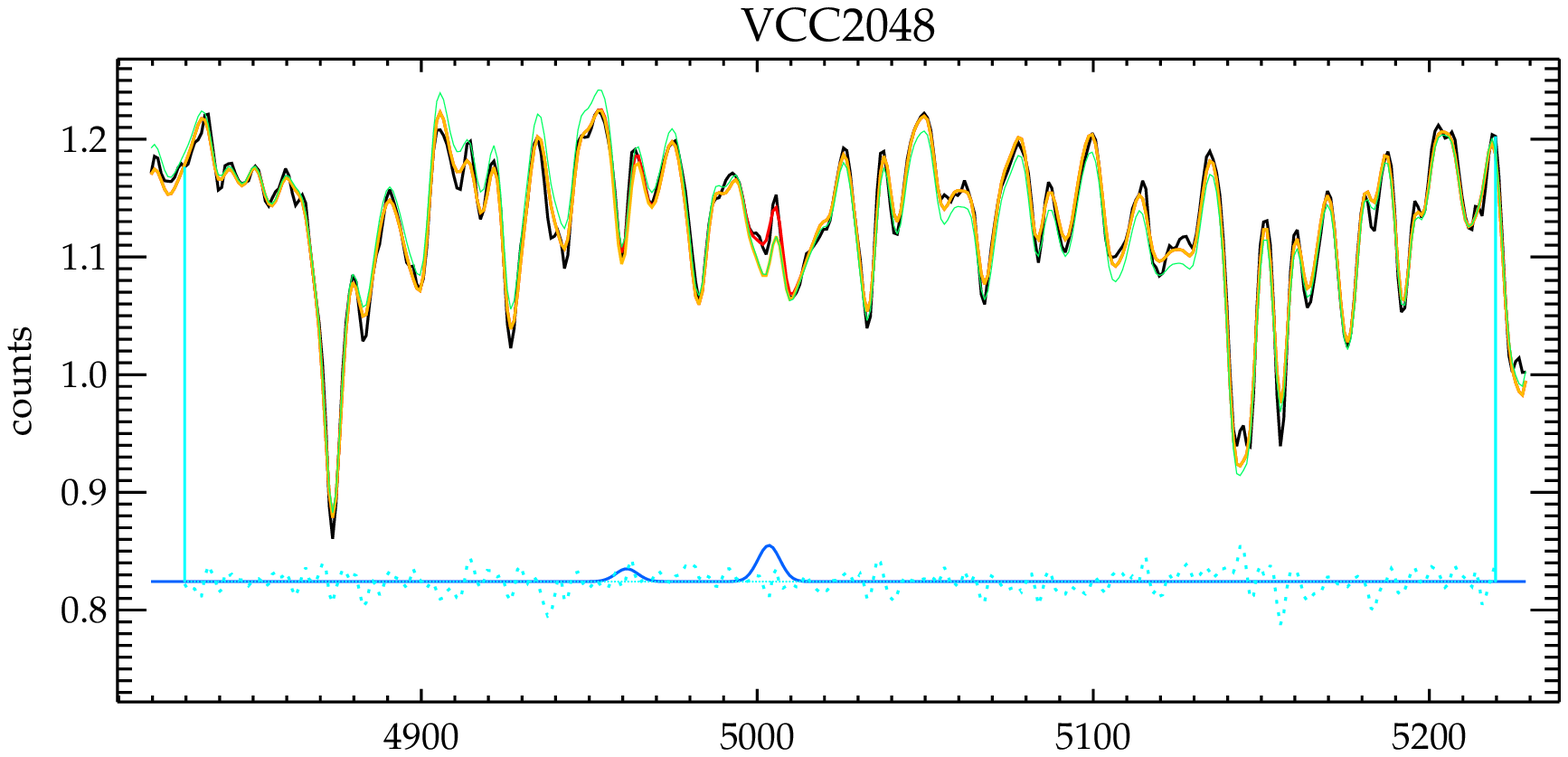}
\begin{scriptsize}
\hspace{12cm} $\lambda$ [\AA] \hspace{8cm} $\lambda$ [\AA] \hspace{-1cm} 
\end{scriptsize}
\caption{Continued.}
\label{example-spectra2}  
\end{figure*}

\section{Comparison of published {\atlas} SSPs and those obtained with the MILES models.}
\label{appendix-comparison}

\begin{figure*}
\centering
\includegraphics[width=1.99\columnwidth]{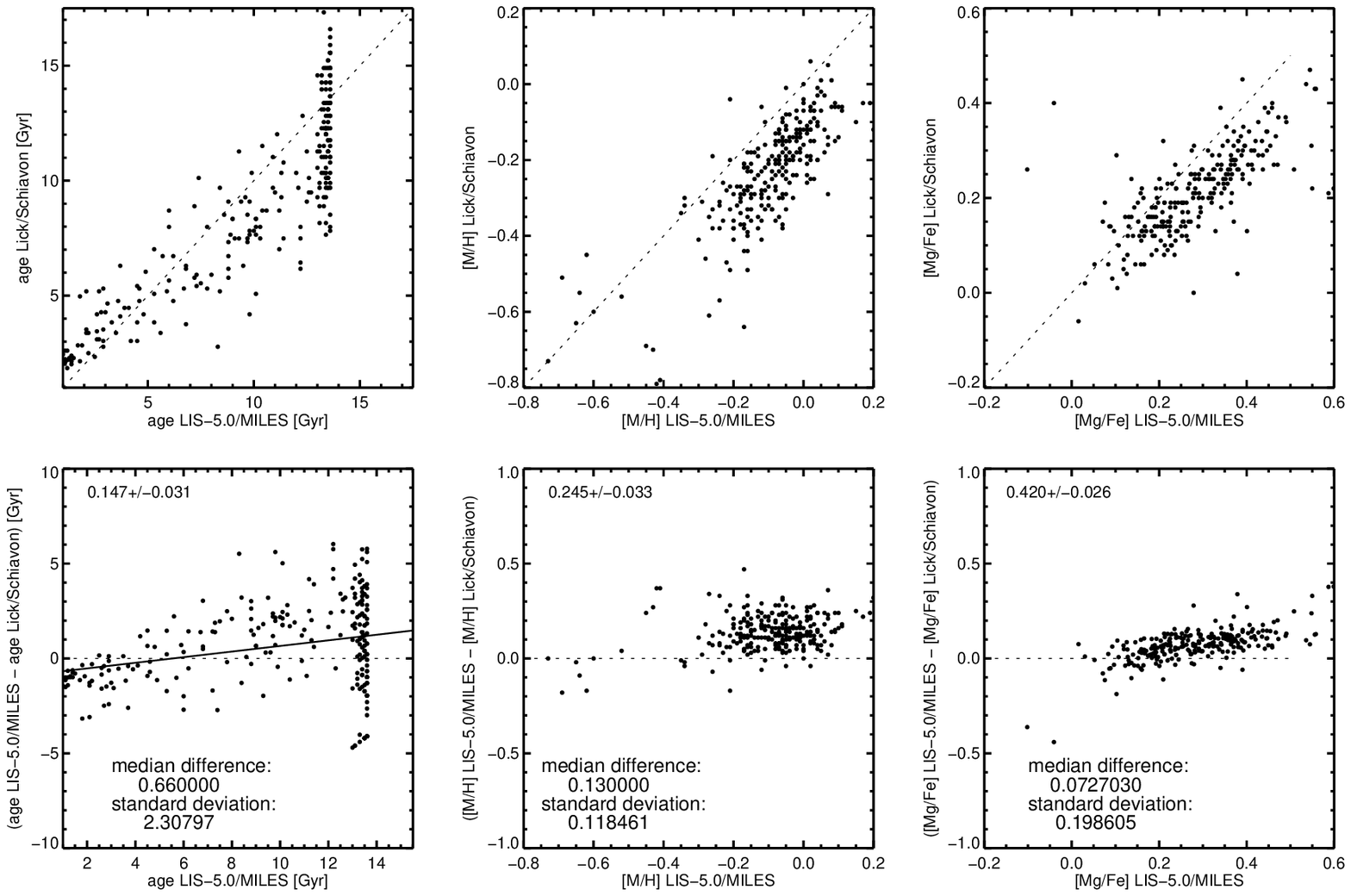}
\caption{Comparison of the {\atlas} SSP parameters derived with the use of MILES models of \protect\cite{vazdekis:2015} using \protect\cite{mcdermid:2015} line strengths transformed into the LIS-5.0 system vs. the published SSP values of \protect\cite{mcdermid:2015}, measured using a customized version of Schiavon models and based on the Lick system line strengths. The bottom row includes best straight-line fits (solid black lines) to the shown relations, together with the lines' slope values (quoted in the upper left corners of the respective panels).}
\label{miles-schiavon}  
\end{figure*}

In Fig.~\ref{miles-schiavon} we show the comparison of SSP ages, metallicities and abundance ratios for the {\atlas} sample coming from the LIS-5.0/MILES vs. Lick/Schiavon models, the latter published in \cite{mcdermid:2015}. 

The two sets of models differ in the allowed age range, with Schiavon going up to 17.8\,Gyr and MILES up to 14.0\,Gyr. The former cover lower {\hb} ranges than the latter, which was the reason \cite{kuntschner:2010} opted for them in their analysis of the SAURON project ETG sample, a sizable part of which occupies low-{\hb} regions. 

Overall, we see in the figure that the parameters derived using the two sets of models/indices agree reasonably well, with a tendency for a discrepancy slightly dependent on the derived value (more positive difference for older ages and higher [M/H] and [Mg/Fe]). Also, the MILES ages saturate at $\sim$\,14\,Gyr, which is a direct consequence of the allowed age range, as well as the fact that in the case of MILES models, the lowest-{\hb} region is affected by an age degeneracy, which translates into larger error bars for the oldest galaxies. The MILES values are on average offset from the Schiavon values by 0.66, 0.13, and 0.07 (with standard deviations of 2.31, 0.12 and 0.20), respectively. 

We conclude that the [MgFe50]' index, used for the calculation of total metallicity, is insensitive to the [Mg/Fe] ratio and that any shift in age between the two sets of models translates into a vertical shift on the {\hbo} vs. [MgFe50]' grid. The age/metallicity degeneracy is therefore responsible for the corresponding  shift in the metallicity estimate. This gives us confidence in the accuracy of the methods, in particular stellar population models, employed and the values derived for the dE sample. 

\section{Tabulated line-strength and SSP integrated values}
\label{app-a}

In Tables~\ref{resultstable} and~\ref{resultstable-ssp} we provide the 1\,{\re} integrated line-strength and SSP values, as well as SSP gradients, presented in the figures in the main part of the paper.

\begin{table*}
\caption{(1) Galaxy names as in Table 1, (2)-(6) integrated within 1\,{\re} stellar velocity dispersion {\sig} and $H\beta$, $H\beta_o$, Fe5015, and {\mgb} index values, (7)-(10) $H\beta$, $H\beta_o$, Fe5015, and {\mgb} index gradients.} 
\begin{threeparttable}
\centering
\begin{tabular}{|r|r|r|r|r|r|r|r|r|r|}
\hline
object    &$\sigma_{Re}$&$H\beta_{Re}$ &$H\beta o_{Re}$ &Fe50$_{Re}$ &Mgb$_{Re}$   &$\nabla H\beta$ &$\nabla H\beta_o$ &$\nabla$ Fe50 &$\nabla$ Mgb  \\
          &    (km/s)     &              &                &              &           &                &                &                &            \\
   (1)    &    (2)        &    (3)       &    (4)         &    (5)       &   (6)     &    (7)         &    (8)         &    (9)        &   (10)     \\
\hline
ID0650&        38.8$\pm$       5.9&       1.99$\pm$      0.31&       3.07$\pm$      0.35&       3.80$\pm$      0.49&       1.84$\pm$      0.24&     -0.30$\pm$      0.23&     -0.20$\pm$      0.18&     -1.07$\pm$      0.17&     -0.47$\pm$     0.11\\
ID0918&        78.8$\pm$       1.2&       1.98$\pm$      0.23&       3.09$\pm$      0.25&       4.49$\pm$      0.29&       2.66$\pm$      0.15&     -0.23$\pm$      0.16&     -0.09$\pm$      0.18&     -1.02$\pm$      0.10&     -1.03$\pm$     0.09\\
NGC3073&       39.5$\pm$       3.7&       4.10$\pm$      0.22&       5.65$\pm$      0.26&       3.25$\pm$      0.31&       1.55$\pm$      0.16&     -1.18$\pm$      0.20&     -2.16$\pm$      0.16&      0.41$\pm$      0.08&      0.44$\pm$     0.05\\
VCC0308&       35.6$\pm$       4.8&       2.32$\pm$      0.33&       3.49$\pm$      0.37&       3.64$\pm$      0.48&       1.76$\pm$      0.23&     -0.56$\pm$      0.26&     -0.75$\pm$      0.22&     -0.60$\pm$      0.13&      0.04$\pm$     0.07\\
VCC0523&       46.8$\pm$       3.6&       2.40$\pm$      0.19&       3.56$\pm$      0.20&       3.85$\pm$      0.23&       1.99$\pm$      0.13&     -0.13$\pm$      0.13&     -0.02$\pm$      0.12&     -0.57$\pm$      0.11&     -0.21$\pm$     0.03\\
VCC0543&       39.6$\pm$       4.9&       2.00$\pm$      0.19&       3.11$\pm$      0.18&       3.59$\pm$      0.18&       1.74$\pm$      0.09&      0.15$\pm$      0.20&      0.18$\pm$      0.13&     -0.74$\pm$      0.30&     -0.86$\pm$     0.11\\
VCC0856&       32.8$\pm$       4.0&       2.01$\pm$      0.21&       3.27$\pm$      0.23&       3.41$\pm$      0.25&       1.81$\pm$      0.14&     -0.25$\pm$      0.25&      0.11$\pm$      0.25&     -0.65$\pm$      0.21&     -0.20$\pm$     0.05\\
VCC0929&       63.1$\pm$       3.3&       2.03$\pm$      0.21&       3.07$\pm$      0.22&       4.06$\pm$      0.22&       2.45$\pm$      0.12&     -0.01$\pm$      0.16&     -0.07$\pm$      0.12&     -0.23$\pm$      0.12&     -0.22$\pm$     0.05\\
VCC1010&       59.3$\pm$       1.8&       1.97$\pm$      0.16&       3.09$\pm$      0.18&       4.40$\pm$      0.15&       2.76$\pm$      0.08&      0.05$\pm$      0.16&     -0.01$\pm$      0.15&     -0.60$\pm$      0.23&     -0.02$\pm$     0.08\\
VCC1036&       57.7$\pm$       2.9&       2.08$\pm$      0.23&       3.32$\pm$      0.26&       4.09$\pm$      0.32&       2.29$\pm$      0.15&     -0.30$\pm$      0.15&     -0.23$\pm$      0.12&     -1.15$\pm$      0.08&     -0.50$\pm$     0.07\\
VCC1087&       39.8$\pm$       4.0&       1.88$\pm$      0.19&       2.93$\pm$      0.22&       3.81$\pm$      0.21&       2.09$\pm$      0.12&     -0.11$\pm$      0.14&     -0.31$\pm$      0.16&     -0.51$\pm$      0.13&     -0.52$\pm$     0.11\\
VCC1183&       42.3$\pm$       3.2&       2.14$\pm$      0.17&       3.20$\pm$      0.17&       3.79$\pm$      0.16&       1.73$\pm$      0.08&     -0.25$\pm$      0.24&     -0.10$\pm$      0.24&     -0.69$\pm$      0.36&     -0.56$\pm$     0.20\\
VCC1261&       53.7$\pm$       3.0&       2.11$\pm$      0.22&       3.07$\pm$      0.23&       3.71$\pm$      0.26&       2.02$\pm$      0.14&     -0.39$\pm$      0.14&     -0.60$\pm$      0.23&     -0.78$\pm$      0.07&     -0.10$\pm$     0.02\\
VCC1407&       42.3$\pm$       6.0&       2.03$\pm$      0.19&       3.17$\pm$      0.24&       3.22$\pm$      0.22&       2.22$\pm$      0.12&     -0.33$\pm$      0.19&     -0.17$\pm$      0.30&     -0.75$\pm$      0.18&     -0.24$\pm$     0.17\\
VCC1422&       49.0$\pm$       3.4&       2.15$\pm$      0.17&       3.22$\pm$      0.18&       3.95$\pm$      0.16&       2.37$\pm$      0.08&     -0.17$\pm$      0.20&     -0.16$\pm$      0.24&     -0.11$\pm$      0.09&     -0.11$\pm$     0.07\\
VCC1431&       60.5$\pm$       3.8&       1.87$\pm$      0.30&       2.99$\pm$      0.38&       4.07$\pm$      0.49&       2.71$\pm$      0.26&     -0.18$\pm$      0.13&     -0.15$\pm$      0.16&     -0.39$\pm$      0.08&     -0.71$\pm$     0.14\\
VCC1528&       48.0$\pm$       3.3&       2.10$\pm$      0.17&       3.18$\pm$      0.18&       4.41$\pm$      0.15&       2.28$\pm$      0.09&     -0.07$\pm$      0.22&     -0.21$\pm$      0.17&     -0.69$\pm$      0.20&     -0.89$\pm$     0.25\\
VCC1545&       50.9$\pm$       3.5&       1.70$\pm$      0.18&       2.72$\pm$      0.20&       4.12$\pm$      0.16&       2.39$\pm$      0.09&     -0.33$\pm$      0.22&     -0.06$\pm$      0.28&     -0.97$\pm$      0.18&     -0.84$\pm$     0.16\\
VCC1861&       30.0$\pm$       3.7&       2.15$\pm$      0.24&       3.41$\pm$      0.27&       4.06$\pm$      0.31&       2.52$\pm$      0.17&     -0.01$\pm$      0.18&     -0.19$\pm$      0.13&     -0.82$\pm$      0.20&     -0.19$\pm$     0.10\\
VCC2048&       45.0$\pm$       4.2&       2.14$\pm$      0.27&       3.29$\pm$      0.31&       3.63$\pm$      0.34&       1.80$\pm$      0.18&     -0.23$\pm$      0.12&     -0.16$\pm$      0.14&     -0.26$\pm$      0.04&     -0.21$\pm$     0.01\\
\hline
\end{tabular}
\label{resultstable}
\end{threeparttable}
\end{table*}

\begin{table*}
\caption{SSP ages, metallicities, abundance ratios and their gradients derived from line-strength measurements, together with measurement uncertainties (except for ages and metallicities where lower and upper value limits are provided instead).}
\begin{threeparttable}
\centering
\begin{tabular}{|r|r|r|r|r|r|r|}
\hline
object    &age$_{Re}$    &[M/H]$_{Re}$ &$[Mg/Fe]_{Re}$  &$\nabla$\,age   &$\nabla$\,[M/H]&$\nabla [Mg/Fe]$ \\
          &              &              &                &                &                &                 \\
   (1)    &    (2)       &    (3)       &   (4)          &    (5)         &   (6)          &   (7)           \\
\hline
ID0650 &        12.55$^{       9.68}_{       14.00}$ &     -0.67$\pm$     0.08 &      0.09$\pm$      0.14 &      0.10$\pm$      0.22&     -0.35$\pm$      0.11&      0.28$\pm$      0.20\\
\rule{0pt}{3ex}
ID0918 &        12.15$^{       10.54}_{       14.00}$ &     -0.37$\pm$     0.05 &      0.06$\pm$     0.08 &    0.01$\pm$      0.15&     -0.40$\pm$     0.08&     0.01$\pm$      0.15\\
\rule{0pt}{3ex}
NGC3073 &        1.25$^{       1.18}_{       1.32}$ &     -0.43$\pm$     0.09 &      0.06$\pm$      0.19 &      0.32$\pm$      0.15&     -0.24$\pm$      0.12&      0.25$\pm$      0.19\\
\rule{0pt}{3ex}
VCC0308 &        6.65$^{       5.32}_{       14.00}$ &     -0.65$\pm$     0.10 &       0.14$\pm$      0.15 &      0.49$\pm$      0.25&     -0.32$\pm$      0.11&      0.27$\pm$      0.20\\
\rule{0pt}{3ex}
VCC0523 &        6.25$^{       5.76}_{       6.88}$ &     -0.54$\pm$     0.05 &       0.12$\pm$     0.07 &     0.04$\pm$      0.28&     -0.16$\pm$     0.09&      0.10$\pm$      0.16\\
\rule{0pt}{3ex}
VCC0543 &        10.85$^{       8.82}_{       14.00}$ &     -0.72$\pm$     0.03 &       0.12$\pm$     0.05 &    -0.05$\pm$      0.16&     -0.25$\pm$     0.09&   0.00$\pm$      0.18\\
\rule{0pt}{3ex}
VCC0856 &        10.55$^{       8.45}_{       14.00}$ &     -0.75$\pm$     0.05 &       0.21$\pm$     0.07 &     0.08$\pm$      0.31&     -0.20$\pm$      0.15&      0.15$\pm$      0.28\\
\rule{0pt}{3ex}
VCC0929 &        12.75$^{       11.21}_{       14.00}$ &     -0.51$\pm$     0.04 &       0.18$\pm$     0.05 &     0.05$\pm$      0.16&    -0.08$\pm$     0.07&     0.01$\pm$      0.14\\
\rule{0pt}{3ex}
VCC1010 &        12.45$^{       11.33}_{       14.00}$ &     -0.38$\pm$     0.03 &       0.13$\pm$     0.03 &     0.05$\pm$      0.18&     -0.13$\pm$     0.09&      0.13$\pm$      0.18\\
\rule{0pt}{3ex}
VCC1036 &        8.75$^{       7.49}_{       14.00}$ &     -0.49$\pm$     0.06 &       0.12$\pm$      0.10 &      0.33$\pm$      0.18&     -0.41$\pm$     0.07&      0.12$\pm$      0.12\\
\rule{0pt}{3ex}
VCC1087 &        11.75$^{       9.44}_{       14.00}$ &     -0.62$\pm$     0.04 &       0.19$\pm$     0.05 &      0.13$\pm$      0.17&     -0.21$\pm$     0.09&     0.04$\pm$      0.14\\
\rule{0pt}{3ex}
VCC1183 &        10.65$^{       8.44}_{       14.00}$ &     -0.68$\pm$     0.03 &      0.06$\pm$     0.04 &     0.06$\pm$      0.30&     -0.21$\pm$      0.15&     0.04$\pm$      0.26\\
\rule{0pt}{3ex}
VCC1261 &        12.85$^{       9.77}_{       14.00}$ &     -0.66$\pm$     0.05 &       0.21$\pm$     0.07 &      0.45$\pm$      0.19&     -0.40$\pm$     0.09&      0.24$\pm$      0.14\\
\rule{0pt}{3ex}
VCC1407 &        10.95$^{       9.41}_{       14.00}$ &     -0.73$\pm$     0.04 &       0.47$\pm$     0.06 &     0.02$\pm$      0.25&     -0.24$\pm$      0.18&      0.22$\pm$      0.27\\
\rule{0pt}{3ex}
VCC1422 &        10.35$^{       9.44}_{       14.00}$ &     -0.53$\pm$     0.03 &       0.22$\pm$     0.05 &     0.09$\pm$      0.21&    -0.06$\pm$      0.15&     0.01$\pm$      0.24\\
\rule{0pt}{3ex}
VCC1431 &        13.15$^{       10.49}_{       14.00}$ &     -0.47$\pm$     0.08 &       0.25$\pm$      0.12 &   0.00$\pm$      0.13&     -0.13$\pm$      0.10&    -0.04$\pm$      0.16\\
\rule{0pt}{3ex}
VCC1528 &        10.95$^{       10.11}_{       14.00}$ &     -0.44$\pm$     0.03 &     -0.02$\pm$     0.04 &      0.16$\pm$      0.30&     -0.35$\pm$      0.14&    -0.07$\pm$      0.25\\
\rule{0pt}{3ex}
VCC1545 &        12.25$^{       9.97}_{       14.00}$ &     -0.50$\pm$     0.03 &       0.13$\pm$     0.04 &    -0.02$\pm$     0.08&     -0.29$\pm$      0.10&     0.08$\pm$      0.20\\
\rule{0pt}{3ex}
VCC1861 &        7.75$^{       6.49}_{       14.00}$ &     -0.44$\pm$     0.06 &       0.21$\pm$     0.09 &      0.13$\pm$      0.28&     -0.22$\pm$      0.12&      0.19$\pm$      0.19\\
\rule{0pt}{3ex}
VCC2048 &        9.45$^{       7.56}_{       14.00}$ &     -0.68$\pm$     0.06 &       0.15$\pm$      0.11 &      0.10$\pm$      0.16&     -0.14$\pm$     0.07&     0.08$\pm$      0.12\\
\hline
\end{tabular}
\label{resultstable-ssp}
\end{threeparttable}
\end{table*}

\section{Line strength and population profiles}
\label{app-pop_profiles}

In Figure~\ref{ls_profiles1} we show line-strength profiles and in  Figure~\ref{population_gradients1} age, [M/H] and [Mg/Fe] profiles, on which gradient calculations presented in this paper are based.

\begin{figure*}
\begin{fullpage}
\centering
\includegraphics[width=1.99\columnwidth]{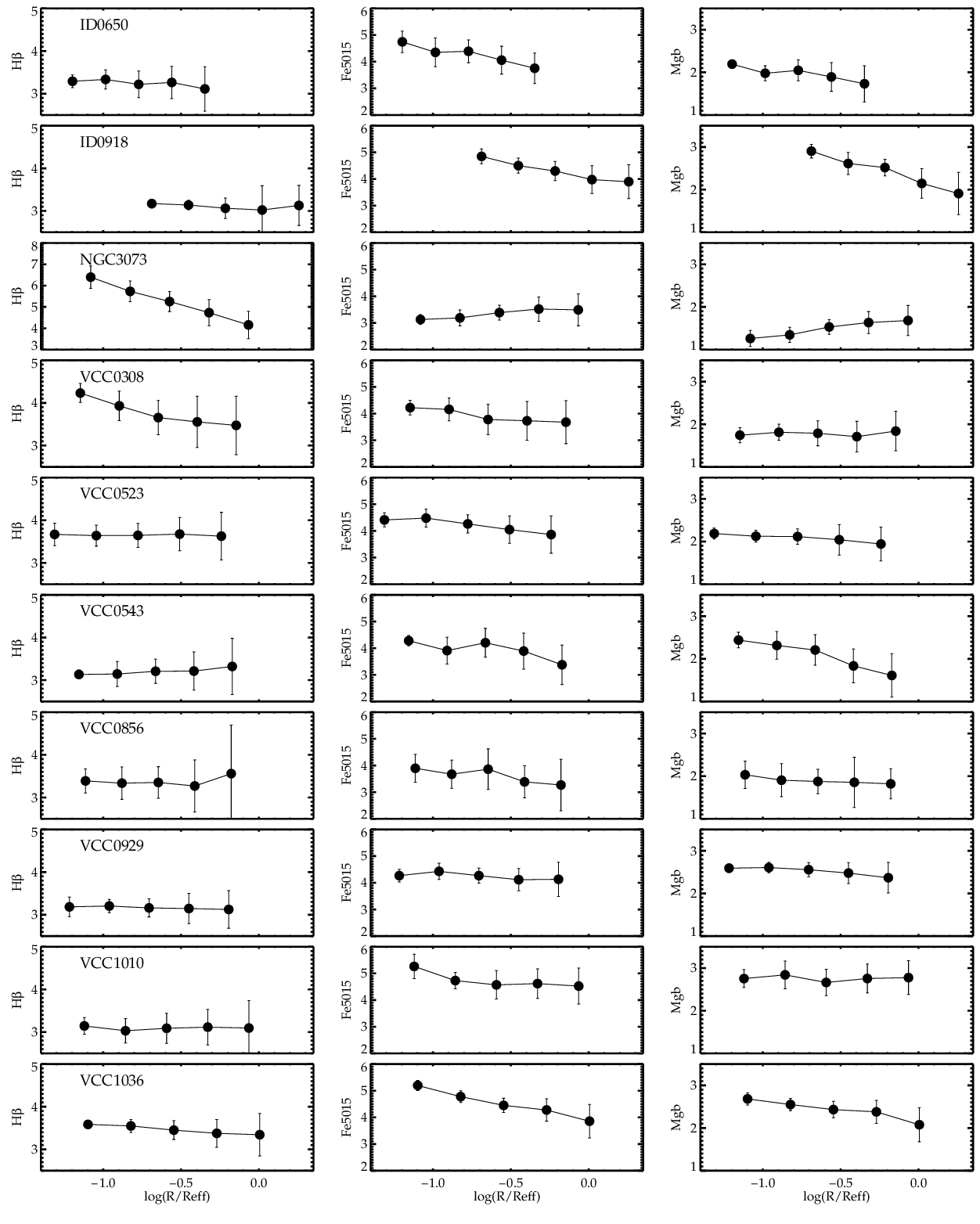}
\caption{{\hbo}, Fe5015 and Mg$b$ profiles.}
\label{ls_profiles1}  
\end{fullpage}
\end{figure*}

\addtocounter{figure}{-1}
\addtocounter{subfigure}{1}

\begin{figure*}
\begin{fullpage}
\centering
\includegraphics[width=1.99\columnwidth]{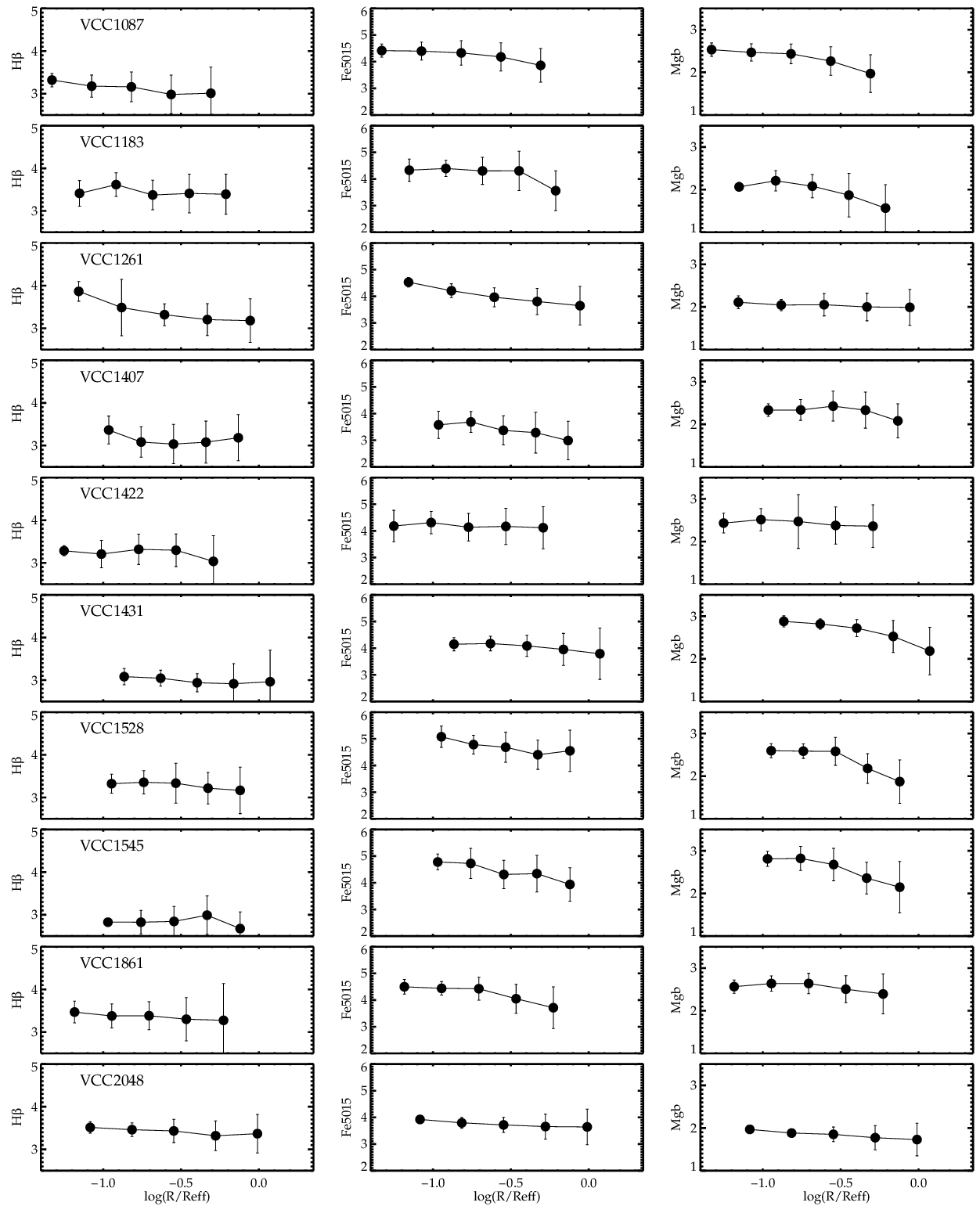}
\caption{Continued.}
\label{ls_profiles2}  
\end{fullpage}
\end{figure*}

\begin{figure*}
\begin{fullpage}
\centering
\includegraphics[width=1.99\columnwidth]{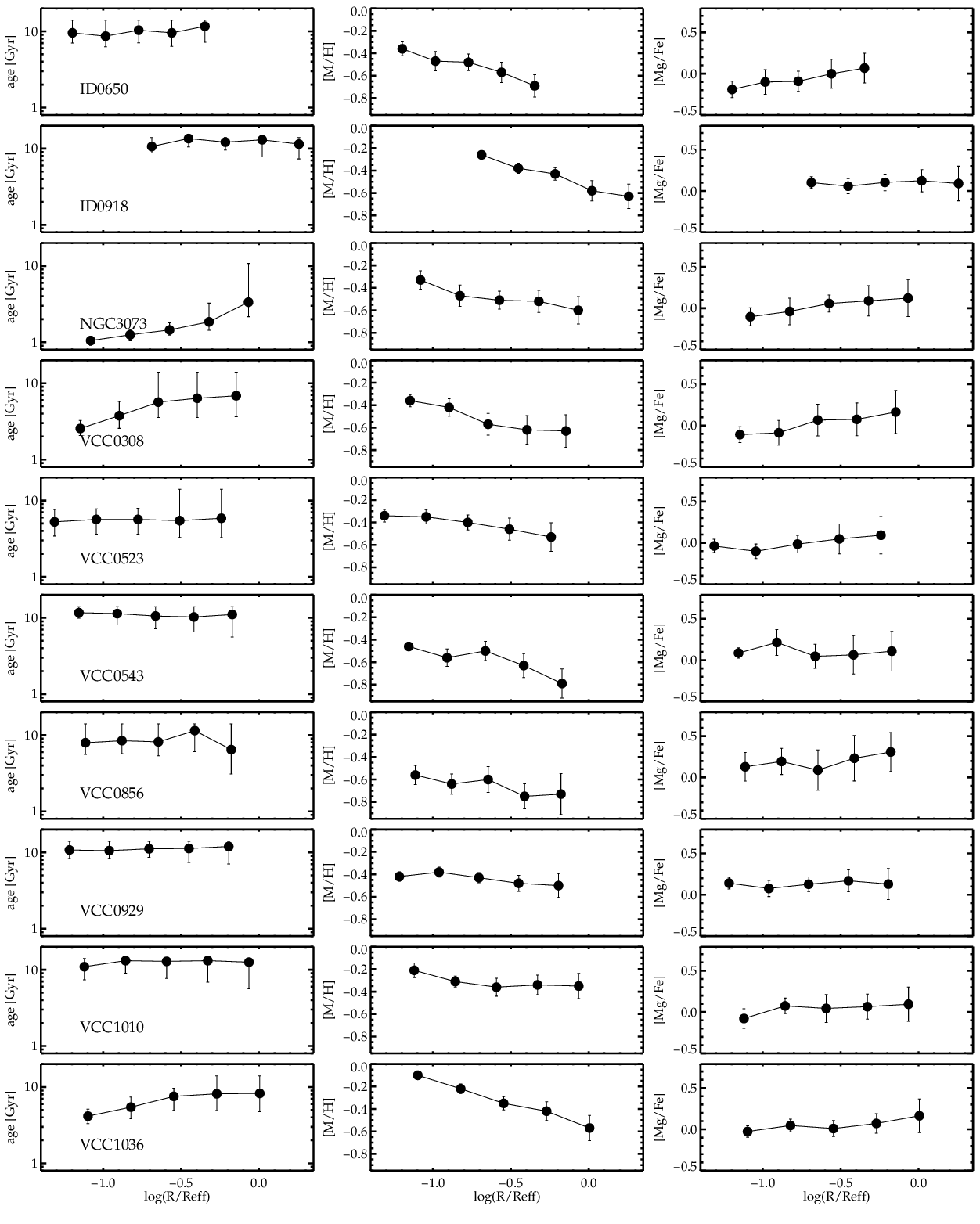}
\caption{Age, [M/H] and [Mg/Fe] profiles.}
\label{population_gradients1}  
\end{fullpage}
\end{figure*}

\addtocounter{figure}{-1}
\addtocounter{subfigure}{1}

\begin{figure*}
\begin{fullpage}
\centering
\includegraphics[width=1.99\columnwidth]{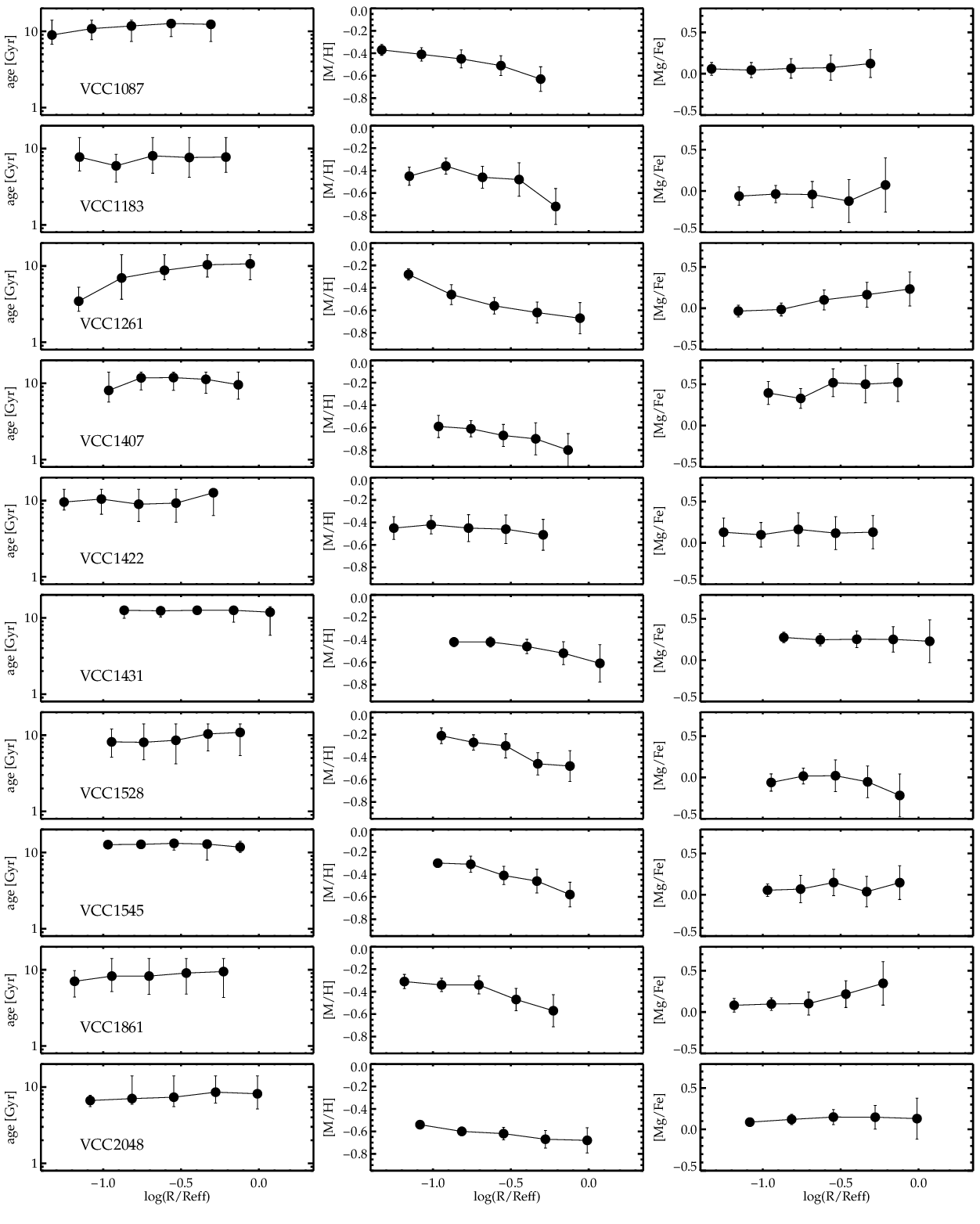}
\caption{Continued.}
\label{population_gradients2}  
\end{fullpage}
\end{figure*}

\section{Line strength-$\sigma$ scaling relations.}
\label{app-lines}

\begin{figure*}
\centering
\includegraphics[width=1.99\columnwidth]{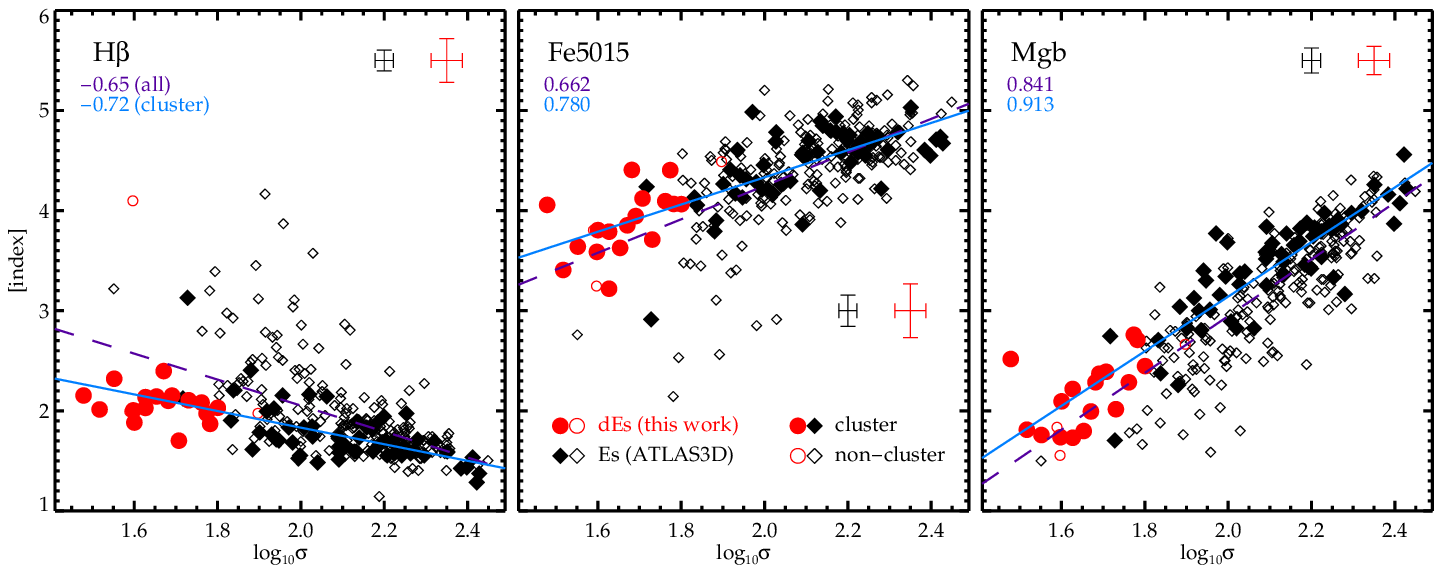}
\includegraphics[width=1.99\columnwidth]{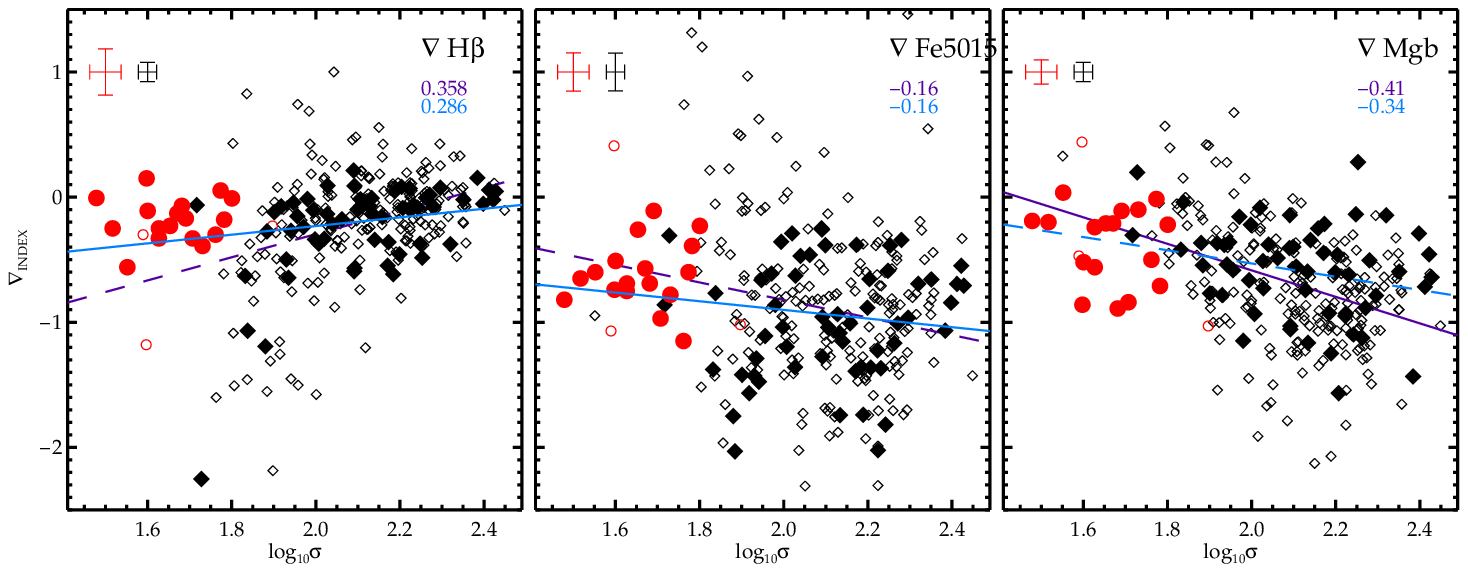}

\caption{Line-strength scaling relations for the combined sample of our dwarf early-type galaxies and massive early-types from the {\atlas} survey (with the latter values transformed to LIS-5.0 from those given in \protect\citealt{mcdermid:2015} and Kuntschner et al., in prep.). \underline{Top}: integrated line-strength values as a function of~{\sig}. \underline{Bottom}: line-strength gradients as a function of {\sig}. In each panel we provide a Spearman correlation coefficient value for a given relation, for cluster (blue, bottom) as well as all objects (violet, top), in order to show its statistical significance. Best-fit lines to the combined cluster-only as well as the entire sample (blue solid and violet dashed lines, respectively) are also shown. Average errors for each sample are depicted with crosses.}

\label{gradients_integrated_relation}  
\end{figure*}

To complement the SSP relations shown in the main part of the paper, we include here line-strength\,--\,{\sig} and line-strength gradients\,--\,{\sig}relations.

\label{lastpage}

\end{document}